\documentclass[preprint,superscriptaddress,floatfix]{revtex4-1}  % chktex 8

\usepackage{preamble}

\usepackage{mathrsfs}
\usepackage{amsmath}
\usepackage{amssymb}
\usepackage{graphicx}

\usepackage{epstopdf}
\usepackage{grffile}
\usepackage{esint}
\usepackage{textcomp}
\usepackage{url}
\usepackage[english]{babel}
\usepackage{todonotes}
\usepackage{makecell}

\usepackage[bookmarksnumbered=true,unicode=true]{hyperref}

\graphicspath{{graphics/}}  % comment this out because JES flattens graphics directory

\begin{document}

% \title{Linear stability analysis of transient electrodeposition in charged porous media}
\title{Linear stability analysis of transient electrodeposition in charged porous media: suppression of dendritic growth by surface conduction}

\date{\today}

\author{Edwin Khoo}
\affiliation{Department of Chemical Engineering, Massachusetts Institute of Technology, Cambridge, Massachusetts 02139, USA}

\author{Hongbo Zhao}
\affiliation{Department of Chemical Engineering, Massachusetts Institute of Technology, Cambridge, Massachusetts 02139, USA}

\author{Martin Z. Bazant}
\email[Corresponding author: ]{bazant@mit.edu}
\affiliation{Department of Chemical Engineering, Massachusetts Institute of Technology, Cambridge, Massachusetts 02139, USA}
\affiliation{Department of Mathematics, Massachusetts Institute of Technology, Cambridge, Massachusetts 02139, USA}

\begin{abstract}
  We study the linear stability of transient electrodeposition in a charged random porous medium, whose pore surface charges can be of any sign, flanked by a pair of planar metal electrodes. Discretization of the linear stability problem results in a generalized eigenvalue problem for the dispersion relation that is solved numerically, which agrees well with the analytical approximation obtained from a boundary layer analysis valid at high wavenumbers. Under galvanostatic conditions in which an overlimiting current is applied, in the classical case of zero surface charges, the electric field at the cathode diverges at Sand's time due to electrolyte depletion. The same phenomenon happens for positive charges but earlier than Sand's time. However, negative charges allow the system to sustain an overlimiting current via surface conduction past Sand's time, keeping the electric field bounded. Therefore, at Sand's time, negative charges greatly reduce surface instabilities and suppress dendritic growth, while zero and positive charges magnify them. We compare theoretical predictions for overall surface stabilization with published experimental data for copper electrodeposition in cellulose nitrate membranes and demonstrate good agreement between theory and experiment. We also apply the stability analysis to how crystal grain size varies with duty cycle during pulse electroplating.
\end{abstract}

\maketitle

\section{Introduction}

Linear stability analysis is routinely applied to nonlinear systems to study how the onset of instability is related to system parameters and to provide physical insights on the conditions and early dynamics of pattern formation~\cite{langer_instabilities_1980,kessler_pattern_1988,cross_pattern_1993}. Some examples in hydrodynamics include the Orr-Sommerfeld equation that predicts the dependence on Reynolds number of the transition from laminar flow to turbulent flow~\cite{orr_stability_1907,orr_stability_1907-1,sommerfeld_beitrag_1908,orszag_accurate_1971} and the electroconvective instability that causes the transition of a quasiequilibrium electric double layer to an nonequilibrium one that contains an additional extended space charge region~\cite{zaltzman_electro-osmotic_2007}. Here, we focus on morphological stability analysis in which linear stability analysis is used to analyze morphological instabilities of interfaces formed between different phases observed in various diverse phenomena such as electrodeposition~\cite{kessler_pattern_1988,ben-jacob_formation_1990,lopez-tomas_quasi-twodimensional_1995,leger_growth_1999,sagues_growth_2000,rosso_shape_2003,schwarzacher_kinetic_2004,rosso_electrodeposition_2007}, solidification~\cite{langer_instabilities_1980,kessler_pattern_1988,ben-jacob_formation_1990,cross_pattern_1993} and morphogenesis~\cite{turing_chemical_1952,cross_pattern_1993}. Some particular examples of morphological stability analysis include the Saffman-Taylor instability (viscous fingering)~\cite{saffman_penetration_1958,saffman_viscous_1986,bensimon_viscous_1986,homsy_viscous_1987}, viscous fingering coupled with electrokinetic effects~\cite{mirzadeh_electrokinetic_2017}, the Mullins-Sekerka instability of a spherical particle during diffusion-controlled or thermally controlled growth~\cite{mullins_morphological_1963} and of a planar interface during solidification of a dilute binary alloy~\cite{mullins_stability_1964,sekerka_stability_1965}, and control of phase separation using electro-autocatalysis or electro-autoinhibition in driven open electrochemical systems~\cite{bazant_theory_2013,bazant_thermodynamic_2017}.

\subsection{Stability of metal electrodeposition}

We focus on electrodeposition as a specific example of an electrochemical system for which morphological stability has been widely researched both theoretically and experimentally. The fundamental aspect of electrodeposition concerns the inherent instability of the governing physics while the practical aspect is about applications such as electroplating of metals and charging of metal batteries. To elucidate the physics behind electrodeposition, in liquid electrolytes, the morphologies of electrodeposits formed and their transitions for metals such as copper, zinc and silver are particularly well studied~\cite{ben-jacob_formation_1990,lopez-tomas_quasi-twodimensional_1995,leger_growth_1999,sagues_growth_2000,rosso_shape_2003,rosso_electrodeposition_2007,schneider_nanoscale_2017}. Depending on conditions such as applied current, applied voltage and electrolyte concentration, a variety of morphological patterns such as diffusion-limited aggregation (DLA) patterns~\cite{witten_diffusion-limited_1981,witten_diffusion-limited_1983,meakin_formation_1983,vicsek_pattern_1984,brady_fractal_1984,matsushita_fractal_1984,ben-jacob_experimental_1985,sawada_dendritic_1986,grier_morphology_1986,trigueros_pattern_1991,erlebacher_computer_1993}, dense branching morphologies (DBM)~\cite{ben-jacob_experimental_1985,sawada_dendritic_1986,grier_morphology_1986,ben-jacob_formation_1986,ben-jacob_interfacial_1987,grier_stability_1987,ben-jacob_characterization_1988,garik_laplace-_1989,trigueros_pattern_1991,fleury_experimental_1991,erlebacher_computer_1993,elezgaray_dense_2000,leger_internal_2000} and dendritic structures~\cite{ben-jacob_dynamics_1983,ben-jacob_boundary-layer_1984,ben-jacob_pattern_1984,ben-jacob_experimental_1985,sawada_dendritic_1986,grier_morphology_1986,ben-jacob_interfacial_1987,ben-jacob_characterization_1988,trigueros_pattern_1991} have been observed. Ion concentration fields~\cite{argoul_interferometric_1996,leger_experimental_1997,leger_dynamical_1998,leger_front_1999,leger_probing_2000,rosso_onset_2002}, electroconvection~\cite{fleury_theory_1992,fleury_coupling_1993,fleury_mechanism_1994,rosso_experimental_1994,huth_role_1995}, gravity-induced convection (buoyancy)~\cite{rosso_experimental_1994,huth_role_1995,chazalviel_quantitative_1996} and the presence of impurities~\cite{fleury_geometrical_1991,fleury_experimental_1991,kuhn_influence_1993,kuhn_spatiotemporal_1994,kuhn_revisited_1994} have also been examined to determine their significant effects on morphology. While the range of possible morphologies of electrodeposits is diverse, for electroplating of metals, it is desirable to electrodeposit layers that are as uniform and homogeneous as possible.

Electrodeposition is also a critical process in the development of rechargeable/secondary lithium metal batteries (LMBs) that use lithium metal for the negative electrode. For negative electrodes that use lithium chemistry, because lithium metal has the lowest standard reduction potential ($-3.04\,\textn{V}$ vs.\ the standard hydrogen electrode), highest theoretical specific ($3860\,\textn{mAh}/\textn{g}$) and volumetric ($2061\,\textn{mAh}/\textn{cm}^3$) capacities, and lowest mass density ($0.53\,\textn{g}/\textn{cm}^3$) out of all possible candidates, it is the most promising choice for achieving high energy densities in LMBs~\cite{brandt_historical_1994,tarascon_issues_2001,aurbach_short_2002,xu_nonaqueous_2004,armand_building_2008,cairns_batteries_2010,scrosati_lithium_2010,etacheri_challenges_2011,xu_lithium_2014,li_review_2014,tu_nanostructured_2015,tikekar_design_2016,choi_promise_2016,blomgren_development_2017,lin_reviving_2017,cheng_toward_2017,lin_nanoscale_2017,albertus_status_2018,lu_high-performance_2018}. However, the charging of LMBs is equivalent to lithium electrodeposition at the negative electrode, which is an inherently unstable process that can cause the formation of dendrites that penetrate the separator and result in internal short circuits and thermal runaway during charging~\cite{brandt_historical_1994,tarascon_issues_2001,aurbach_short_2002,xu_nonaqueous_2004,armand_building_2008,cairns_batteries_2010,scrosati_lithium_2010,etacheri_challenges_2011,xu_lithium_2014,li_review_2014,tu_nanostructured_2015,tikekar_design_2016,choi_promise_2016,blomgren_development_2017,lin_reviving_2017,cheng_toward_2017,lin_nanoscale_2017,albertus_status_2018,lu_high-performance_2018}. This process has been especially well investigated in lithium polymer batteries that use a polymer electrolyte~\cite{brissot_situ_1998,brissot_dendritic_1999,brissot_situ_1999,brissot_concentration_2001,rosso_onset_2001,teyssot_inter-electrode_2005,rosso_dendrite_2006,stone_resolution_2012}. Detailed studies of various growth modes of lithium in liquid electrolytes during charging have been recently performed~\cite{bai_transition_2016,kushima_liquid_2017,bai_interactions_2018}, which will aid in the development of better models for lithium electrodeposition. Modern lithium-ion batteries (LIBs)~\cite{brandt_historical_1994,tarascon_issues_2001,aurbach_short_2002,whittingham_lithium_2004,xu_nonaqueous_2004,armand_building_2008,bruce_nanomaterials_2008,cairns_batteries_2010,scrosati_lithium_2010,goodenough_challenges_2010,etacheri_challenges_2011,goodenough_li-ion_2013,li_review_2014,blomgren_development_2017,lu_high-performance_2018} partially mitigate this problem of dendrite formation and propagation by employing lithium intercalating materials such as graphite for the negative electrode where only lithium ions and not reduced lithium atoms are involved in the intercalation reactions, which is also known as the ``rocking chair technology''~\cite{tarascon_issues_2001}. Nonetheless, lithium plating still occurs in LIBs when they are charged at high rates or low temperatures~\cite{vetter_ageing_2005,goodenough_challenges_2010,goodenough_li-ion_2013,li_review_2014,palacin_why_2016}. Although the root causes of the widely publicized LIB failures in two Boeing 787 Dreamliners in January 2013 were not conclusively identified~\cite{williard_lessons_2013}, there is no doubt that safety is of paramount importance in both LMBs and LIBs, which requires a thorough understanding of dendrite formation.

For both electroplating of metals and charging of high energy density LMBs, it would be advantageous to perform them at as large a current or voltage as possible without causing dendrite formation. It is therefore important to understand the possible mechanisms for the electrochemical system to sustain a high current or voltage and how these mechanisms interact with the metal electrodeposition and LMB charging processes. In a neutral channel or porous medium containing an electrolyte, when ion transport is governed by diffusion and electromigration, which is collectively termed electrodiffusion, the maximum current that can be attained by the electrochemical system is called the diffusion-limited current~\cite{newman_electrochemical_2004,bard_electrochemical_2000}. In practice, overlimiting current (OLC) beyond the electrodiffusion limit has been observed experimentally in ion-exchange membranes~\cite{rubinstein_voltage_1979,rosler_ion_1992,krol_concentration_1999,krol_chronopotentiometry_1999,rubinshtein_experimental_2002,rubinstein_direct_2008,deng_overlimiting_2013,schlumpberger_scalable_2015,schlumpberger_shock_2016,nikonenko_intensive_2010,nikonenko_desalination_2014,strathmann_electrodialysis_2010} and microchannels and nanochannels~\cite{kim_concentration_2007,yossifon_selection_2008,zangle_propagation_2009,zangle_theory_2010,zangle_effects_2010,nam_experimental_2015,schiffbauer_probing_2015,sohn_surface_2018}. Depending on the length scale of the pores or channel, some possible physical mechanisms for OLC~\cite{dydek_overlimiting_2011} are surface conduction~\cite{zangle_propagation_2009,zangle_theory_2010,zangle_effects_2010,mani_propagation_2009,mani_deionization_2011,dydek_nonlinear_2013}, electroosmotic flow~\cite{yaroshchuk_coupled_2011,rubinstein_convective_2013} and electroosmotic instability~\cite{rubinstein_electro-osmotically_2000,zaltzman_electro-osmotic_2007}. Some chemical mechanisms for OLC include water splitting~\cite{nikonenko_intensive_2010,nikonenko_desalination_2014} and current-induced membrane discharge~\cite{andersen_current-induced_2012}. In this paper, we focus on porous media consisting of pores with a nanometer length scale, therefore the dominant OLC mechanism is expected to be surface conduction~\cite{dydek_overlimiting_2011}. When a sufficiently large current or voltage is applied across a porous medium whose pore surfaces are charged, the bulk electrolyte eventually gets depleted at an ion-selective interface such as an electrode. In order to sustain the current beyond the electrodiffusion limit, the counterions in the electric double layers (EDLs) next to the charged pore surfaces migrate under the large electric field generated in the depletion region. This phenomenon is termed surface conduction and results in the formation and propagation of deionization shocks away from the ion-selective interface in porous media~\cite{mani_deionization_2011,dydek_nonlinear_2013,yaroshchuk_over-limiting_2012} and microchannels and nanochannels~\cite{zangle_propagation_2009,zangle_theory_2010,zangle_effects_2010,dydek_overlimiting_2011,mani_propagation_2009,nielsen_concentration_2014}. The deionization shock separates the ``front'' electrolyte-rich region, in which bulk electrodiffusion dominates, from the ``back'' electrolyte-poor region, in which electromigration in the EDLs dominates.

\subsection{Theories of pattern formation}

Morphological stability analysis of electrodeposition is typically performed in the style of the pioneering Mullins-Sekerka stability analysis~\cite{mullins_morphological_1963,mullins_stability_1964}. The destabilizing effect arises from the amplification of surface protrusions by diffusive fluxes while the main stabilizing effect arises from the surface energy penalty incurred in creating additional surface area. The balance between these two effects, which is influenced by system parameters, sets a characteristic length scale or wavenumber for the surface instability. In 1963, by applying an infinitesimally small spherical harmonic perturbation to the surface of a spherical particle undergoing growth by solute diffusion or heat diffusion, Mullins and Sekerka derived a dispersion relation that related growth rates of the eigenmodes to particle radius and degree of supersaturation~\cite{mullins_morphological_1963}. Similarly, in 1964, Mullins and Sekerka imposed a infinitesimally small sinusoidal perturbation on a planar liquid-solid interface during the solidification of a dilute binary alloy and obtained a dispersion relation relating the surface perturbation growth rate to system parameters such as temperature and concentration gradients~\cite{mullins_stability_1964}. In the spirit of the Mullins-Sekerka stability analysis, about 16 years later in 1980, Aogaki, Kitazawa, Kose and Fueki applied linear stability analysis to study electrodeposition with a steady-state base state in the presence of a supporting electrolyte, i.e., electromigration of the minor species can be ignored, and without explicitly accounting for electrochemical reaction kinetics~\cite{aogaki_theory_1980}. Following up on this work, from 1981 to 1982, Aogaki and Makino changed the steady-state base state to a time-dependent base state under galvanostatic conditions while keeping other assumptions intact~\cite{aogaki_theory_1981,aogaki_image_1982,aogaki_image_1982-1}. In 1984, Aogaki and Makino extended their previous work to account for surface diffusion of adsorbed metal atoms under galvanostatic~\cite{aogaki_morphological_1984,aogaki_morphological_1984-1} and potentiostatic conditions~\cite{aogaki_morphological_1984-2,aogaki_morphological_1984-3}. In the same year, Makino, Aogaki and Niki also used such a linear stability analysis to extract surface parameters of metals under galvanostatic and potentiostatic conditions~\cite{makino_determination_1984} and applied it to study how hydrogen adsorption affects these extracted parameters under galvanostatic conditions~\cite{makino_application_1984}. Later work by Barkey, Muller and Tobias in 1989~\cite{barkey_roughness_1989,barkey_roughness_1989-1}, and Chen and Jorne in 1991~\cite{chen_dynamics_1991} additionally assumed the presence of a diffusion boundary layer next to the electrode.

Subsequent developments in linear stability analysis of electrodeposition relaxed some assumptions made in the past literature and added more physics and electrochemistry. Butler-Volmer reaction kinetics was first explicitly considered by Pritzker and Fahidy in 1992 for a steady-state base state with a diffusion boundary layer next to the electrode~\cite{pritzker_morphological_1992}. Also considering Butler-Volmer reaction kinetics with a steady-state base state, in 1995, Sundstr\"om and Bark used the Nernst-Planck equations for ion transport without assuming the existence of a diffusion boundary layer, numerically solved for the dispersion relation and performed extensive parameter sweeps over key parameters of interest such as surface energy and exchange current density~\cite{sundstrom_morphological_1995}. Extending these two papers in 1998, Elezgaray, L\'eger and Argoul used a time-dependent base state under galvanostatic conditions, numerically solved for both the time-dependent base state and perturbed state to obtain the dispersion relation and demonstrated good agreement between their theory and experiments on copper electrodeposition in a thin gap cell~\cite{elezgaray_linear_1998}. The role of electrolyte additives in stabilizing electrodeposition was examined in the linear stability analysis performed by Haataja, Srolovitz and Bocarsly in 2002 and 2003~\cite{haataja_morphological_2002,haataja_morphological_2003,haataja_morphological_2003-1}, and McFadden et al.\ in 2003~\cite{mcfadden_mechanism_2003}. By demonstrating that the effects of the anode can be ignored under certain conditions when deriving the dispersion relation, BuAli, Johns and Narayanan in 2006 simplified Sundstr\"om and Bark's analysis to obtain an analytical expression for the dispersion relation~\cite{buali_growth_2006}. In 2004 and 2005, Monroe and Newman included additional mechanical effects such as pressure, viscous stress and deformational stress to the linear stability analysis of electrodeposition, which provided more stabilization beyond that provided by surface energy~\cite{monroe_effect_2004,monroe_impact_2005}. For a steady-state base state, in 2014, Tikekar, Archer and Koch studied how tethered immobilized anions provide additional stabilization to electrodeposition by reducing the electric field at the cathode and, after making some approximations, derived analytical expressions for the dispersion relation for small and large current densities~\cite{tikekar_stability_2014}. Tikekar, Archer and Koch then extended this work in 2016 by accounting for elastic deformations that provide further stabilization~\cite{tikekar_stabilizing_2016}. Subsequently in 2018, Tikekar, Li, Archer and Koch looked at using polymer electrolyte viscosity to suppress morphological instabilities driven by electroconvection~\cite{tikekar_electroconvection_2018}. Building on Monroe and Newman's 2004 and 2005 work on interfacial deformation effects~\cite{monroe_effect_2004,monroe_impact_2005}, Ahmad and Viswanathan identified a new mechanism for stabilization driven by the difference of the metal density in the metal electrode and solid electrolyte in 2017~\cite{ahmad_stability_2017}, and further generalized this work in the same year to account for anisotropy~\cite{ahmad_role_2017}. Natsiavas, Weinberg, Rosato and Ortiz in 2016 also investigated the stabilizing effect of prestress and showed good agreement between theory and experiment~\cite{natsiavas_effect_2016}. Relaxing the usual assumption of electroneutrality, in 2015, Nielsen and Bruus performed linear stability analysis for a steady-state base state that accounts for the extended space charge region that is formed when the electric double layer becomes nonequilibrium at an overlimiting current~\cite{nielsen_morphological_2015}.

Without performing a linear stability analysis, some models focus on describing the initiation and subsequent propagation of dendrites. The classic work in this class of models is by Chazalviel in 1990 in which they used the Poisson's equation for electrostatics, i.e., electroneutrality is not assumed, and showed that the initiation of ramified electrodeposits is caused by the creation of a space charge layer upon anion depletion at the cathode, the induction time for initiation is the time needed for building up this space charge layer, and the velocity of the ramified growth is equal to the electromigration velocity of the anions~\cite{chazalviel_electrochemical_1990}; some experimental results were also obtained by Fleury, Chazalviel, Rosso and Sapoval in support of this model~\cite{fleury_role_1990}, and some of the numerical results of the original analysis were subsequently improved by Rosso, Chazalviel and Chassaing~\cite{rosso_calculation_2006}. Via an asymptotic analysis of the Poisson-Nernst-Planck equations for ion transport, Bazant also showed that the velocity of the ramified growth is approximately equal to the anion electromigration velocity and estimated the induction time for the onset of ramified growth~\cite{bazant_regulation_1995}. Building on past theoretical and experimental work done on silver electrodeposition by Barton and Bockris~\cite{barton_electrolytic_1962}, and zinc electrodeposition by Despic, Diggle and Bockris~\cite{despic_mechanism_1968,diggle_mechanism_1969}, Monroe and Newman investigated the propagation velocity and length of a dendrite tip that grows via Butler-Volmer kinetics~\cite{monroe_dendrite_2003}. By examining the thermodynamics and kinetics of heterogeneous nucleation and growth, which is assumed to proceed via the linearized Butler-Volmer equation valid for small overpotentials, Ely and Garc\'{\i}a identified five different regimes of nucleus behavior~\cite{ely_heterogeneous_2013}. Assuming a concentration-dependent electrolyte diffusivity and the existence of a hemispherical dendrite ``precursor'' that grows with Tafel kinetics, Akolkar studied the subsequent propagation velocity and length of the dendrite~\cite{akolkar_mathematical_2013} and how they are affected by temperature~\cite{akolkar_modeling_2014}.

\subsection{Contributions of this work}

In this paper, we perform linear stability analysis of electrodeposition inside a charged random porous medium, whose pore surface charges can generally be of any sign, that is filled with a liquid electrolyte and flanked on its sides by a pair of planar metal electrodes. The linear stability analysis is carried out with respect to a time-dependent base state and focuses on overlimiting current carried by surface conduction. By doing so, we combine and generalize previous work done in~\cite{sundstrom_morphological_1995,elezgaray_linear_1998,tikekar_stability_2014}. For simplicity, we ignore bulk convection, electroosmotic flow, surface adsorption, surface diffusion of adsorbed species~\cite{aogaki_morphological_1984,aogaki_morphological_1984-1,aogaki_morphological_1984-2,aogaki_morphological_1984-3} and additional mechanical effects such as pressure, viscous stress and deformational stress~\cite{monroe_effect_2004,monroe_impact_2005,tikekar_stabilizing_2016,ahmad_stability_2017,ahmad_role_2017,natsiavas_effect_2016}. We expect surface diffusion of adsorbed species, which alleviates electrodiffusion limitations, and interfacial deformation effects to stabilize electrodeposition, hence our work here can be considered a worst-case analysis. The only electrochemical reaction considered here is metal electrodeposition, therefore in the context of LMBs and LIBs, electrochemical and chemical reactions between lithium and the electrolyte that cause the formation of the solid electrolyte interphase (SEI) layer~\cite{verma_review_2010,cheng_review_2016,peled_reviewsei:_2017,wang_review_2018} are not included. We first derive governing equations for the full model that consists of coupling ion transport with electrochemical reaction kinetics, followed by applying linear stability analysis on the full model via the imposition of sinusoidal spatial perturbations around the time-dependent base state. For the dispersion relation, we perform a boundary layer analysis on the perturbed state to derive an accurate approximation for it and a convergence analysis of its full numerical solution. To better understand the physics of the dispersion relation, we carry out parameter sweeps over the pore surface charge density, Damk\"ohler number and applied current density under galvanostatic conditions. We also compare the numerical and approximate solutions of the maximum wavenumber, maximum growth rate and critical wavenumber in order to verify the accuracy of these approximations. Subsequently, we apply the linear stability analysis to compare theoretical predictions and experimental data for copper electrodeposition in cellulose nitrate membranes~\cite{han_dendrite_2016}, and also use the stability analysis as a tool for investigating the dependence of crystal grain size on duty cycle during pulse electroplating.

\section{Full model}\label{sec:Full model}

\subsection{Transport}\label{sec:Transport}

\begin{figure}
  \centering
  \includegraphics[scale=0.67]{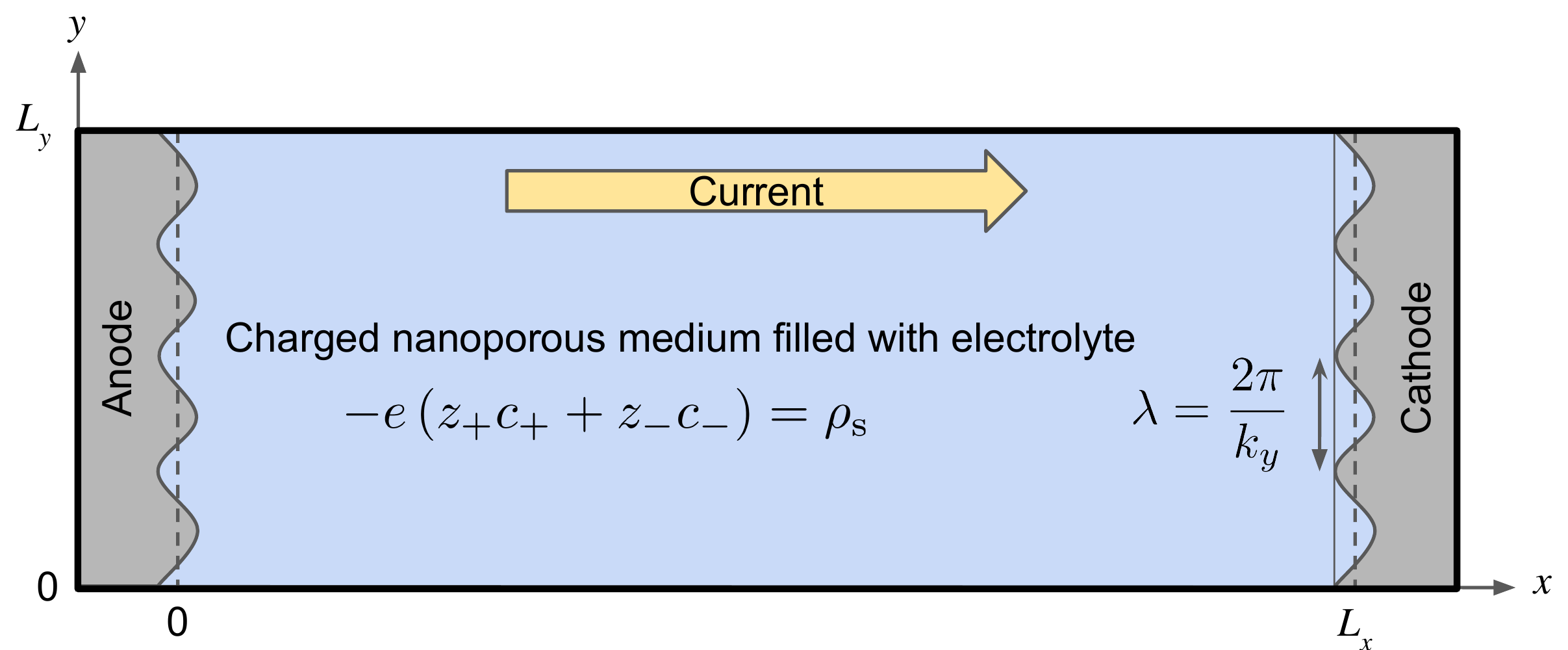}
  \caption{2D projection of 3D system considered: charged random nanoporous medium filled with binary asymmetric liquid electrolyte where anode is on the left at $x = 0$ and cathode is on the right at $x = L_x$ along the $x$ axis, which is the direction of the inter-electrode spacing, and $y$ axis is the direction of the sinusoidal perturbation. $\lambda = \frac{2\pi}{k_y}$ is the perturbation wavelength where $k_y$ is the wavenumber in the $y$ direction. The linear stability analysis is actually performed in 3D in which the sinusoidal perturbation is applied in the $y$ and $z$ directions, and the extension of this 2D projection to the 3D case is straightforward. The current in the electrolyte flows from left to right. The equation shown on the left describes macroscopic electroneutrality given by Equation~\ref{eq:Macroscopic electroneutrality} where $\rho_\textn{s}$ is the volume-averaged background charge density.}\label{fig:Schematic}
\end{figure}

The starting point for modeling ion transport is the leaky membrane model that is able to predict overlimiting current carried by surface conduction, which we have previously coupled with Butler-Volmer reaction kinetics for analyzing steady state current-voltage relations and linear sweep voltammetry~\cite{khoo_theory_2018}. The system under consideration is a binary asymmetric liquid electrolyte in a finite 3D charged random nanoporous medium where $x\in\left[0,L_x\right]$, $y\in\left[0,L_y\right]$ and $z\in\left[0,L_z\right]$, whose 2D projection is illustrated in Figure~\ref{fig:Schematic}. We assume that the nanoporous medium is random with well connected pores such as cellulose nitrate membranes so that we can investigate macroscopic electrode-scale morphological instabilities~\cite{han_dendrite_2016}. The cations are electroactive and the anions are inert. Initially, at $t=0$, the anode surface is located at $x=0$ while the cathode surface is located at $x=L_x$. As is typical for linear stability analysis of electrodeposition~\cite{sundstrom_morphological_1995,elezgaray_linear_1998,nielsen_morphological_2015}, we pick a moving reference frame with a velocity $u\left(t\right) = u_x\left(t\right)e_x$ that is equal to the velocity of the electrode/electrolyte interface so that the average positions of the dissolving anode and growing cathode remain stationary. For the porous medium, we denote its porosity, tortuosity, internal pore surface area per unit volume and pore surface charge per unit area as $\epsilon_\textn{p}$, $\tau$, $a_\textn{p}$ and $\sigma_\textn{s}$ respectively. We assume that there are no homogeneous reactions and all material properties such as $\epsilon_\textn{p}$, $\tau$, $a_\textn{p}$ and $\sigma_\textn{s}$ are constant and uniform. We also assume that dilute solution theory holds true, hence all activity coefficients are $1$ and the cation and anion macroscopic diffusivities $D_{\pm0}$, where the $0$ subscript indicates dilute limit, are constant, uniform and independent of concentrations. Accounting for corrections due to the tortuosity of the porous medium, the macroscopic diffusivity $D_{\pm0}$ is related to the molecular (free solution) diffusivity $D_{\pm0}^\textn{m}$ by $D_{\pm0} = \frac{D_{\pm0}^\textn{m}}{\tau}$~\cite{ferguson_nonequilibrium_2012}. The assumption of dilute solution theory further implies that the convective flux in the moving reference frame is negligible and the effect of the moving reference frame velocity $u\left(t\right)= u_x\left(t\right)e_x$ is only significant in the equation describing electrode surface growth or dissolution~\cite{sundstrom_morphological_1995,elezgaray_linear_1998,nielsen_morphological_2015}, which we will discuss in Section~\ref{sec:Boundary conditions, constraints and initial conditions}. Under these assumptions, for ion transport, the Nernst-Planck equations describing species conservation, charge conservation equation and macroscopic electroneutrality constraint are given by
\begin{align}
  \epsilon_\textn{p}\frac{\partial c_\pm}{\partial t} + \nabla\cdot F_\pm &= 0, \quad F_\pm = -\epsilon_\textn{p}D_{\pm0}\left(\nabla c_\pm + \frac{z_\pm ec_\pm}{k_\textn{B}T}\nabla\phi\right), \\
  \nabla\cdot J &= 0, \quad J = e\left(z_+F_+ + z_-F_-\right), \\
  \rho_\textn{s} = \frac{\sigma_\textn{s}}{h_\textn{p}} &= \frac{a_\textn{p}\sigma_\textn{s}}{\epsilon_\textn{p}} = -e\left(z_+c_+ + z_-c_-\right), \label{eq:Macroscopic electroneutrality}
\end{align}
where $c_\pm$, $F_\pm$, $z_\pm$, are the cation and anion concentrations, fluxes and charge numbers respectively, $\phi$ is the electrolyte electric potential, $J$ is the electrolyte current density, $h_\textn{p} = \frac{\epsilon_\textn{p}}{a_\textn{p}}$ is the effective pore size and $\rho_\textn{s}$ is the volume-averaged background charge density. Denoting the numbers of cations and anions that are formed from complete dissociation of $1$ molecule of neutral salt as $\nu_\pm$, electroneutrality requires that $z_+\nu_+ + z_-\nu_- = 0$. We will use the macroscopic electroneutrality constraint given by Equation~\ref{eq:Macroscopic electroneutrality} to eliminate $c_+$ as a dependent variable, therefore leaving $c_-$ and $\phi$ as the remaining dependent variables.

For a classical system with an uncharged nanoporous medium, i.e., $\rho_\textn{s} = 0$, the maximum current density that the system can possibly attain under electrodiffusion is called the diffusion-limited or limiting current density $J_\limn$, which is given by~\cite{khoo_theory_2018}
\begin{equation}
  J_\limn = \frac{2\left(z_+-z_-\right)e\epsilon_\textn{p}D_{+0}\nu_-c_0}{L_x}
\end{equation}
where $c_0$ is the neutral salt bulk concentration. The limiting current $I_\limn$ is then given by $I_\limn = J_\limn A$ where $A = L_y L_z$ is the surface area of the electrode. Under galvanostatic conditions, when a current density $J_\textn{a}$ larger than $J_\limn$ is applied, the cation and anion concentrations at the cathode reach $0$ and the electrolyte electric potential and electric field there diverge in finite time, which is called Sand's time~\cite{sand_concentration_1899} denoted as $t_\textn{s}$; see~\cite{van_soestbergen_diffuse-charge_2010} for a discussion of some subtlety associated with this transition time when $J_\textn{a}$ is exactly equal to $J_\limn$. If we have not assumed macroscopic electroneutrality and instead modeled electric double layers that can go out of equilibrium at high currents and voltages, then the electric field would be large but finite at and past $t_\textn{s}$~\cite{zaltzman_electro-osmotic_2007,nielsen_morphological_2015,bazant_current-voltage_2005,chu_electrochemical_2005}. Defining the dimensionless Sand's time $\tilde{t}_\textn{s} = \frac{D_\textn{amb0}t_\textn{s}}{L_x^2}$ and dimensionless applied current density $\tilde{J}_\textn{a} = \frac{J_\textn{a}}{J_\limn}$ where $D_\textn{amb0} = \frac{\left(z_+-z_-\right)D_{+0}D_{-0}}{z_+D_{+0}-z_-D_{-0}}$~\cite{newman_electrochemical_2004,khoo_theory_2018} is the ambipolar diffusivity of the neutral salt in the dilute limit and $\frac{L_x^2}{D_\textn{amb0}}$ is the diffusion time scale, $\tilde{t}_\textn{s}$ is given by~\cite{van_soestbergen_diffuse-charge_2010}
\begin{equation}
  \tilde{t}_\textn{s} = \frac{\pi}{16\tilde{J}_\textn{a}^2}, \quad \tilde{J}_\textn{a} > 1. \label{eq:Sand's time}
\end{equation}
For galvanostatic conditions, $t_\textn{s}$ is a critically important time scale because the formation of dendrites often occurs near or at $t_\textn{s}$, therefore it will be central to the linear stability analysis results discussed in Section~\ref{sec:Results}.

Unlike the classical case of $\rho_\textn{s} = 0$, for $\rho_\textn{s} < 0$, the system can sustain an overlimiting current $\tilde{J}_\textn{a} > 1$ via surface conduction that is the electromigration of the counterions in the electric double layers (EDLs), which are formed next to the charged pore surfaces, under a large electric field generated in the depletion region next to the cathode. This additional surface conductivity thus enables the system to go beyond $t_\textn{s}$ and eventually reach a steady state, in stark contrast to the finite time divergence of the classical case at $t_\textn{s}$. On the other hand, for $\rho_\textn{s} > 0$, the counterions in the EDLs are the inert anions instead of the electroactive cations, which contribute a surface current that flows in the opposite direction from that of the bulk current. Because of this ``negative'' surface conductivity conferred by $\rho_\textn{s} > 0$ relative to $\rho_\textn{s} = 0$, at the cathode, the bulk electrolyte concentration vanishes and the electric field diverges earlier than $t_\textn{s}$; in other words, $\rho_\textn{s} > 0$ effectively reduces $t_\textn{s}$. A more quantitative way of explaining this is that the ``negative'' surface conductivity causes the maximum current density that can be achieved, which is denoted as $J_\maxn$, to be smaller than $J_\limn$, and $J_\maxn$ decreases as $\rho_\textn{s}$ increases. In effect, a more positive $\rho_\textn{s}$ decreases $J_\limn$, which thus leads to a smaller $t_\textn{s}$ for a given $J_\textn{a}$ according to Equation~\ref{eq:Sand's time}. Details of how to numerically compute $J_\maxn$ are found in~\cite{khoo_theory_2018}; note that $J_\maxn$ here corresponds to $I_\maxn^\textn{BV}$ in~\cite{khoo_theory_2018}.

\subsection{Electrochemical reaction kinetics}\label{sec:Electrochemical reaction kinetics}

In order to analyze how spatial perturbations of an electrode surface affect its linear stability, we need to account for the effects of surface curvature and surface energy in the electrochemical reaction kinetics model. The mean curvature of the electrode/electrolyte interface $\mathscr{H}$ is given by $\mathscr{H} = -\frac{1}{2}\nabla_\textn{s}\cdot\hat{n}$ where $\nabla_\textn{s}$ is the surface gradient operator and $\hat{n}$ is the unit normal that points outwards from the electrolyte~\cite{deen_analysis_2011}. In this paper, we consider a charge transfer reaction that involves only the cations and electrons while the anions are inert. More concretely, we suppose the following charge transfer reaction consuming $n$ electrons $\textn{O}^{z_\textn{O}} + n\textn{e}^- \rightleftharpoons \textn{R}^{z_\textn{R}}$ where $\textn{O}^{z_\textn{O}}$ is the oxidized species $\textn{O}$ with charge $z_\textn{O}$, $\textn{e}^-$ is the electron $\textn{e}$ with charge $-1$, $\textn{R}^{z_\textn{R}}$ is the reduced species $\textn{R}$ with charge $z_\textn{R}$, and $z_\textn{O} - n = z_\textn{R}$ because of charge conservation. If the reduced species is solid metal, i.e., $z_\textn{R} = 0$, as is the case in metal electrodeposition, the creation of additional electrode/electrolyte interfacial area results in a surface energy penalty that appears in the electrochemical potential of the reduced species. Therefore, the electrochemical potentials $\mu_i$ of the oxidized species $\textn{O}$, electron $\textn{e}$ and reduced species $\textn{R}$ for $i \in \left \{\textn{O},\textn{e},\textn{R}\right \}$ are given by
\begin{align}
  \mu_\textn{O} &= k_\textn{B}T\ln a_\textn{O} + z_\textn{O}e\phi + \mu_\textn{O}^\Theta, \\
  \mu_\textn{e} &= k_\textn{B}T\ln a_\textn{e} - e\phi_\textn{e} + \mu_\textn{e}^\Theta, \\
  \mu_\textn{R} &= k_\textn{B}T\ln a_\textn{R} + z_\textn{R}e\phi + \mu_\textn{R}^\Theta + 2\Omega\gamma\mathscr{H},
\end{align}
where the surface energy term $2\Omega\gamma\mathscr{H}$~\cite{sundstrom_morphological_1995,elezgaray_linear_1998,monroe_dendrite_2003,monroe_effect_2004,monroe_impact_2005,deen_analysis_2011,tikekar_stability_2014,nielsen_morphological_2015} is included in $\mu_\textn{R}$ when the reduced species is solid metal ($z_\textn{R} = 0$) and the $\Theta$ superscript indicates standard state. The activity of species $i$ is given by $a_i = \gamma_i\hat{c}_i$ where $\gamma_i$ is the activity coefficient of species $i$ and $\hat{c}_i = \frac{c_i}{c_i^\Theta}$ is the concentration of species $i$ normalized by its standard concentration $c_i^\Theta$. $\mu_i^\Theta$ is the standard electrochemical potential of species $i$, $\phi_\textn{e}$ is the electrode electric potential, $\Omega = \frac{M_\textn{m}}{\rho_\textn{m}}$ is the atomic volume of the solid metal where $M_\textn{m}$ and $\rho_\textn{m}$ are the atomic mass and mass density of the metal respectively, and $\gamma$ is the isotropic surface energy of the metal/electrolyte interface. The quantity $\frac{\Omega\gamma}{k_\textn{B}T}$ is the capillary constant that has units of length~\cite{mullins_morphological_1963,mullins_stability_1964,sekerka_stability_1965}. The interfacial electric potential difference $\Delta\phi$ is defined as $\Delta\phi = \phi_\textn{e} - \phi$. At equilibrium when $\mu_\textn{O} + n\mu_\textn{e} = \mu_\textn{R}$, we obtain the Nernst equation
\begin{equation}
  \Delta\phi^\textn{eq} = \frac{k_\textn{B}T}{ne}\ln\frac{a_\textn{O}a_\textn{e}^n}{a_\textn{R}} + E^\Theta - \frac{2\Omega\gamma\mathscr{H}}{ne}, \quad E^\Theta = \frac{\mu_\textn{O}^\Theta + n\mu_\textn{e}^\Theta - \mu_\textn{R}^\Theta}{ne}, \label{eq:Nernst equation}
\end{equation}
where the ``eq'' superscript denotes equilibrium and $E^\Theta$ is the standard electrode potential. When the system is driven out of equilibrium so that $\mu_\textn{O} + n\mu_\textn{e} \neq \mu_\textn{R}$, the system generates a Faradaic current density $J_\textn{F}$ that is given by~\cite{bazant_theory_2013,bazant_thermodynamic_2017,ferguson_nonequilibrium_2012}
\begin{equation}
  J_\textn{F} = nek_0\left[\exp\left(-\frac{\mu_\ddagger^\textn{r,ex}-\mu_\textn{O}-n\mu_\textn{e}}{k_\textn{B}T}\right) - \exp\left(-\frac{\mu_\ddagger^\textn{r,ex}-\mu_\textn{R}}{k_\textn{B}T}\right)\right]
\end{equation}
where $k_0$ is the overall reaction rate constant and $\mu_\ddagger^\textn{r,ex}$ is the excess electrochemical potential of the transition state for the Faradaic reaction. Using the Butler-Volmer hypothesis, $\mu_\ddagger^\textn{r,ex}$ consists of a chemical contribution $k_\textn{B}T\ln\gamma_\ddagger^\textn{r}$, where $\gamma_\ddagger^\textn{r}$ is the activity coefficient of the transition state for the Faradaic reaction, and a convex combination of the electrostatic energies, surface energies (only for the reduced species) and standard electrochemical potentials weighted by $\alpha$, which is the charge transfer coefficient. Therefore, $\mu_\ddagger^\textn{r,ex}$ is given by
\begin{equation}
  \mu_\ddagger^\textn{r,ex} = k_\textn{B}T\ln\gamma_\ddagger^\textn{r} + \left(1-\alpha\right)\left(z_\textn{O}e\phi-ne\phi_\textn{e}+\mu_\textn{O}^\Theta+n\mu_\textn{e}^\Theta\right) + \alpha\left(z_\textn{R}e\phi+\mu_\textn{R}^\Theta+2\Omega\gamma\mathscr{H}\right). \label{eq:Excess electrochemical potential of TS for Faradaic reaction}
\end{equation}
Defining the overpotential $\eta$ as $\eta = \frac{\mu_\textn{R}-\mu_\textn{O}-n\mu_\textn{e}}{ne} = \Delta\phi - \Delta\phi^\textn{eq}$, we rewrite $J_\textn{F}$ as
\begin{equation}
  J_\textn{F} = j_0\left \{\exp\left(-\frac{\alpha ne\eta}{k_\textn{B}T}\right) - \exp\left[\frac{\left(1-\alpha\right)ne\eta}{k_\textn{B}T}\right]\right \}, \quad j_0 = \frac{k_0ne\left(a_\textn{O}a_\textn{e}^n\right)^{1-\alpha}a_\textn{R}^\alpha}{\gamma_\ddagger^\textn{r}}, \label{eq:J_F}
\end{equation}
where $j_0$ is the exchange current density. In this form, we can identify the cathodic and anodic charge transfer coefficients, which are denoted as $\alpha_\textn{c}$ and $\alpha_\textn{a}$ respectively, as $\alpha_\textn{c} = \alpha$ and $\alpha_\textn{a} = 1 - \alpha$ such that $\alpha_\textn{c} + \alpha_\textn{a} = 1$. We note that our particular choice of $\mu_\ddagger^\textn{r,ex}$ in Equation~\ref{eq:Excess electrochemical potential of TS for Faradaic reaction} corresponds to choosing the ``mechanical transfer coefficient'' $\alpha_\textn{m}$ defined in~\cite{monroe_effect_2004} to be equal to $\alpha_\textn{a}$, causing $j_0$ to not depend explicitly on $\mathscr{H}$.

In this paper, we assume that the only charge transfer reaction occurring is metal electrodeposition that happens via the electrochemical reduction of cations in the electrolyte to solid metal on the electrode. The activity of solid metal is $1$ while we assume that the activity of electrons is $1$. In addition, like in Section~\ref{sec:Transport}, we assume that dilute solution theory is applicable, therefore the activity coefficients of the cation, anion and transition state for the Faradaic reaction are $1$ and we replace activities of the cation and anion with their normalized concentrations $\hat{c}_\pm$. Therefore, $\Delta\phi^\textn{eq}$ and $j_0$ simplify to
\begin{equation}
  \Delta\phi^\textn{eq} = \frac{k_\textn{B}T}{ne}\ln\hat{c}_+ + E^\Theta - \frac{2\Omega\gamma\mathscr{H}}{ne}, \quad j_0 = k_0ne\hat{c}_+^{1-\alpha}. \label{eq:Simplified Nernst equation and exchange current density}
\end{equation}
To compare the reaction and diffusion rates, we define the Damk\"ohler number $\textn{Da}$ as the ratio of the Faradaic current density scale $e\epsilon_\textn{p}k_0$ and limiting current density $J_\limn$ given by
\begin{equation}
  \textn{Da} = \frac{e\epsilon_\textn{p}k_0}{J_\limn}.\label{eq:Da}
\end{equation}
When $\textn{Da}$ is large, i.e., $\textn{Da} \gg 1$, the system is diffusion-limited but when $\textn{Da}$ is small, i.e., $\textn{Da} \ll 1$, the system is reaction-limited.

\subsection{Boundary conditions, constraints and initial conditions}\label{sec:Boundary conditions, constraints and initial conditions}

We use ``a'' and ``c'' superscripts to denote the anode and cathode respectively, $r = \left[x, y, z\right]^\mathrm{T}$ to denote the position vector, and $r_\textn{m}^\textn{a,c} = \left[x_\textn{m}^\textn{a,c}, y_\textn{m}^\textn{a,c}, z_\textn{m}^\textn{a,c}\right]^\mathrm{T}$ to denote the positions of the anode and cathode. We ground the anode at all times, i.e., $\phi_\textn{e}^\textn{a}=0$. Because the cations are electroactive, we impose no-flux boundary conditions for the cation flux on all boundaries except the anode and cathode where Faradaic reactions involving the cations occur. On the other hand, because the anions are inert, we impose no-flux boundary conditions for the anion flux on all boundaries. At the anode and cathode, we require the conservation of charges across the electrode/electrolyte interfaces. In summary, the boundary conditions are given by $\hat{n}\cdot F_-\left(r=r_\textn{m}^\textn{a,c}\right) = 0$, $\hat{n}\cdot J\left(r=r_\textn{m}^\textn{a,c}\right) = \epsilon_\textn{p}J_\textn{F}^\textn{a,c}$, $\hat{n}\cdot F_-\left(r=r_\textn{other}\right) = 0$ and $\hat{n}\cdot J\left(r=r_\textn{other}\right) = 0$ where $r_\textn{other}$ refers to all boundaries except the anode and cathode.

The velocity of the electrode/electrolyte interface $v_\textn{I}^\textn{a,c}$ is defined as $v_\textn{I}^\textn{a,c} = \epsilon_\textn{p}\frac{\dd r_\textn{m}^\textn{a,c}}{\dd t}$ and its normal component $v_\textn{In}^\textn{a,c}$ is given by $v_\textn{In}^\textn{a,c} = \hat{n}\cdot v_\textn{I}^\textn{a,c} = \epsilon_\textn{p}\hat{n}\cdot \frac{\dd r_\textn{m}^\textn{a,c}}{\dd t}$. Because the growth (dissolution) of the electrode surface is caused by metal electrodeposition (electrodissolution), taking into account the moving reference frame velocity $u\left(t\right) = u_x\left(t\right)e_x$, the normal interfacial velocity is related to the normal current density by $v_\textn{In}^\textn{a,c} = -\frac{\Omega\hat{n}\cdot J\left(r=r_\textn{m}^\textn{a,c}\right)}{z_+e} - \hat{n}\cdot u\left(r=r_\textn{m}^\textn{a,c}\right)$ and therefore, $\epsilon_\textn{p}\hat{n}\cdot \frac{\dd r_\textn{m}^\textn{a,c}}{\dd t} = -\frac{\Omega\hat{n}\cdot J\left(r=r_\textn{m}^\textn{a,c}\right)}{z_+e} - \hat{n}\cdot u\left(r=r_\textn{m}^\textn{a,c}\right)$.

For galvanostatic conditions in which we apply a current $I_\textn{a}$ on the system, we require $\int\hat{n}\cdot J\left(r=r_\textn{m}^\textn{c}\right)\ud S^\textn{c} = \int-\hat{n}\cdot J\left(r=r_\textn{m}^\textn{a}\right)\ud S^\textn{a} = I_\textn{a}$ to satisfy charge conservation whereas for potentiostatic conditions in which we apply an electric potential $V$ on the cathode, we set $\phi_\textn{e}^\textn{c} = V$. For initial conditions, we set $c_-\left(t=0\right) = \nu_-c_0 - \frac{\rho_\textn{s}+\left\lvert\rho_\textn{s}\right\rvert}{2z_-e} \equiv \beta_1$ where $c_0$ is the initial neutral salt bulk concentration~\cite{khoo_theory_2018}, and $x_\textn{m}^\textn{a}\left(t=0\right) = 0$ and $x_\textn{m}^\textn{c}\left(t=0\right) = L_x$, i.e., the anode and cathode are initially planar.

\section{Linear stability analysis}

\subsection{Perturbations and linearization}\label{sec:Perturbations and linearization}

Linear stability analysis generally involves imposing a spatial perturbation around a base state, keeping constant and linear terms of the perturbed state, and determining the dispersion relation that relates the growth rate of the perturbation to its wavenumber or wavelength. For electrodeposition specifically, the objective is to impose a spatial perturbation on a planar electrode surface and determine the effects of key system parameters on the linear stability of the surface in response to this perturbation. In this paper, we will choose a time-dependent base state, therefore the dispersion relation is also time-dependent. In 3D, the electrode/electrolyte interface can be written explicitly as $x = h\left(y,z,t\right)$ where $h$ is the electrode surface height. Given $h$, we can derive explicit expressions for surface variables such as the curvature $\mathscr{H}$ and normal interfacial velocity $v_\textn{In}$ in terms of $h$ and its spatial and temporal derivatives~\cite{deen_analysis_2011,crank_free_1987}, which are provided in Section I of Supplementary Material. For brevity, we let $k = \left[k_y, k_z\right]^\mathrm{T}$ and $\xi = \left[y, z\right]^\mathrm{T}$ where $k$ is the wavevector, and $k_y$ and $k_z$ are the wavenumbers in the $y$ and $z$ directions respectively. Therefore, $k \cdot \xi = k_y y + k_z z$, $k^2 = \left\lVert k \right\rVert_2^2 = k_y^2 + k_z^2$ where $\left\lVert \cdot \right\rVert_2$ is the $L^2$-norm and $\left\lVert k \right\rVert_2$ is the overall wavenumber, and the wavelength $\lambda$ is given by $\lambda = \frac{2\pi}{\left\lVert k \right\rVert_2}$. For brevity again, we write the overall wavenumber as $k$ and it is obvious from context whether $k$ refers to the wavevector or overall wavenumber. The perturbation that will be imposed is sinusoidal in the $y$ and $z$ directions given by
\begin{equation}
  h\left(\xi,t\right) = h^{\left(0\right)}\left(t\right) + \epsilon\Re\left[h^{\left(1\right)}\exp\left(ik\cdot\xi+\omega t\right)\right] + \mathcal{O}\left(\epsilon^2\right)
\end{equation}
where $\epsilon \ll 1$ is a dimensionless small parameter, the $\left(0\right)$ and $\left(1\right)$ superscripts denote the base and perturbed states respectively, $\Re\left(\cdot\right)$ gives the real part of a complex number, $h^{\left(1\right)}$ is the complex-valued perturbation amplitude of the electrode surface height, and $\omega$ is the complex-valued growth rate of the perturbation. In response to such an electrode surface perturbation, we assume that the perturbations to $c_-$ and $\phi$ are similarly given by
\begin{align}
  c_-\left(x,\xi,t\right) &= c_-^{\left(0\right)}\left(x,t\right) + \epsilon\Re\left[c_-^{\left(1\right)}\left(x\right)\exp\left(ik\cdot\xi+\omega t\right)\right] + \mathcal{O}\left(\epsilon^2\right), \\
  \phi\left(x,\xi,t\right) &= \phi^{\left(0\right)}\left(x,t\right) + \epsilon\Re\left[\phi^{\left(1\right)}\left(x\right)\exp\left(ik\cdot\xi+\omega t\right)\right] + \mathcal{O}\left(\epsilon^2\right),
\end{align}
where $c_-^{\left(1\right)}$ and $\phi^{\left(1\right)}$ are the complex-valued perturbation amplitudes of anion concentration and electrolyte electric potential respectively.

To evaluate $c_-$ and $\phi$ and their gradients $\nabla c_-$ and $\nabla\phi$ at the interface at $x = h$, we require their Taylor series expansions around the base state interface at $x = h^{\left(0\right)}$. Letting $\hat{\epsilon} = \epsilon\exp\left(ik\cdot\xi+\omega t\right)$ and $\theta \in \left \{c_-, \phi\right \}$, these expansions are given by~\cite{sundstrom_morphological_1995,elezgaray_linear_1998,nielsen_morphological_2015}
\begin{align}
  \theta\left(x=h\right) &= \theta^{\left(0\right)}\left(x=h^{\left(0\right)}\right) + \left.\hat{\epsilon}\left(h^{\left(1\right)}\frac{\partial\theta^{\left(0\right)}}{\partial x}+\theta^{\left(1\right)}\right)\right\rvert_{x=h^{\left(0\right)}} + \mathcal{O}\left(\epsilon^2\right), \\
  \frac{\partial\theta}{\partial x}\left(x=h\right) &= \frac{\partial\theta^{\left(0\right)}}{\partial x}\left(x=h^{\left(0\right)}\right) + \left.\hat{\epsilon}\left(h^{\left(1\right)}\frac{\partial^2\theta^{\left(0\right)}}{\partial x^2}+\frac{\dd\theta^{\left(1\right)}}{\dd x}\right)\right\rvert_{x=h^{\left(0\right)}} + \mathcal{O}\left(\epsilon^2\right), \\
  \frac{\partial\theta}{\partial y}\left(x=h\right) &= \hat{\epsilon}ik_y\theta^{\left(1\right)}\left(x=h^{\left(0\right)}\right) + \mathcal{O}\left(\epsilon^2\right), \quad \frac{\partial\theta}{\partial z}\left(x=h\right) = \hat{\epsilon}ik_z\theta^{\left(1\right)}\left(x=h^{\left(0\right)}\right) + \mathcal{O}\left(\epsilon^2\right), \\
  \nabla\theta\left(x=h\right) &= \frac{\partial\theta}{\partial x}\left(x=h\right)e_x +  \frac{\partial\theta}{\partial y}\left(x=h\right)e_y +  \frac{\partial\theta}{\partial z}\left(x=h\right)e_z + \mathcal{O}\left(\epsilon^2\right).
\end{align}

After substituting these perturbation expressions into the full model in Section~\ref{sec:Full model}, we obtain the base and perturbed states by matching the $\mathcal{O}\left(1\right)$ and $\mathcal{O}\left(\epsilon\right)$ terms respectively. The dispersion relation $\omega\left(k\right)$ is subsequently computed by solving these $\mathcal{O}\left(1\right)$ and $\mathcal{O}\left(\epsilon\right)$ equations. The growth rate $\omega$ is generally complex-valued and for a particular $k$ value, there is an infinite discrete spectrum of $\omega$ values. However, for linear stability analysis, we are primarily interested in the maximum of the real parts of the $\omega$ values, which is denoted as $\max\left \{\Re\left(\omega\right)\right \}$, that corresponds to the most unstable eigenmode. If $\max\left \{\Re\left(\omega\right)\right \} < 0$, the perturbation decays exponentially in time and the base state is linearly stable. On the other hand, if $\max\left \{\Re\left(\omega\right)\right \} > 0$, the perturbation grows exponentially in time and the base state is linearly unstable. Lastly, if $\max\left \{\Re\left(\omega\right)\right \} = 0$, the perturbation does not decay nor grow and the base state is marginally stable. The imposition of the boundary conditions at $r = r_\textn{other}$ described in Section~\ref{sec:Boundary conditions, constraints and initial conditions} results in the quantization of the set of valid wavenumbers, which is explained in detail in Section IIIC of Supplementary Material.

\subsection{Nondimensionalization}\label{sec:Nondimensionalization}

To make the equations more compact and derive key dimensionless parameters, in Table~\ref{tab:Scales used for nondimensionalization}, we define the scales that are used for nondimensionalizing the full model in Section~\ref{sec:Full model} and the perturbation expressions in Section~\ref{sec:Perturbations and linearization}. $\tilde{L}_y$ and $\tilde{L}_z$ are the aspect ratios in the $y$ and $z$ directions respectively. For convenience, we also define $\beta_\textn{D} = -\frac{z_-D_{-0}}{2\left(z_+D_{+0}-z_-D_{-0}\right)}$ (weighted ratio of cation and anion diffusivities), $\beta_\textn{m} = \nu_+c_0\Omega$ (ratio of atomic volume of solid metal to reciprocal electrolyte concentration), $\beta_\textn{v} = \frac{\beta_\textn{m}}{\beta_\textn{D}}$ and $\xi_+ = \frac{\nu_+c_0}{c_+^\Theta}$ (ratio of electrolyte concentration to standard cation concentration), and note that $\hat{c}_+ = \xi_+\left(\tilde{c}_--\tilde{\rho}_\textn{s}\right)$. Two important dimensionless parameters emerge from this nondimensionalization process, namely the Damk\"ohler number $\textn{Da} = \tilde{k}_0$ that is described earlier in Equation~\ref{eq:Da} and the capillary number $\textn{Ca}$ that is given by
\begin{equation}
  \textn{Ca} = \tilde{\gamma} = \frac{\Omega\gamma}{L_x k_\textn{B}T}, \label{eq:Ca}
\end{equation}
which is the ratio of the capillary constant $\frac{\Omega\gamma}{k_\textn{B}T}$~\cite{mullins_morphological_1963,mullins_stability_1964,sekerka_stability_1965} to the inter-electrode distance $L_x$, and $\tilde{\gamma}$ is the dimensionless isotropic surface energy of the metal/electrolyte interface.

\begingroup
\begin{table} \caption{Scales used for nondimensionalization.}\label{tab:Scales used for nondimensionalization}
  \centering
  \begin{ruledtabular}
  \begin{tabular}{cc}
    Variables and parameters & Scale \\ \hline  % chktex -2
    $x$, $y$, $z$, $L_y$, $L_z$, $r$, $r_\textn{m}$, $h$, $\lambda$, $\xi$ & $L_x$ \\
    $t$ & $\frac{L_x^2}{D_\textn{amb0}}$ (diffusion time) \\
    $c_\pm$ & $\nu_\pm c_0$ \\
    $\phi$, $\phi_\textn{e}$, $\Delta\phi^\textn{eq}$, $E^\Theta$, $\Delta\phi$, $\eta$ & $\frac{k_\textn{B}T}{e}$ (thermal voltage) \\
    $D_{\pm0}$ & $D_\textn{amb0}$ \\
    $\rho_\textn{s}$ & $z_+\nu_+ec_0 = -z_-\nu_-ec_0$ \\
    $F_\pm$ & $\frac{\epsilon_\textn{p}D_\textn{amb0}\nu_\pm c_0}{L_x}$ \\
    $u$, $u_x$, $v_\textn{I}$, $v_\textn{In}$ & $\frac{\epsilon_\textn{p}D_\textn{amb0}}{L_x}$ \\
    $J$ & $J_\limn$ \\
    $I$ & $I_\limn$ \\
    $j_0$, $J_\textn{F}$ & $\frac{J_\limn}{\epsilon_\textn{p}}$ \\
    $k_0$ & $\frac{J_\limn}{e\epsilon_p}$ \\
    $\gamma$ & $\frac{L_x k_\textn{B}T}{\Omega}$ \\
    $\mathscr{H}$, $k_y$, $k_z$, $k$ & $\frac{1}{L_x}$ \\
    $\omega$ & $\frac{D_\textn{amb0}}{L_x^2}$ (reciprocal diffusion time) \\
  \end{tabular}
  \end{ruledtabular}
\end{table}
\endgroup

To avoid cluttering the notation, we drop tildes for all dimensionless variables and parameters, and all variables and parameters are dimensionless in the following sections unless otherwise stated. We also rewrite the $\left(0\right)$ and $\left(1\right)$ superscripts, which denote the base and perturbed states respectively, as $0$ and $1$ subscripts respectively. Similarly, we drop the $0$ subscript for diffusivities and the $-$ subscript for anion-related variables and parameters. As shorthand, we use subscripts to denote partial derivatives with respect to $x$ (with the exception that $u_x$ denotes the $x$ component of the moving reference frame velocity $u$), $y$, $z$ and $t$, primes to denote total derivatives with respect to $x$, and an overhead dot to denote the total derivative with respect to $t$. All equations for the dimensionless full model are provided in Section II of Supplementary Material. Details for deriving the dimensionless equations for the base and perturbed states are provided in Section III of Supplementary Material, and we summarize them in Sections~\ref{sec:Base state} and~\ref{sec:Perturbed state} below.

\subsection{Base state}\label{sec:Base state}

The equations for the base state are obtained by substituting the perturbation expressions in Section~\ref{sec:Perturbations and linearization} into the full model in Section~\ref{sec:Full model} and matching terms at $\mathcal{O}\left(1\right)$. Equivalently, the base state is simply the full model specialized to 1D in the $x$ direction with the curvature-related terms dropped, which only appear at $\mathcal{O}\left(\epsilon\right)$. Therefore, at $\mathcal{O}\left(1\right)$, the governing PDEs (partial differential equations) are given by
\begin{align}
  c_{0,t} - D\left[c_{0,xx} + z\left(c_0\phi_{0,x}\right)_x\right] &= 0, \label{eq:O(1) species conservation} \\
  \beta_\textn{D}\left[\left(D-D_+\right)c_{0,xx} + z_+D_+\rho_\textn{s}\phi_{0,xx} - \left(z_+D_+-zD\right)\left(c_0\phi_{0,x}\right)_x\right] &= 0, \label{eq:O(1) charge conservation}
\end{align}
where the first PDE is the Nernst-Planck equation describing species conservation of anions and the second PDE is the charge conservation equation. The boundary conditions at the anode at $x=h_0^\textn{a}$ are given by
\begin{align}
  \phi_\textn{e}^\textn{a} &= 0, \\
  -D\left(-c_{0,x} - zc_0\phi_{0,x}\right) &= 0, \\
  \hat{n}\cdot J_0 &= j_{0,0}\left \{\exp\left(-\alpha n\eta_0\right)-\exp\left[\left(1-\alpha\right)n\eta_0\right]\right \}, \label{eq:O(1) Butler-Volmer BC at anode} \\
  -\dot{h}_0 &= -\beta_\textn{v}\hat{n}\cdot J_0 + u_x,
\end{align}
where $\eta_0 = \phi_\textn{e} - \phi_0 - \frac{1}{n}\ln\left[\xi_+\left(c_0-\rho_\textn{s}\right)\right] - E^\Theta$, $j_{0,0} = \textn{Da}n\left[\xi_+\left(c_0-\rho_\textn{s}\right)\right]^{1-\alpha}$ and $\hat{n}\cdot J_0 = \beta_\textn{D}\left[-\left(D-D_+\right)c_{0,x} - z_+D_+\rho_\textn{s}\phi_{0,x} + \left(z_+D_+-zD\right)c_0\phi_{0,x}\right]$. Since the unit normal at the cathode points in the opposite direction from that at the anode, the signs of the expressions involving $\hat{n}$ at the cathode are opposite to that at the anode. Therefore, the boundary conditions at the cathode at $x=h_0^\textn{c}$ are given by
\begin{align}
  -D\left(c_{0,x} + zc_0\phi_{0,x}\right) &= 0, \\
  \hat{n}\cdot J_0 &= j_{0,0}\left \{\exp\left(-\alpha n\eta_0\right)-\exp\left[\left(1-\alpha\right)n\eta_0\right]\right \}, \label{eq:O(1) Butler-Volmer BC at cathode} \\
  \dot{h}_0 &= -\beta_\textn{v}\hat{n}\cdot J_0 - u_x,
\end{align}
where $\eta_0 = \phi_\textn{e} - \phi_0 - \frac{1}{n}\ln\left[\xi_+\left(c_0-\rho_\textn{s}\right)\right] - E^\Theta$, $j_{0,0} = \textn{Da}n\left[\xi_+\left(c_0-\rho_\textn{s}\right)\right]^{1-\alpha}$ and $\hat{n}\cdot J_0 = \beta_\textn{D}\left[\left(D-D_+\right)c_{0,x} + z_+D_+\rho_\textn{s}\phi_{0,x} - \left(z_+D_+-zD\right)c_0\phi_{0,x}\right]$.

We pick $u_x\left(x=h_0^\textn{a}\right)$ and $u_x\left(x=h_0^\textn{c}\right)$ such that the positions of the anode and cathode in the base state remain stationary, i.e., $\dot{h}_0^\textn{a} = \dot{h}_0^\textn{c} = 0$. Therefore, $u_x = \beta_\textn{v}\hat{n}\cdot J_0\left(x=h_0^\textn{a}\right) = -\beta_\textn{v}\hat{n}\cdot J_0\left(x=h_0^\textn{c}\right)$ where the second equality automatically holds true because of charge conservation in the 1D $\mathcal{O}\left(1\right)$ base state. Physically, $u_x$ is equal to the velocity of the growing planar cathode/electrolyte interface or the dissolving planar anode/electrolyte interface in the base state. The initial conditions are given by
\begin{align}
  c_0\left(t=0\right) &= \beta_1, \quad h_0^\textn{a}\left(t=0\right) = 0, \quad h_0^\textn{c}\left(t=0\right) = 1.
\end{align}
Since $\dot{h}_0^\textn{a} = \dot{h}_0^\textn{c} = 0$, $h_0^\textn{a}\left(t\right) = 0$ and $h_0^\textn{c}\left(t\right) = 1$ at all $t$. For galvanostatic conditions in which we apply a current density $J_\textn{a}$ on the system, we impose
\begin{align}
  J_\textn{a} &= \left.\beta_\textn{D}\left[\left(D-D_+\right)c_{0,x} + z_+D_+\rho_\textn{s}\phi_{0,x} - \left(z_+D_+-zD\right)c_0\phi_{0,x}\right]\right\rvert_{x=h_0^\textn{c}} \\
  &= \left.\beta_\textn{D}\left[\left(D-D_+\right)c_{0,x} + z_+D_+\rho_\textn{s}\phi_{0,x} - \left(z_+D_+-zD\right)c_0\phi_{0,x}\right]\right\rvert_{x=h_0^\textn{a}}.
\end{align}
For potentiostatic conditions in which we apply an electric potential $V$ on the cathode, we impose $\phi_\textn{e}^\textn{c}=V$.

The equations for the time-dependent base state cannot generally be solved analytically, therefore we would have to solve them numerically. However, at steady state, the base state admits semi-analytical solutions for any $\rho_\textn{s}$~\cite{khoo_theory_2018}. Specifically, $c_0$, $\phi_{0,x}$ and their spatial derivatives can be analytically expressed in terms of the Lambert W function~\cite{corless_lambertw_1996}. On the other hand, $\phi_0$ is known semi-analytically because it can be analytically expressed in terms of the Lambert W function up to an additive constant, which is a function of $J_\textn{a}$ and $\rho_\textn{s}$ and is found by numerically solving the algebraic Butler-Volmer equations given by Equations~\ref{eq:O(1) Butler-Volmer BC at anode} and~\ref{eq:O(1) Butler-Volmer BC at cathode} with MATLAB's $\mathtt{fsolve}$ or $\mathtt{fzero}$ function.

\subsection{Perturbed state}\label{sec:Perturbed state}

To derive the equations for the perturbed state at $\mathcal{O}\left(\epsilon\right)$, we substitute the perturbation expressions in Section~\ref{sec:Perturbations and linearization} into the full model in Section~\ref{sec:Full model} and match terms at $\mathcal{O}\left(\epsilon\right)$. One important outcome is that the curvature-related terms appear as functions of $k^2$ because they are associated with second-order spatial partial derivatives in the $y$ and $z$ directions. At $\mathcal{O}\left(\epsilon\right)$, the governing ODEs (ordinary differential equations) are given by
\begin{align}
  & D\left \{c_1''-k^2c_1 + z\left[\left(c_0\phi_1'+\phi_{0,x}c_1\right)_x-k^2c_0\phi_1\right]\right \} = \omega c_1, \label{eq:ODE for c_1} \\
  & \left(D-D_+\right)\left(c_1''-k^2c_1\right) + z_+D_+\rho_\textn{s}\left(\phi_1''-k^2\phi_1\right) - \left(z_+D_+-zD\right)\left[\left(c_0\phi_1'+\phi_{0,x}c_1\right)_x-k^2c_0\phi_1\right] = 0, \label{eq:ODE for phi_1}
\end{align}
where the first ODE describes the perturbation in species conservation of anions and the second ODE describes the perturbation in charge conservation. For brevity, we define $\alpha_3 = -\alpha\exp\left(-\alpha n\eta_0\right) - \left(1-\alpha\right)\exp\left[\left(1-\alpha\right)n\eta_0\right]$. The boundary conditions at the anode at $x=h_0^\textn{a}$ are given by
\begin{align}
  & c_{0,t}h_1^\textn{a} -D\left[-c_1' - z\left(c_0\phi_1'+\phi_{0,x}c_1\right)\right] = 0, \\
  & \beta_\textn{v}j_{0,0}\left(\hat{D}_1h_1^\textn{a} + \hat{D}_2c_1 + \hat{D}_3\phi_1\right) = \omega h_1^\textn{a}, \\
  & \beta_\textn{m}\left[-\left(D-D_+\right)c_1' - z_+D_+\rho_\textn{s}\phi_1' + \left(z_+D_+-zD\right)\left(c_0\phi_1'+\phi_{0,x}c_1\right)\right] = \omega h_1^\textn{a},
\end{align}
where the $\hat{D}_1$, $\hat{D}_2$ and $\hat{D}_3$ parameters are
\begin{align}
  \hat{D}_1 &= \alpha_3n\left(-\phi_{0,x}+\frac{\gamma k^2}{n}\right) + \frac{\exp\left(-\alpha n\eta_0\right)c_{0,x}}{c_0-\rho_\textn{s}}, \quad \hat{D}_2 = \frac{\exp\left(-\alpha n\eta_0\right)}{c_0-\rho_\textn{s}}, \quad \hat{D}_3 = -\alpha_3n.
\end{align}
Because the unit normal at the cathode is in the opposite direction from that at the anode, the signs of the expressions involving $\hat{n}$ at the cathode are opposite to that at the anode. Hence, the boundary conditions at the cathode at $x=h_0^\textn{c}$ are given by
\begin{align}
  & -c_{0,t}h_1^\textn{c} -D\left[c_1' + z\left(c_0\phi_1'+\phi_{0,x}c_1\right)\right] = 0, \\
  & \beta_\textn{v}j_{0,0}\left(\hat{G}_1h_1^\textn{c} + \hat{G}_2c_1 + \hat{G}_3\phi_1\right) = -\omega h_1^\textn{c}, \label{eq:O(epsilon) BV BC at cathode} \\
  & \beta_\textn{m}\left[\left(D-D_+\right)c_1' + z_+D_+\rho_\textn{s}\phi_1' - \left(z_+D_+-zD\right)\left(c_0\phi_1'+\phi_{0,x}c_1\right)\right] = -\omega h_1^\textn{c},
\end{align}
where the $\hat{G}_1$, $\hat{G}_2$ and $\hat{G}_3$ parameters are
\begin{align}
  \hat{G}_1 &= \alpha_3n\left(-\phi_{0,x}-\frac{\gamma k^2}{n}\right) + \frac{\exp\left(-\alpha n\eta_0\right)c_{0,x}}{c_0-\rho_\textn{s}}, \quad \hat{G}_2 = \frac{\exp\left(-\alpha n\eta_0\right)}{c_0-\rho_\textn{s}}, \quad \hat{G}_3 = -\alpha_3n.
\end{align}

The capillary number $\textn{Ca} = \gamma$ appears in the $\hat{D}_1$ and $\hat{G}_1$ parameters in the form of $\gamma k^2$, which is the source of the surface stabilizing effect that arises from the surface energy penalty incurred in creating additional surface area. The competition between this surface stabilizing effect and the surface destabilizing effect arising from the $c_0$, $c_{0,x}$ and $\phi_{0,x}$ fields sets the scale for the critical wavenumber $k_\textn{c}$, which is the wavenumber at which the perturbation growth rate $\omega$ is $0$ and the electrode surface is marginally stable.

\subsection{Discretization of perturbed state}\label{sec:Discretization of perturbed state}

Without making further approximations, the equations for the perturbed state do not admit analytical solutions, thus we have to resort to numerical methods to solve them. To do so, the equations for the perturbed state are spatially discretized over a uniform grid with $N$ grid points and a grid spacing $\Delta x = \frac{1}{N-1}$ using second-order accurate finite differences~\cite{leveque_finite_2007}. Details of this discretization process are provided in Section IV of Supplementary Material. In summary, the discretized equations can be written as a generalized eigenvalue problem given by
\begin{align}
  Yv &= \omega Zv, \quad v = \begin{bmatrix}
                               h_1^\textn{a} & c_{1,1} & \phi_{1,1} & c_{1,2} & \phi_{1,2} & \cdots & c_{1,N-1} & \phi_{1,N-1} & c_{1,N} & \phi_{1,N} & h_1^\textn{c}
                             \end{bmatrix}^\mathrm{T},
\end{align}
where $Y, Z \in \mathbb{R}^{\left(2N+2\right)\times\left(2N+2\right)}$, $v \in \mathbb{C}^{2N+2}$, $\omega \in \mathbb{C}$ and the second subscript in $c_{1,i}$ and $\phi_{1,i}$ for $i=1,2,\ldots,N$ denotes the grid point index. In the context of a generalized eigenvalue problem, the eigenvector $v$ consists of the complex-valued amplitudes $c_1$, $\phi_1$, $h_1^\textn{a}$ and $h_1^\textn{c}$ evaluated at the grid points, and the eigenvalue is the complex-valued growth rate $\omega$. Although $Y$ is non-singular, the time-independent terms in the equations for the perturbed state introduce rows of zeros in $Z$, therefore $Z$ is singular and the generalized eigenvalue problem cannot be reduced to a standard eigenvalue problem. Specifically, $Y$ is non-singular with rank $2N+2$ while $Z$ is singular with rank $N$, and the total number of eigenvalues is $2N+2$.

Because $Z$ is singular with rank $N$, there are $N$ finite eigenvalues and $N+2$ infinite eigenvalues. This mathematical property is not always consistently noted in past literature on linear stability analysis of electrodeposition, although Sundstr\"om and Bark did mention that $N$ different eigenvalues are obtained with $N$ grid points that give rise to $2N+2$ equations without explicitly stating that the other $N+2$ eigenvalues are infinite~\cite{sundstrom_morphological_1995}. The infinite eigenvalues are physically irrelevant to the linear stability analysis~\cite{valerio_filtering_2007,kawano_decoupling_2013}, therefore we would want to focus on solving for the finite eigenvalues. This can be achieved by mapping the infinite eigenvalues to other arbitrarily chosen points in the complex plane via simple matrix transformations~\cite{goussis_removal_1989}. Details of how these transformations are performed are given in Section IV of Supplementary Material. There are methods for directly removing the infinite eigenvalues such as the ``reduced'' method~\cite{gary_matrix_1970,peters_$ax_1970,valerio_filtering_2007} but they are more intrusive and require more extensive matrix manipulations as compared to the mapping technique~\cite{goussis_removal_1989} that we use.

The modified generalized eigenvalue problem that results from these transformations can then be solved using any eigenvalue solver. For linear stability analysis, we only need to find the eigenvalue with the largest real part instead of all the finite eigenvalues. Since the time complexity of finding all the eigenvalues typically scales as $\mathcal{O}\left(N^3\right)$ while that for finding $k \leq N$ of them, where $k=1$ in our case, scales as $\mathcal{O}\left(kN^2\right)$, the computational cost is dramatically reduced by a factor of $\mathcal{O}\left(N\right)$ if we use an eigenvalue solver that can find subsets of eigenvalues and eigenvectors such as MATLAB's $\mathtt{eigs}$ solver.

\subsection{Numerical implementation}

The equations for the time-dependent base state in Section~\ref{sec:Base state} are numerically solved using the finite element method in COMSOL Multiphysics 5.3a. The eigenvalue with the largest real part and its corresponding eigenvector from the generalized eigenvalue problem for the perturbed state in Section~\ref{sec:Discretization of perturbed state} are then solved for using the $\mathtt{eigs}$ function in MATLAB R2018a. When the $\mathtt{eigs}$ function occasionally fails to converge for small values of the wavenumber $k$, we use Rostami and Xue's eigenvalue solver based on the matrix exponential~\cite{rostami_robust_2018,xue_solver_nodate}, which is more robust than the $\mathtt{eigs}$ function. The colormaps used for some of the plots in Section~\ref{sec:Results} are obtained from BrewerMap~\cite{cobeldick_colorbrewer:_nodate}, which is a MATLAB program available in the MATLAB File Exchange that implements the ColorBrewer colormaps~\cite{harrower_colorbrewer.org:_2003}.

\section{Results}\label{sec:Results}

Because of the large number of dimensionless parameters present, the parameter space is too immense to be explored thoroughly in this paper. Instead, the key dimensionless parameters that we focus on and vary are $\rho_\textn{s}$, $\textn{Da}$ and $J_\textn{a}$ under galvanostatic conditions. $\rho_\textn{s} = 0$ corresponds to the classical case of an uncharged nanoporous medium while $\rho_\textn{s} \neq 0$ allows us to depart from this classical case and study its effects on the linear stability of the electrode surface. Experimentally, $\rho_\textn{s}$ can be tuned via layer-by-layer deposition of polyelectrolytes~\cite{han_over-limiting_2014,han_dendrite_2016,hammond_form_2004} or tethered immobilized anions~\cite{tu_nanostructured_2015}. $\textn{Da}$ is very sensitive to the specific reactions considered and varies significantly in practice. We focus on galvanostatic conditions instead of potentiostatic conditions because when an overlimiting current $J_\textn{a} > 1$ is applied on a classical system with $\rho_\textn{s} = 0$, as discussed in Section~\ref{sec:Transport}, the Sand's time $t_\textn{s}$ provides a time scale at which the electric field at the cathode diverges that causes the perturbation growth rate to diverge too. This allows us to focus the linear stability analysis on times immediately before, at and immediately after $t_\textn{s}$.

For the results discussed in Sections~\ref{sec:Convergence analysis},~\ref{sec:Parameter sweeps},~\ref{sec:Comparison between numerical and approximate solutions} and~\ref{sec:Pulse electroplating and pulse charging} below, we assume the following dimensional quantities for a typical electrolyte in a typical nanoporous medium: $T = 298\,\textn{K}$, $M_\textn{m} = 6.941\,\textn{g}/\textn{mol}$ (arbitrarily pick lithium metal)~\cite{rumble_crc_2017}, $\rho_\textn{m} = 0.534\,\textn{g}/\textn{cm}^3$ (arbitrarily pick lithium metal)~\cite{rumble_crc_2017}, $L_x = 60\,\mu\textn{m}$, $L_y = L_z = 100 L_x = 6\,\textn{mm}$ ($L_x \ll L_y = L_z$ to model a thin gap cell that reduces effects of gravity-induced convection (buoyancy)~\cite{elezgaray_linear_1998}), $c_0 = 10\,\textn{mM}$ (note that $c_0$ here is the dimensional initial neutral salt bulk concentration, not the dimensionless base state anion concentration), $c_+^\Theta = 1\,\textn{M} = 10^3\,\textn{mol}/\textn{m}^3$ (standard concentration) and $\gamma = 1\,\textn{J}/\textn{m}^2$ (typical surface energy of metal/electrolyte interface)~\cite{sundstrom_morphological_1995}. Corresponding to these dimensional quantities, all dimensionless parameters that are used for the results in Sections~\ref{sec:Convergence analysis},~\ref{sec:Parameter sweeps},~\ref{sec:Comparison between numerical and approximate solutions} and~\ref{sec:Pulse electroplating and pulse charging} are given in Table~\ref{tab:Dimensionless parameters}.

\begingroup
\begin{table} \caption{Dimensionless parameters that are used for results in Sections~\ref{sec:Convergence analysis},~\ref{sec:Parameter sweeps},~\ref{sec:Comparison between numerical and approximate solutions} and~\ref{sec:Pulse electroplating and pulse charging} for a typical electrolyte in a typical nanoporous medium.}\label{tab:Dimensionless parameters}
  \centering
  \begin{ruledtabular}
  \begin{tabular}{ccc}
    Dimensionless parameter & Description & Value \\ \hline  % chktex -2
    $\nu_+$ & \makecell{Number of cations formed from complete \\ dissolution of 1 molecule of neutral salt} & $1$ \\
    $\nu$ & \makecell{Number of anions formed from complete \\ dissolution of 1 molecule of neutral salt} & $1$ \\
    $z_+$ & Cation charge number & $1$ \\
    $z$ & Anion charge number & $-1$ \\
    $D_+$ & Cation diffusivity & $1$ \\
    $D$ & Anion diffusivity & $1$ \\
    $n$ & \makecell{Number of electrons transferred in \\ charge transfer reaction} & $1$ \\
    $\alpha$ & Charge transfer coefficient & $0.5$ \\
    $\textn{Ca} = \gamma$ & \makecell{Capillary number, ratio of capillary constant to \\ inter-electrode distance (Equation~\ref{eq:Ca})} & $8.74 \times 10^{-5}$ \\
    $\beta_\textn{m}$ & \makecell{Ratio of atomic volume of solid metal to \\ reciprocal electrolyte concentration} & $1.30 \times 10^{-4}$ \\
    $\beta_\textn{D}$ & Weighted ratio of cation and anion diffusivities & $0.25$ \\
    $\beta_\textn{v}$ & Ratio of $\beta_\textn{m}$ to $\beta_\textn{D}$ & $5.20 \times 10^{-4}$ \\
    $\xi_+$ & \makecell{Ratio of electrolyte concentration to \\ standard cation concentration} & $0.01$ \\
    $E^\Theta$ & Standard electrode potential & $0$ \\
    $L_y$ & Aspect ratio in $y$ direction & $100$ \\
    $L_z$ & Aspect ratio in $z$ direction & $100$ \\
    $\rho_\textn{s}$ & \makecell{Ratio of background charge density to \\ electrolyte concentration} & \makecell{$-1, -0.75, -0.5, -0.25$, \\ $-0.05, 0, 0.05$} \\
    $\textn{Da}$ & \makecell{Damk\"ohler number, ratio of reaction rate to \\ diffusion rate (Equation~\ref{eq:Da})} & $0.1, 1, 10$ \\
  \end{tabular}
  \end{ruledtabular}
\end{table}
\endgroup

\subsection{Approximations}\label{sec:Approximations}

At the heart of the linear stability analysis is the competition between the destabilizing effect that arises from the amplification of surface protrusions by diffusive fluxes in a positive feedback loop and the stabilizing effect that arises from the surface energy penalty incurred in the creation of additional surface area. Therefore, in the dispersion relation $\omega\left(k\right)$, we expect to see some local maxima or possibly just a single global maximum, which we denote as $\left \{k_\maxn, \omega_\maxn\right \}$, where the electrode surface is maximally unstable. We also expect to see a critical wavenumber $k_\textn{c}$ corresponding to $\omega = 0$ where the electrode surface is marginally stable. When $k$ is larger than $k_\textn{c}$, $\omega$ is always negative because the surface energy stabilizing effect always dominates when the wavenumber is sufficiently large. We note that $k_\textn{c}$ is always greater than $k_\maxn$. Corresponding to $k_\maxn$ and $k_\textn{c}$ are the maximum wavelength $\lambda_\maxn = \frac{2\pi}{k_\maxn}$ and critical wavelength $\lambda_\textn{c} = \frac{2\pi}{k_\textn{c}}$ respectively. In a porous medium, the characteristic pore size $h_\textn{c} = 2d_\textn{p}$, where $d_\textn{p}$ is the pore diameter, sets a threshold or cutoff for overall electrode surface stabilization: we should observe stabilization if $h_\textn{c}$ is smaller than $\lambda_\textn{c}$~\cite{tikekar_stability_2014}. If $h_\textn{c}$ is larger than $\lambda_\textn{c}$, then the most unstable eigenmode dominates the electrode surface growth with a growth rate of $\omega_\maxn$ and the characteristic length scale of this instability is $\lambda_\maxn$. Therefore, $\left \{k_\maxn, \omega_\maxn\right \}$ and $k_\textn{c}$ are the most physically informative points of the dispersion relation. We now derive an approximation for the dispersion relation $\omega\left(k\right)$ that is valid at high values of $k$ relative to $k_\textn{c}$ and will be useful for computing $\left \{k_\maxn, \omega_\maxn\right \}$ and $k_\textn{c}$ quickly and accurately because $k_\maxn$ and $k_\textn{c}$ tend to be large. The approximation is also useful for verifying the full numerical solution at high $k$, which will be discussed in Section~\ref{sec:Convergence analysis}.

When $k$ is sufficiently large compared to $k_\textn{c}$, at the cathode at $x = h_0^\textn{c} = 1$, we expect $k^2c_1$ to balance $c_1''$, and $k^2\phi_1$ to balance $\phi_1''$ in Equations~\ref{eq:ODE for c_1} and~\ref{eq:ODE for phi_1} respectively. Therefore, $k^{-2}$ is a small parameter multiplying the highest order spatial derivative terms $c_1''$ and $\phi_1''$, and the spatial profiles for $c_1$ and $\phi_1$ form a boundary layer with characteristic thickness $k^{-1}$. Hence, as an ansatz for the boundary layer analysis, we assume
\begin{align}
  c_1 &= A\exp\left[k\left(x-1\right)\right], \quad \phi_1 = B\exp\left[k\left(x-1\right)\right],
\end{align}
where $A$ and $B$ are arbitrary constants that are determined from the boundary conditions at $x = h_0^\textn{c} = 1$. By assuming such an ansatz, the cathode is effectively decoupled from the anode and the perturbation growth rate is entirely dependent on the boundary conditions at the cathode. The validity of this ansatz is corroborated by our observations that $h_1^\textn{c} \gg h_1^\textn{a}$ generally in our numerical simulations, especially at large values of $k$, which was also observed by Sundstr\"om and Bark~\cite{sundstrom_morphological_1995}. Imposing the boundary conditions at $x = h_0^\textn{c} = 1$, we obtain
\begin{align}
  \xi_1\left(k\right) &= \frac{c_{0,t}}{zc_0Dk}, \quad \xi_2\left(k\right) = -\frac{z\phi_{0,x}+k}{zc_0k}, \quad B\left(k\right) = -\xi_1h_1 + \xi_2A, \\
  A\left(k\right) &= -\frac{\beta_\textn{v}j_{0,0}\left(\hat{G}_1-\xi_1\hat{G}_3\right) - \beta_\textn{m}\alpha_5\xi_1k}{\beta_\textn{v}j_{0,0}\left(\hat{G}_2+\xi_2\hat{G}_3\right) - \beta_\textn{m}\left[\left(\alpha_1-\alpha_5\xi_2\right)k-\alpha_2\phi_{0,x}\right]}h_1, \\
  \omega\left(k\right) &= \beta_\textn{m}\left \{\frac{\left[\left(\alpha_1-\alpha_5\xi_2\right)k-\alpha_2\phi_{0,x}\right]\left[\beta_\textn{v}j_{0,0}\left(\hat{G}_1-\xi_1\hat{G}_3\right)-\beta_\textn{m}\alpha_5\xi_1k\right]}{\beta_\textn{v}j_{0,0}\left(\hat{G}_2+\xi_2\hat{G}_3\right)-\beta_\textn{m}\left[\left(\alpha_1-\alpha_5\xi_2\right)k-\alpha_2\phi_{0,x}\right]} - \alpha_5\xi_1k\right \}, \label{eq:omega}
\end{align}
where we define $\alpha_1 = D - D_+$, $\alpha_2 = z_+D_+ - zD$ and $\alpha_5 = \alpha_2c_0 - z_+D_+\rho_\textn{s}$ for brevity.

Approximate values of $\left \{k_\maxn, \omega_\maxn\right \}$ can be obtained by solving $\omega'\left(k\right) = 0$ and requiring $\omega''\left(k\right) < 0 $ where the primes indicate total derivatives with respect to $k$. In addition, by solving $\omega\left(k\right) = 0$, we can obtain approximate values of $k_\textn{c}$. However, this process is tedious because the first term inside the braces in Equation~\ref{eq:omega} is a rational function that consists of polynomials in $k$ of relatively high degrees. Specifically, after multiplying the numerator and denominator of this term by $k$, it becomes a rational function with a numerator that is a polynomial in $k$ of degree $4$ and a denominator that is a polynomial in $k$ of degree $2$. Therefore, for the purpose of quickly approximating $\left \{k_\maxn, \omega_\maxn\right \}$ and $k_\textn{c}$, we first find a simpler and yet still accurate analytical approximation for $k_\textn{c}$, which can then used as an initial guess for numerically solving for $\left \{k_\maxn, \omega_\maxn\right \}$ using Equation~\ref{eq:omega} with MATLAB's $\mathtt{fminbnd}$ optimizer. Such an approximation can be obtained by assuming $k_\textn{c}$ is large enough that $\hat{G}_2c_1 \ll \hat{G}_1h_1$ and $\hat{G}_3\phi_1 \ll \hat{G}_1h_1$ and then setting $\omega = 0$ in Equation~\ref{eq:O(epsilon) BV BC at cathode}, thus resulting in
\begin{equation}
  k_\textn{c} = \left \{\frac{1}{\alpha_3\gamma}\left[-\alpha_3n\phi_{0,x}+\frac{\exp\left(-\alpha n\eta_0\right)c_{0,x}}{c_0-\rho_\textn{s}}\right]\right \}^\frac{1}{2}. \label{eq:k_c}
\end{equation}
We observe that $k_\textn{c}$ scales as $\textn{Ca}^{-\frac{1}{2}} = \gamma^{-\frac{1}{2}}$, which is expected because the surface energy stabilizing effect appears in the form of $\gamma k^2$ in $\hat{G}_1$ in Equation~\ref{eq:O(epsilon) BV BC at cathode}, and this scaling agrees with that obtained in previous work done on linear stability analysis of electrodeposition~\cite{sundstrom_morphological_1995,elezgaray_linear_1998,krishnan_instability_2002,tikekar_stability_2014,nielsen_morphological_2015}.

\subsection{Convergence analysis}\label{sec:Convergence analysis}

Before analyzing the physical significance of the linear stability analysis results, we would want to first establish the accuracy and convergence of the full numerical solution of the dispersion relation $\omega\left(k\right)$. To this end, we perform a numerical convergence analysis in which we examine the convergence of the numerical solution as the number of grid points $N$ increases. At the same time, we also compute the approximate $\omega\left(k\right)$ given by Equation~\ref{eq:omega} because we expect the numerical and approximate solutions to agree well at high values of $k$; this therefore provides another way of checking the accuracies of both the numerical and approximate solutions.

\begin{figure}
  \centering
  \includegraphics[scale=0.09]{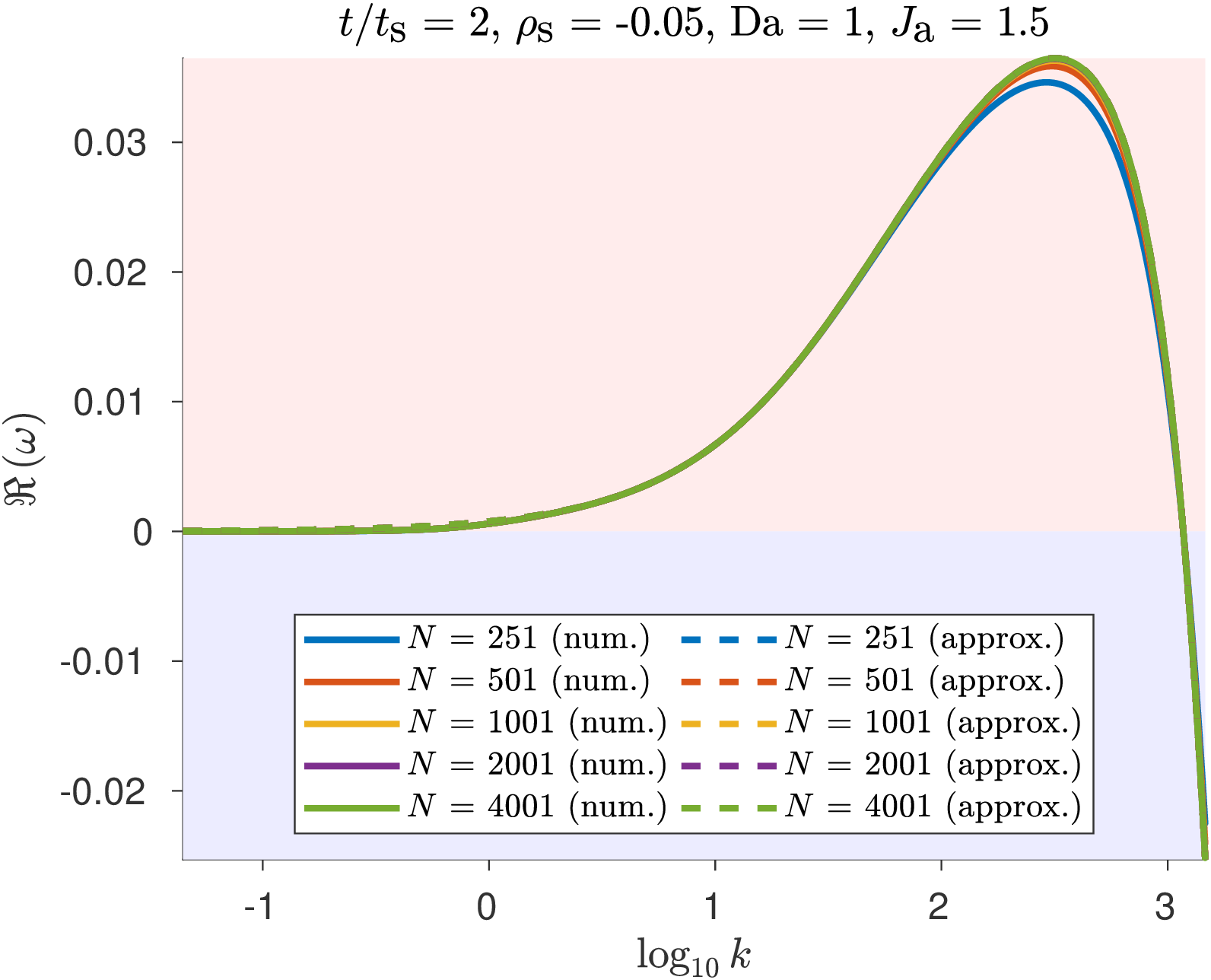}
  \includegraphics[scale=0.09]{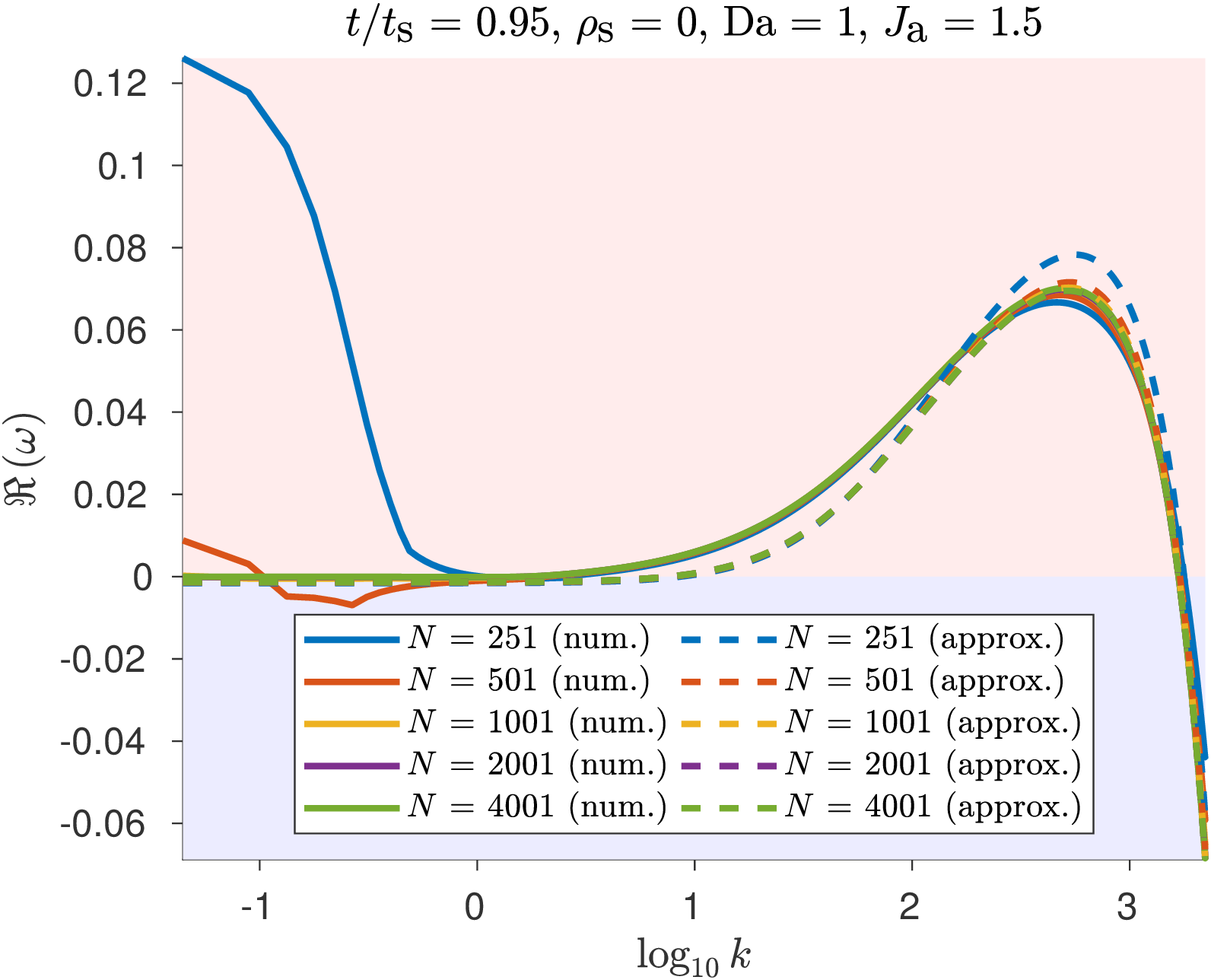}
  \includegraphics[scale=0.09]{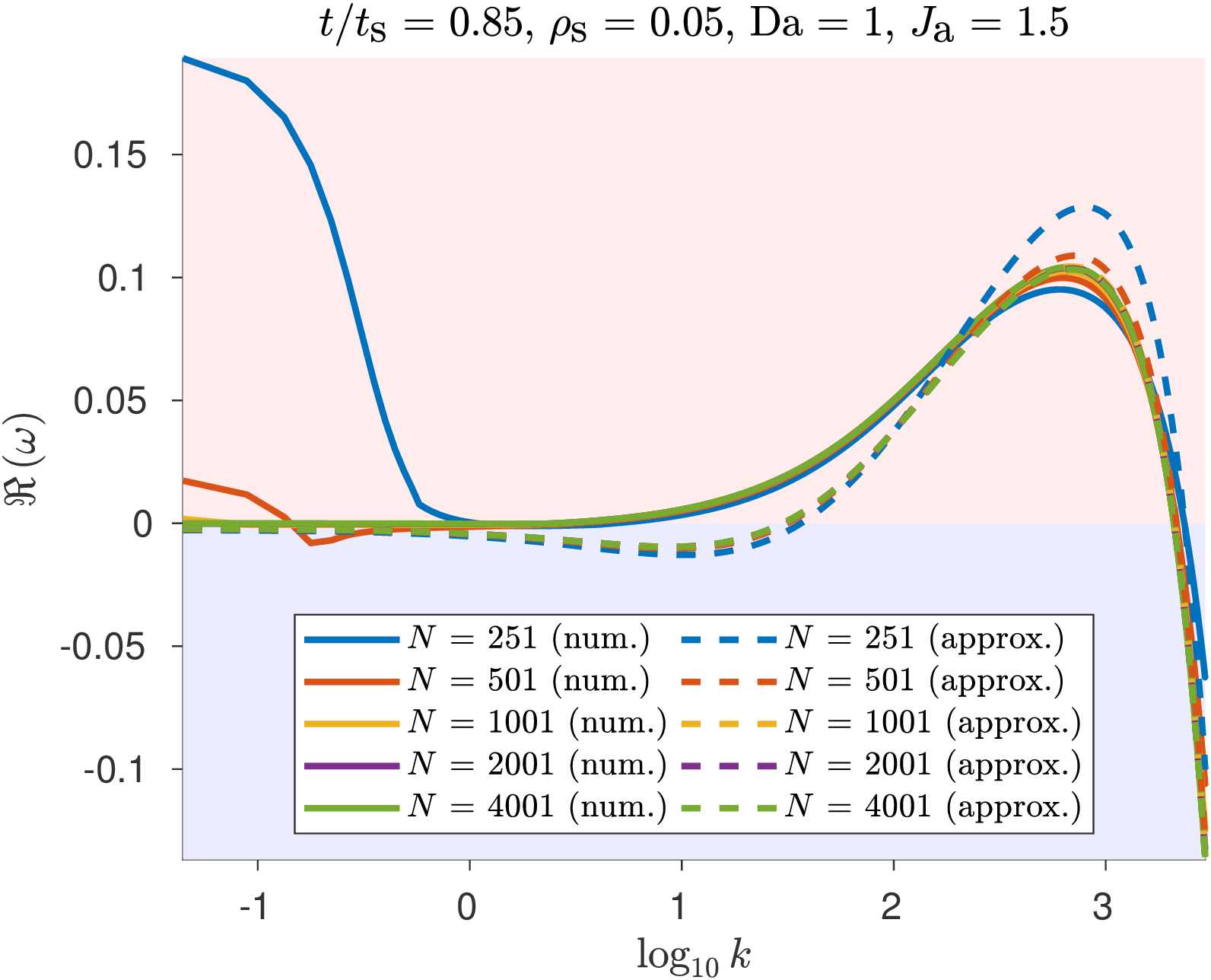}
  \caption{Convergence plots of $\Re\left(\omega\right)$ against $\log_{10}k$ with $\textn{Da} = 1$ and $J_\textn{a} = 1.5$ (overlimiting current) for $\rho_\textn{s} \in \left \{-0.05, 0, 0.05\right \}$ and $N \in \left \{251, 501, 1001, 2001, 4001\right \}$ used in convergence analysis. The $\frac{t}{t_\textn{s}}$ values to which the curves correspond are indicated in the figure titles. In the legends, ``num.'' refers to numerical solutions while ``approx.'' refers to approximate solutions.}\label{fig:Convergence analysis}
\end{figure}

To demonstrate how the numerical dispersion relation $\omega\left(k\right)$ changes with $N$, we fix $\textn{Da} = 1$ and $J_\textn{a} = 1.5$ (overlimiting current) and plot numerically computed $\Re\left(\omega\right)$ against $\log_{10}k$ for $\rho_\textn{s} \in \left \{-0.05, 0, 0.05\right \}$ and $N \in \left \{251, 501, 1001, 2001, 4001\right \}$ at specific $\frac{t}{t_\textn{s}}$ values in Figure~\ref{fig:Convergence analysis}. As expected, the numerical solutions converge quickly as $N$ increases from $251$ to $4001$. For $\rho_\textn{s} = 0$ and $\rho_\textn{s} = 0.05$ at small values of $k$, when the value of $N$ is small at $251$ or $501$, we observe that there are anomalously large values of $\Re\left(\omega\right)$ that vanish at larger values of $N$. This is because when $N$ is too small, the grid is not sufficiently fine to accurately resolve the base state variables, in particular the rapidly increasing electric field at the cathode near $t_\textn{s}$, thus leading to an overestimation of the destabilizing effect caused by electrodiffusion and an underestimation of the stabilizing effect caused by surface energy. The numerical and approximate solutions also expectedly agree well with each other at large values of $k$ and this agreement improves as $N$ increases, thus confirming that the approximations are accurate at high $k$.

Because we are mostly interested in the $k_\maxn$, $\omega_\maxn$ and $k_\textn{c}$ points on the $\omega\left(k\right)$ curve, we plot them against $N$ in Figure~\ref{fig:k_max, omega_max and k_c against N}. We observe that the numerically computed $k_\maxn$, $\omega_\maxn$ and $k_\textn{c}$ curves rapidly level off and converge to constant values as $N$ increases. The numerical and approximate solutions also agree very well as $N$ increases, which is expected because $k_\maxn$ and $k_\textn{c}$ are large and the approximations are accurate at high $k$. As a compromise between numerical accuracy and computational time, we pick $N = 1001$ for all numerical and approximate solutions computed in the following sections.

\begin{figure}
  \centering
  \includegraphics[scale=0.37]{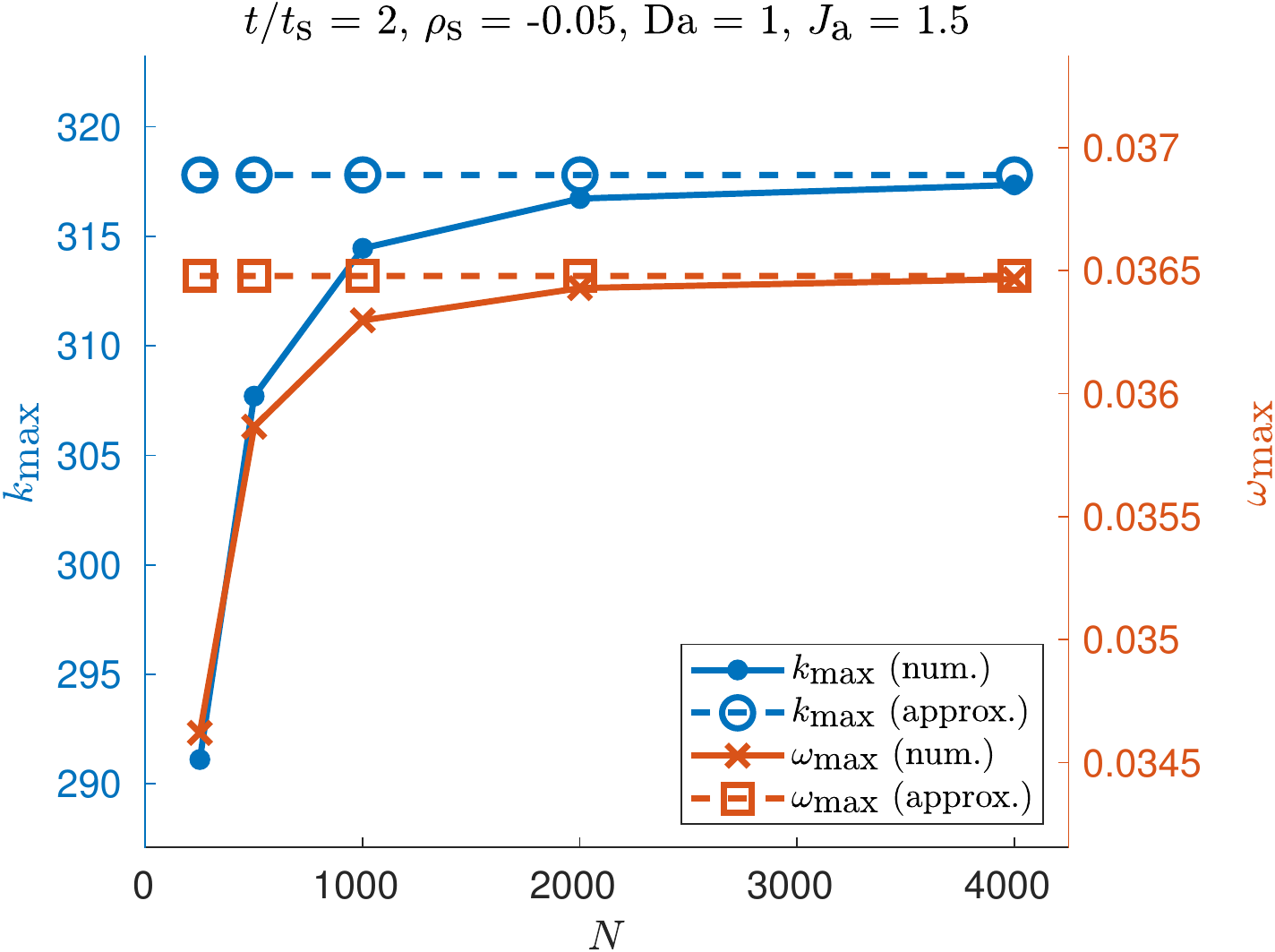}
  \includegraphics[scale=0.37]{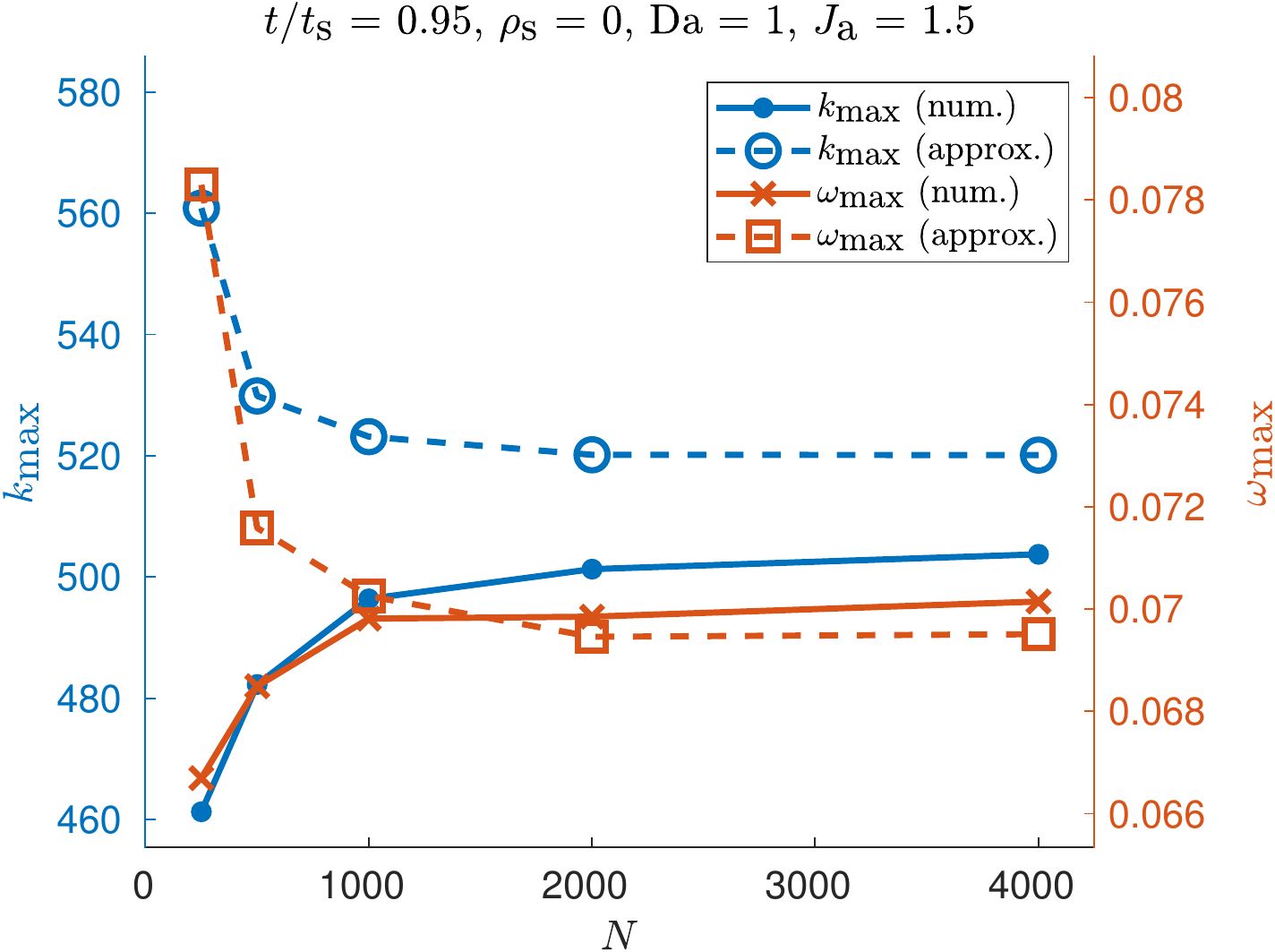}
  \includegraphics[scale=0.37]{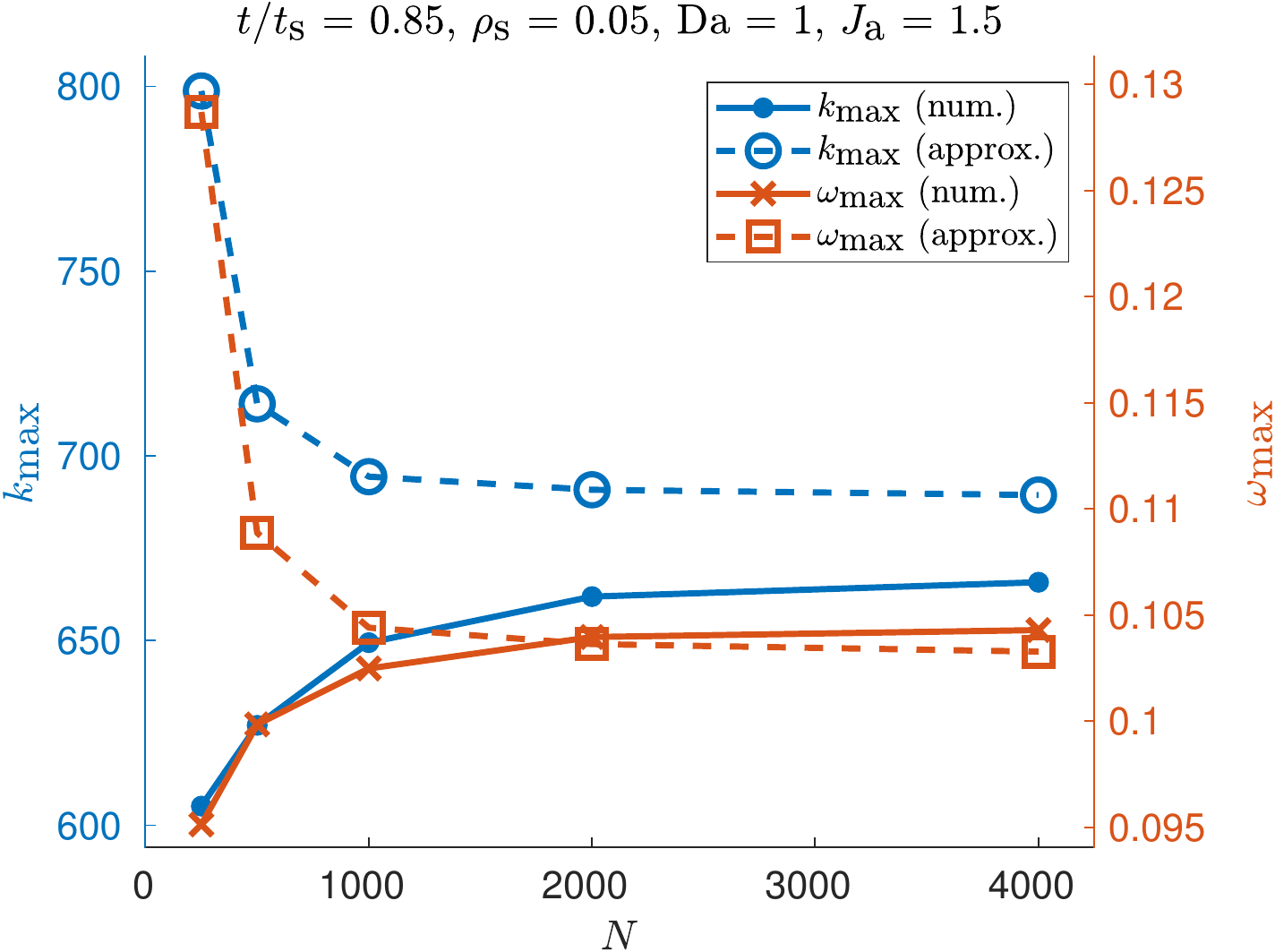}
  \includegraphics[scale=0.37]{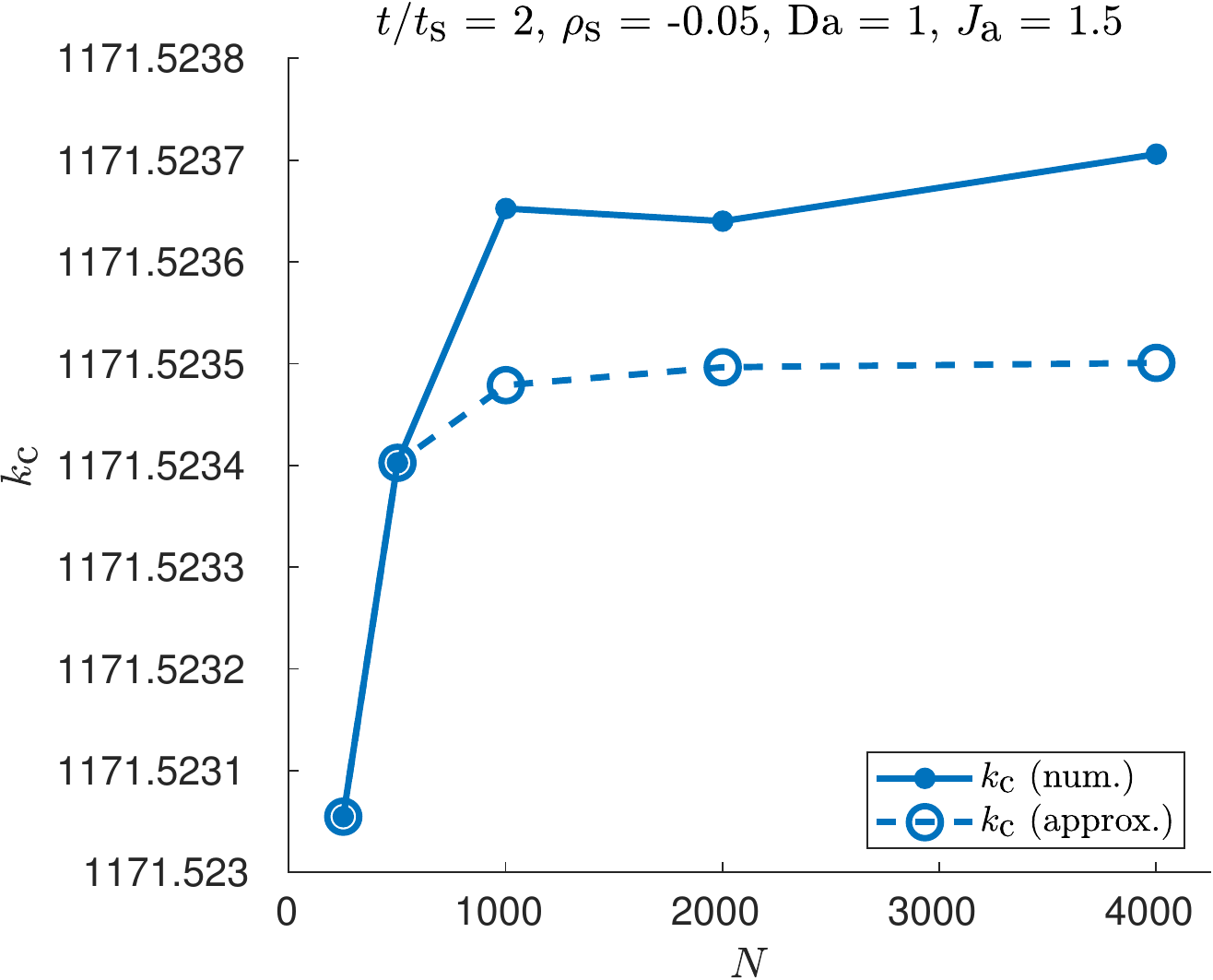}
  \includegraphics[scale=0.37]{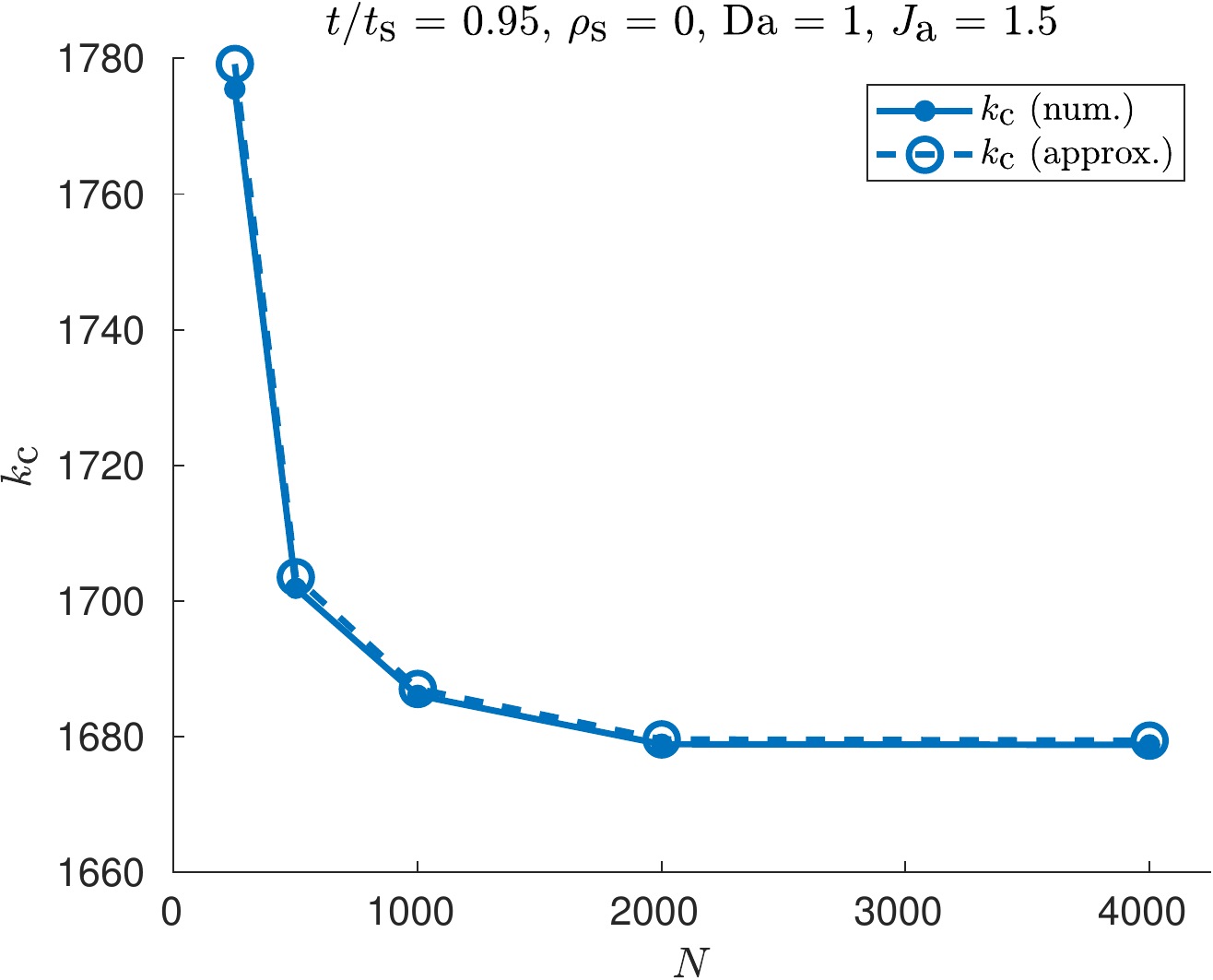}
  \includegraphics[scale=0.37]{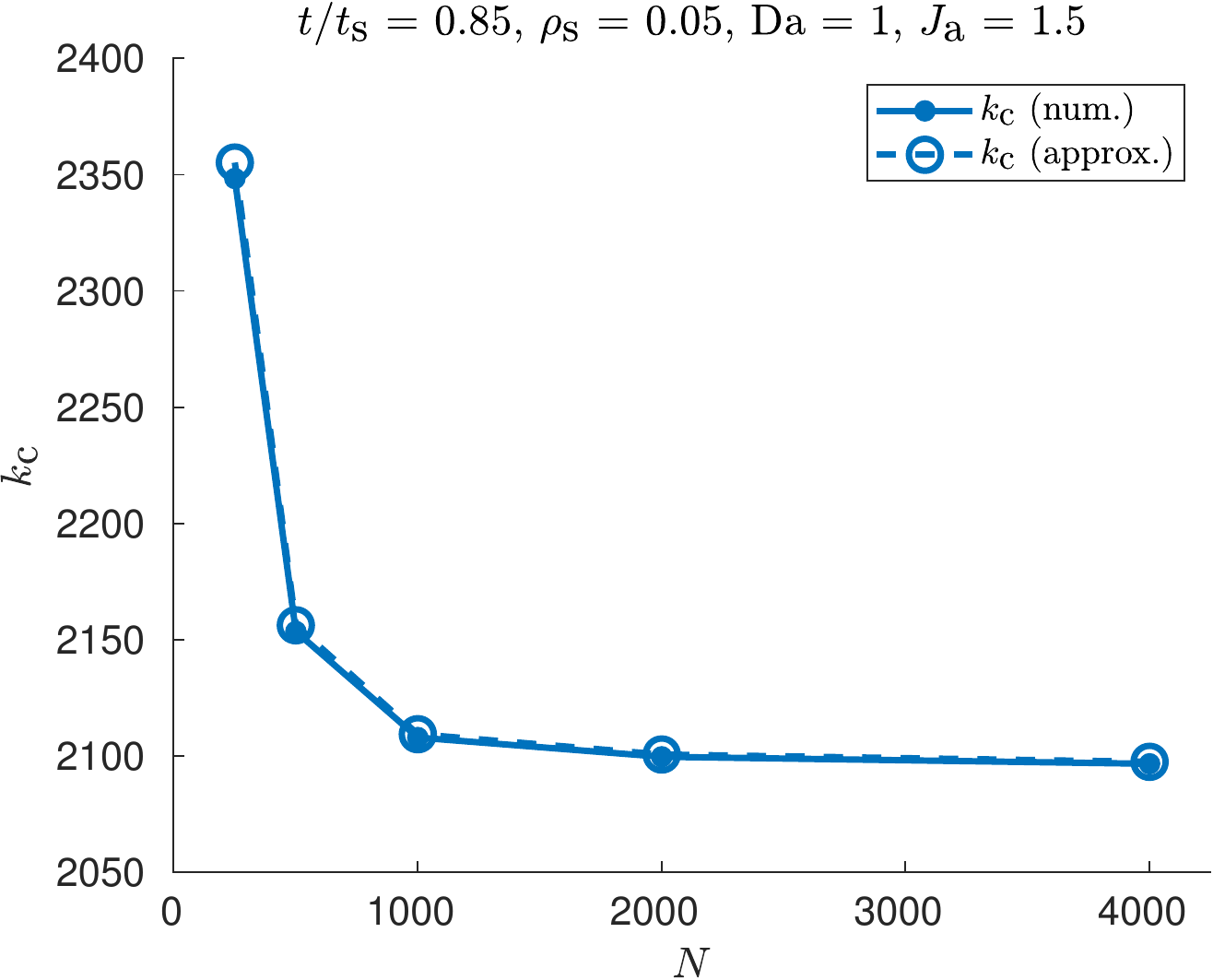}
  \caption{Convergence plots of $k_\maxn$, $\omega_\maxn$ and $k_\textn{c}$ against $N$ with $\textn{Da} = 1$ and $J_\textn{a} = 1.5$ (overlimiting current) for $\rho_\textn{s} \in \left \{-0.05, 0, 0.05\right \}$ used in convergence analysis. Top row: Convergence plots of $k_\maxn$ and $\omega_\maxn$. Bottom row: Convergence plots of $k_\textn{c}$. The $\frac{t}{t_\textn{s}}$ values to which the curves correspond are indicated in the figure titles. In the legends, ``num.'' refers to numerical solutions while ``approx.'' refers to approximate solutions.}\label{fig:k_max, omega_max and k_c against N}
\end{figure}

\subsection{Parameter sweeps}\label{sec:Parameter sweeps}

\begin{figure}
  \centering
  \includegraphics[scale=0.4]{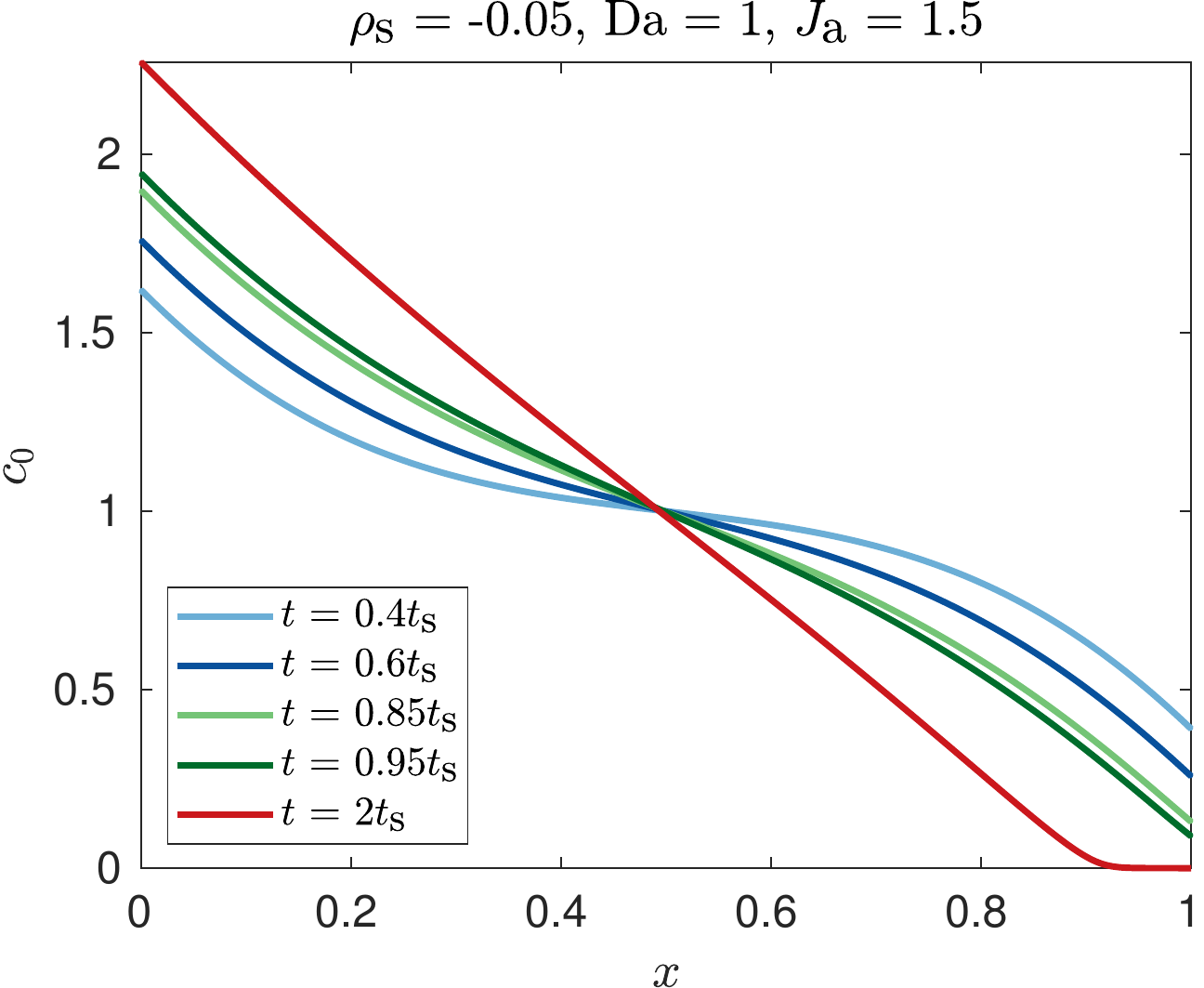}
  \includegraphics[scale=0.4]{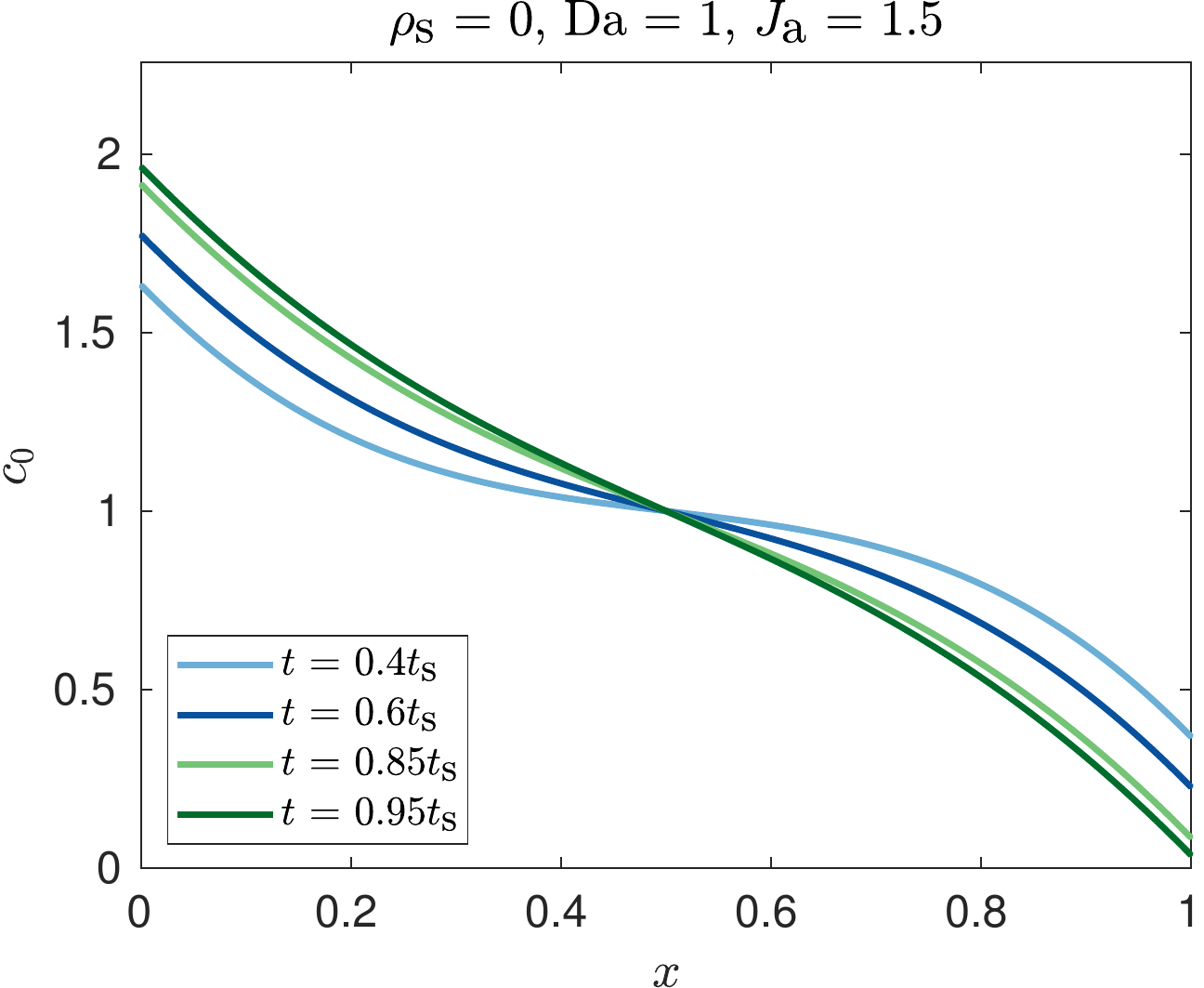}
  \includegraphics[scale=0.4]{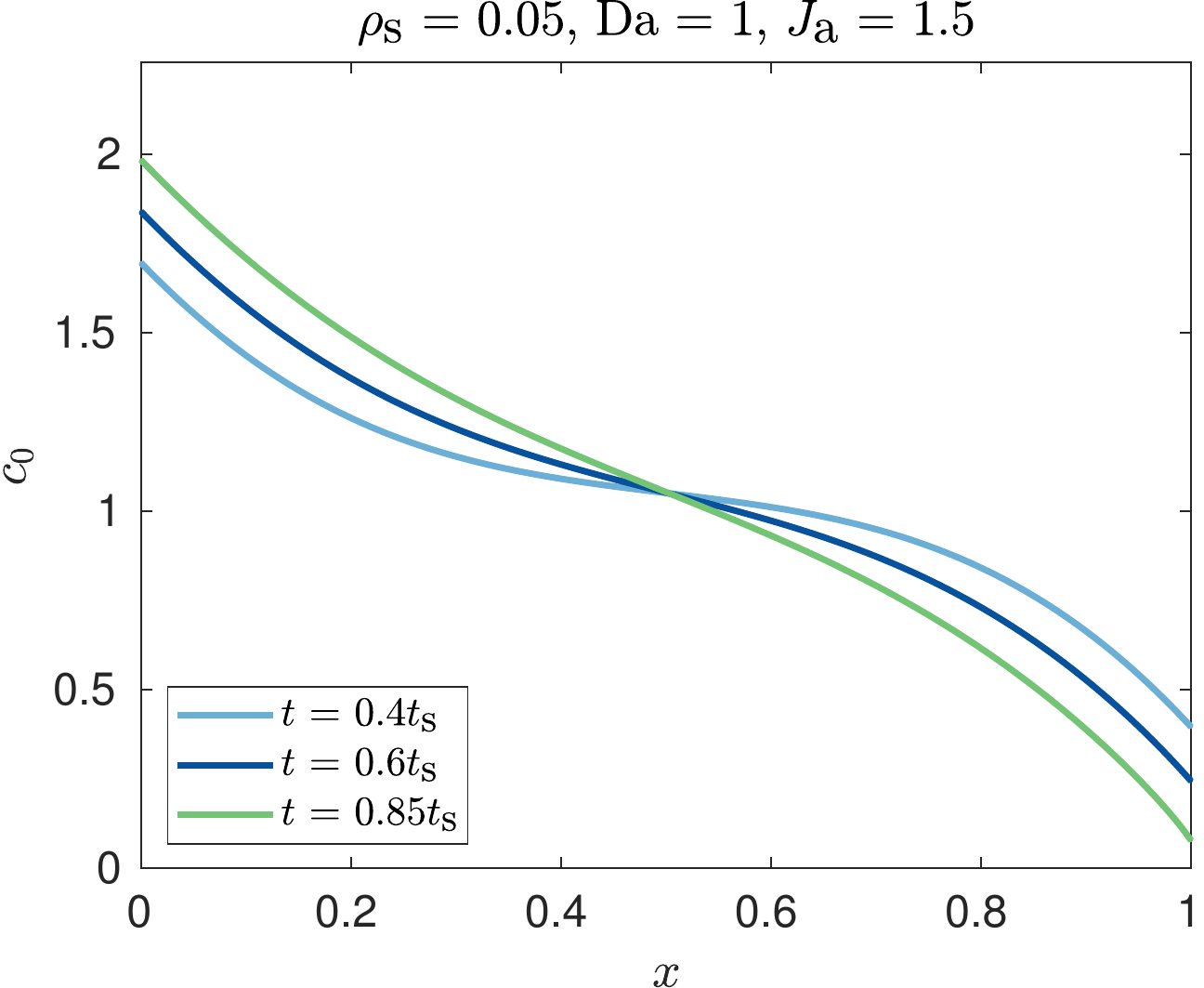}
  \includegraphics[scale=0.4]{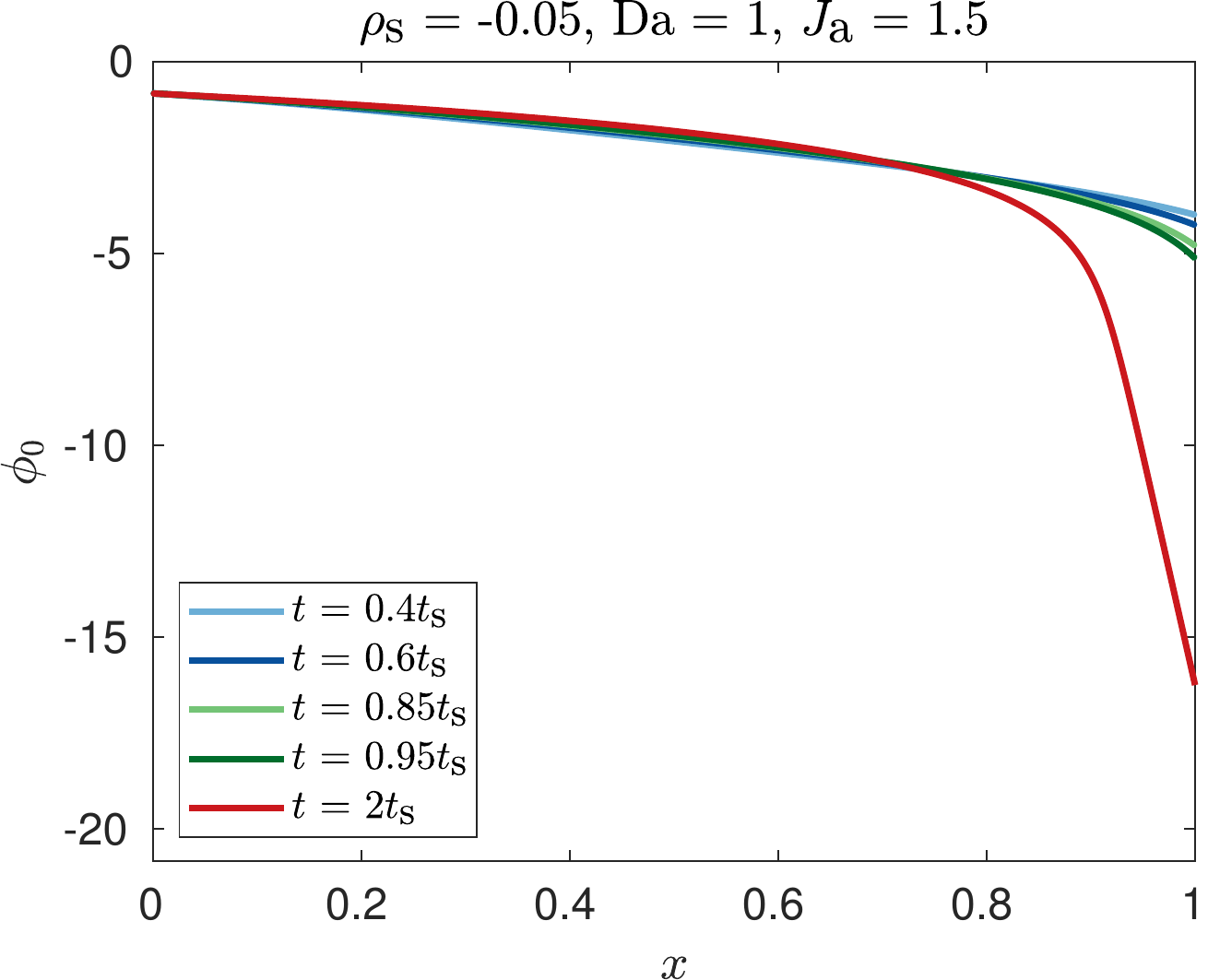}
  \includegraphics[scale=0.4]{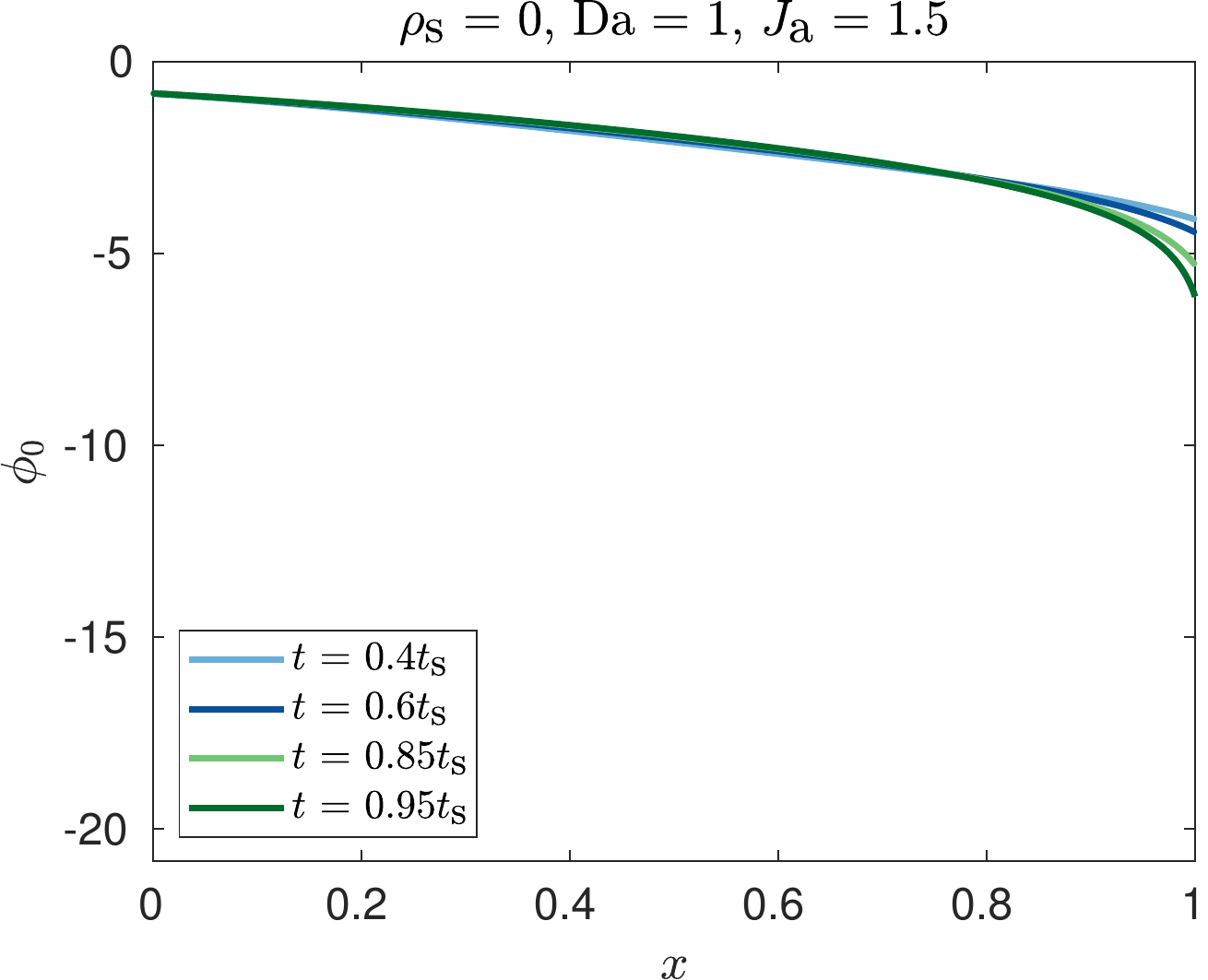}
  \includegraphics[scale=0.4]{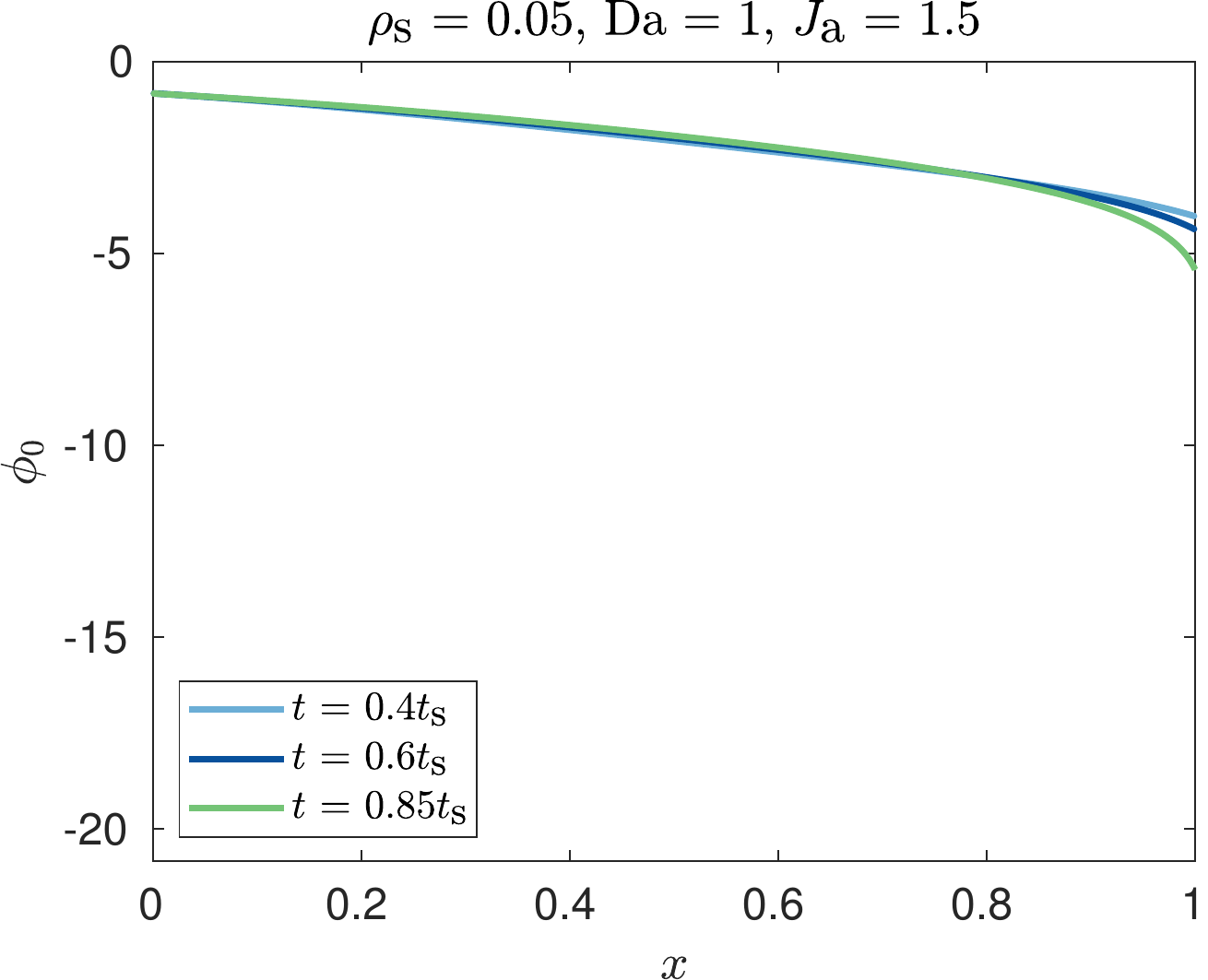}
  \includegraphics[scale=0.4]{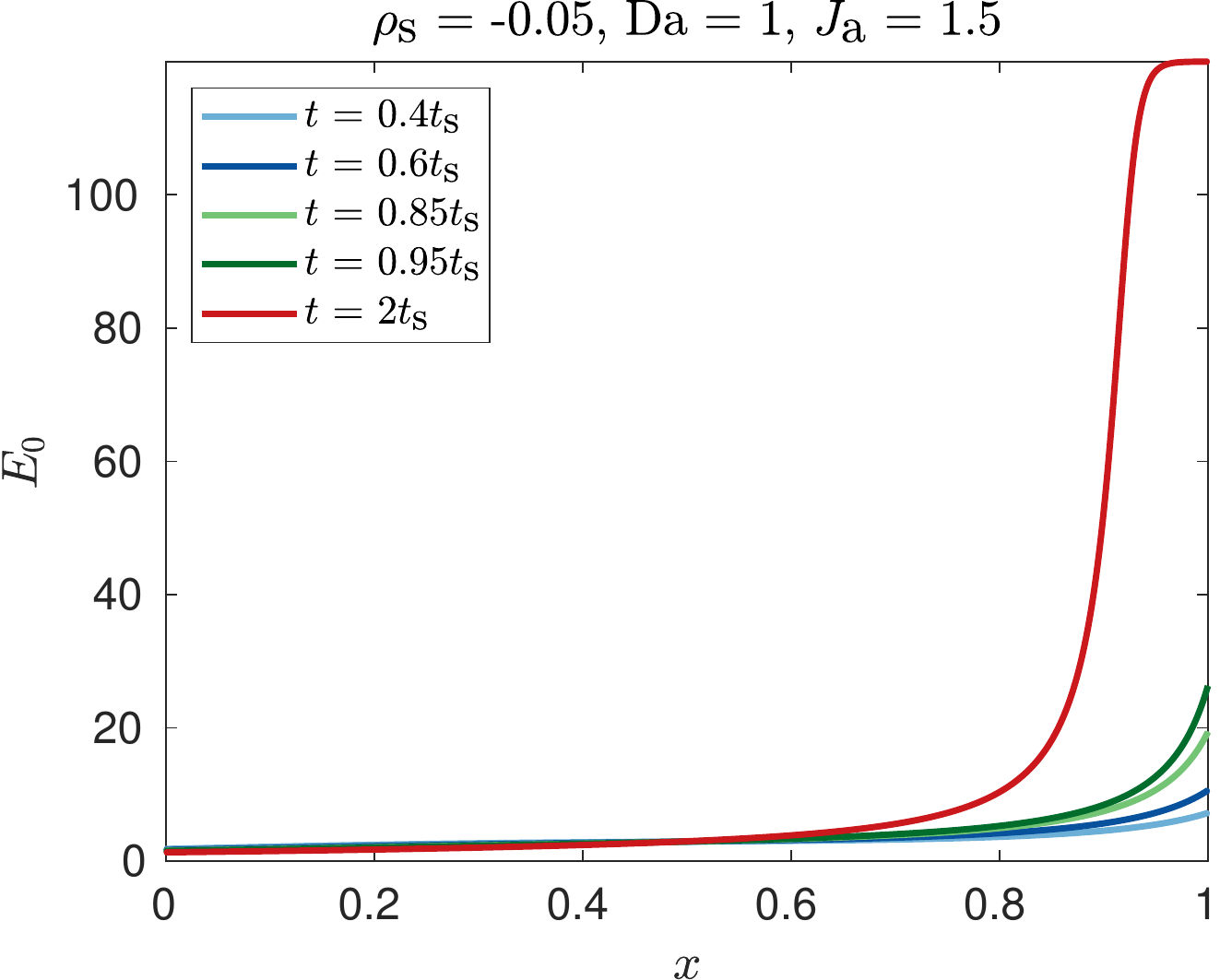}
  \includegraphics[scale=0.4]{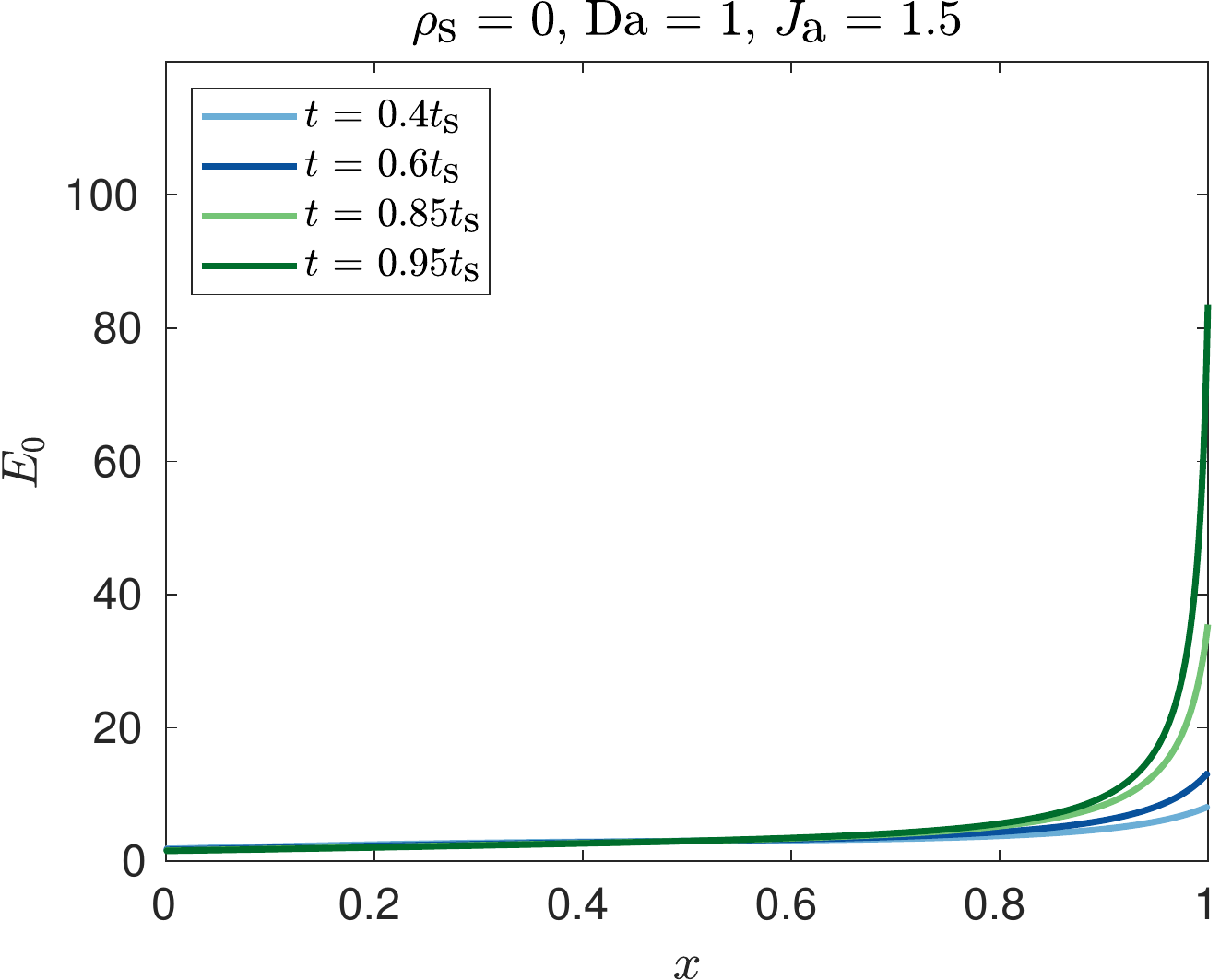}
  \includegraphics[scale=0.4]{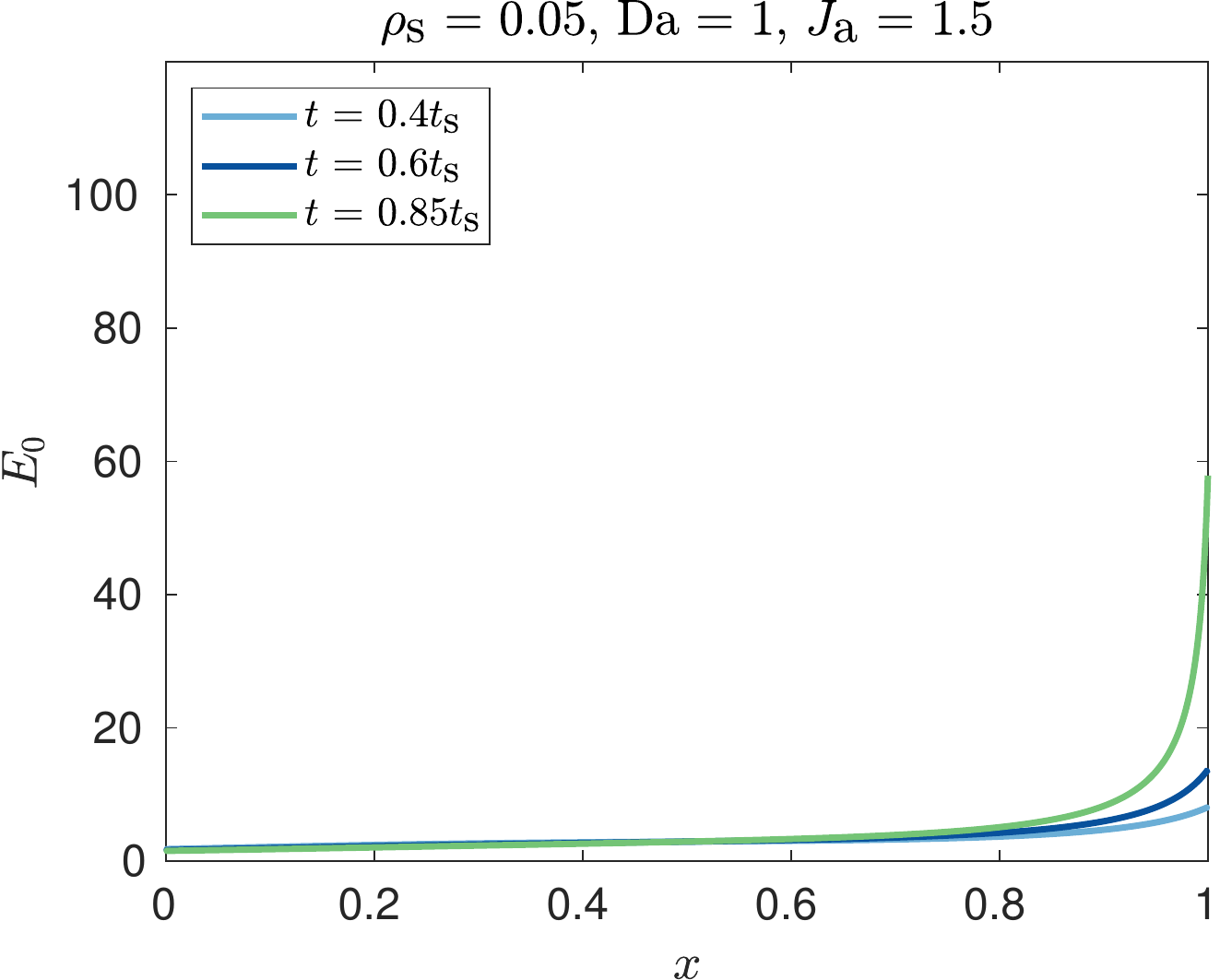}
  \caption{Plots of dimensionless base state anion concentration $c_0$, electrolyte electric potential $\phi_0$ and electric field $E_0$ against $x$ for various $\frac{t}{t_\textn{s}}$ values with $\textn{Da} = 1$ and $J_\textn{a} = 1.5$ (overlimiting current) for $\rho_\textn{s} \in \left \{-0.05,0,0.05\right \}$. First row: Plots of $c_0$ against $x$. Second row: Plots of $\phi_0$ against $x$. Third row: Plots of $E_0$ against $x$. Blue lines correspond to early times $t = 0.4t_\textn{s}$ and $t = 0.6t_\textn{s}$, green lines correspond to times near Sand's time $t = 0.85t_\textn{s}$ and $t = 0.95t_\textn{s}$, and red line corresponds to time beyond Sand's time $t = 2t_\textn{s}$. For each color, intensity increases in the direction of increasing $t$.}\label{fig:Base states}
\end{figure}

The base state anion concentration field $c_0$, electrolyte electric potential field $\phi_0$ and electric field $E_0 = -\phi_{0,x}$ possess salient features that are useful for understanding the linear stability analysis results. We focus on galvanostatic conditions under an overlimiting current $J_\textn{a} > 1$ because as explained in Section~\ref{sec:Transport}, doing so provides us with Sand's time $t_\textn{s}$ as a time scale at which the bulk electrolyte is depleted at the cathode. Depending on the sign of $\rho_\textn{s}$, the $c_0$, $\phi_0$ and $E_0$ fields behave differently at $t = t_\textn{s}$ and beyond. Fixing $\textn{Da} = 1$ and $J_\textn{a} = 1.5$ (overlimiting current), we plot $c_0$, $\phi_0$ and $E_0$ against $x$ for various $\frac{t}{t_\textn{s}}$ values for $\rho_\textn{s} \in \left \{-0.05, 0, 0.05\right \}$ in Figure~\ref{fig:Base states}. For $\rho_\textn{s} = -0.05$, because the system can go beyond $t_\textn{s}$ and eventually reach a steady state, we show plots up to $t = 2t_\textn{s}$. For $\rho_\textn{s} = 0$, since $\phi_0$ and $E_0$ at the cathode diverge at $t_\textn{s}$, which cause the numerical solver to stop converging, we can only show plots up to $t = 0.95t_\textn{s}$. For $\rho_\textn{s} = 0.05$, because $\rho_\textn{s} > 0$ effectively reduces $t_\textn{s}$ as discussed in Section~\ref{sec:Transport}, we show plots up to $t = 0.85t_\textn{s}$.

For $\rho_\textn{s} = -0.05 < 0$, the distinguishing features of running the system at an overlimiting current carried by surface conduction are the anion depletion region at the cathode and the bounded and constant electric field $E_0$ in this depletion region after $t = t_\textn{s}$. Because the anion concentration gradient almost vanishes in the depletion region, the current in this region is predominantly not carried by electrodiffusion but by electromigration of the counterions in the electric double layers (EDLs) under the aforementioned bounded and constant electric field $E_0$, i.e., surface conduction. Moreover, because of this additional surface conductivity, when compared to $\rho_\textn{s} = 0$ and $\rho_\textn{s} = 0.05$, $E_0$ is always smaller at all $x$ for a given $t$. On the other hand, for the classical case of $\rho_\textn{s} = 0$, $E_0$ at the cathode quickly increases near $t_\textn{s}$ and eventually diverges at $t_\textn{s}$. Relative to this classical case, for $\rho_\textn{s} = 0.05 > 0$, $E_0$ is always greater at all $x$ for a given $t$ and eventually diverges at the cathode earlier than $t_\textn{s}$ because of the ``negative'' surface conductivity conferred by the positive background charge as discussed in Section~\ref{sec:Transport}.

\begin{figure}
  \centering
  \includegraphics[scale=0.38]{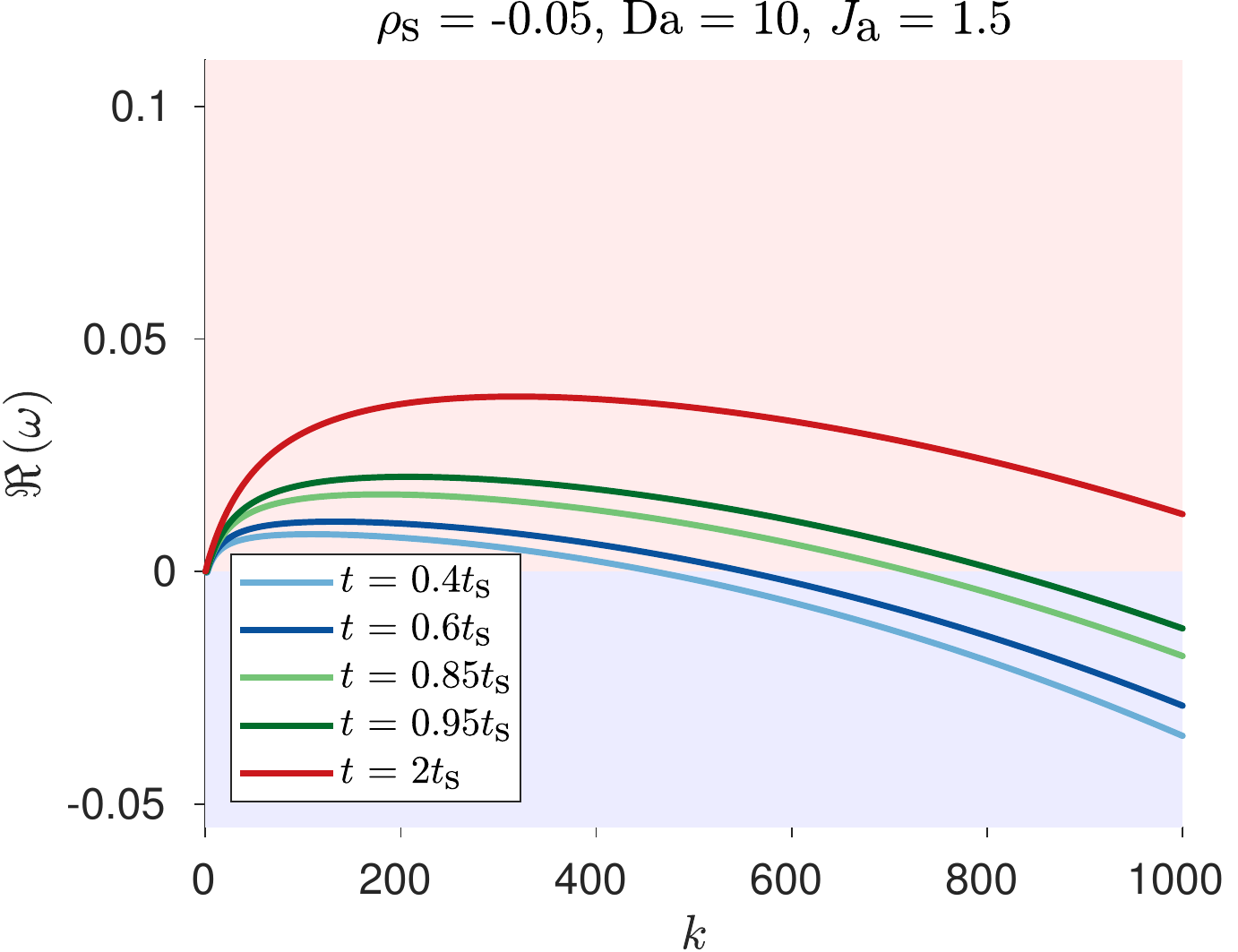}
  \includegraphics[scale=0.38]{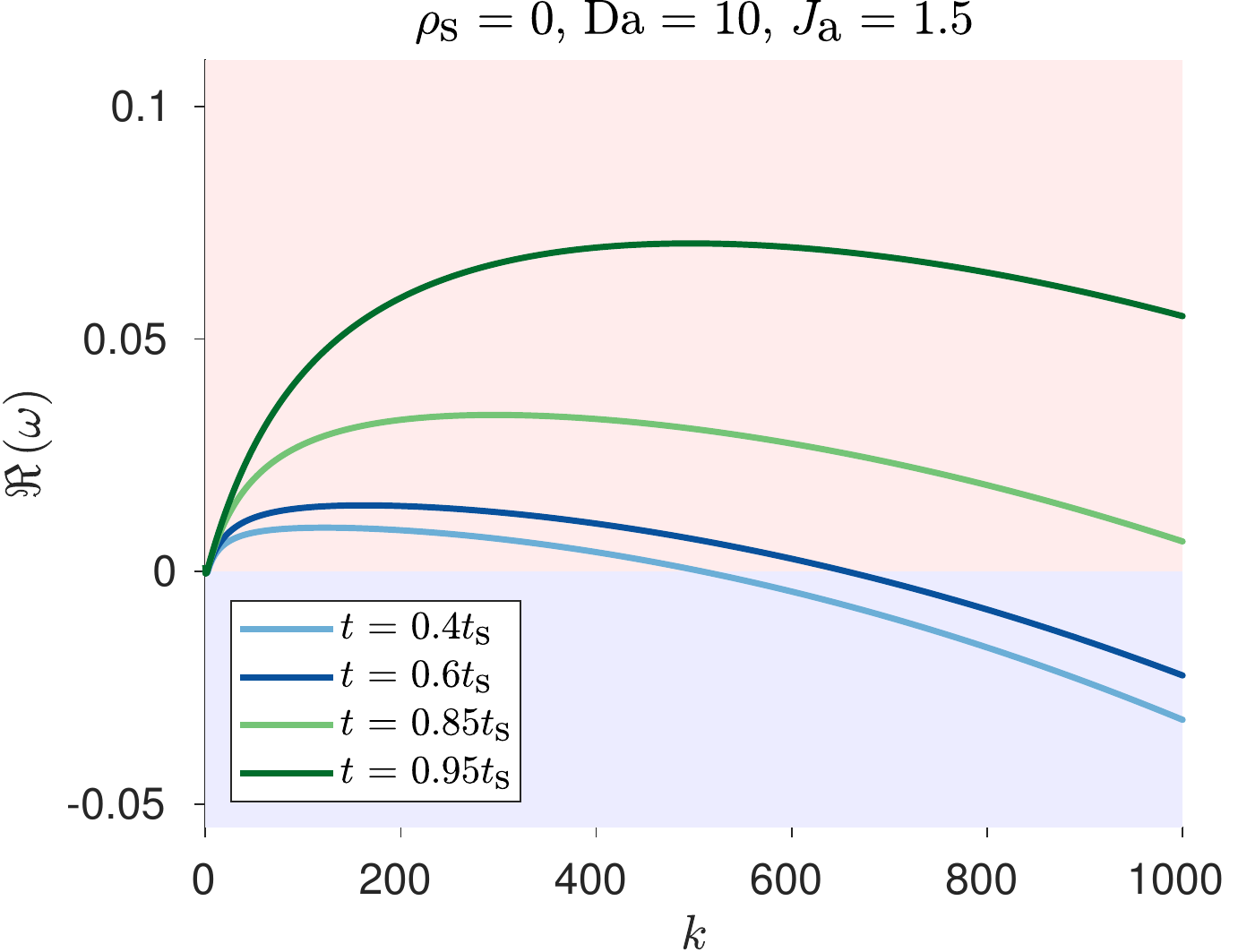}
  \includegraphics[scale=0.38]{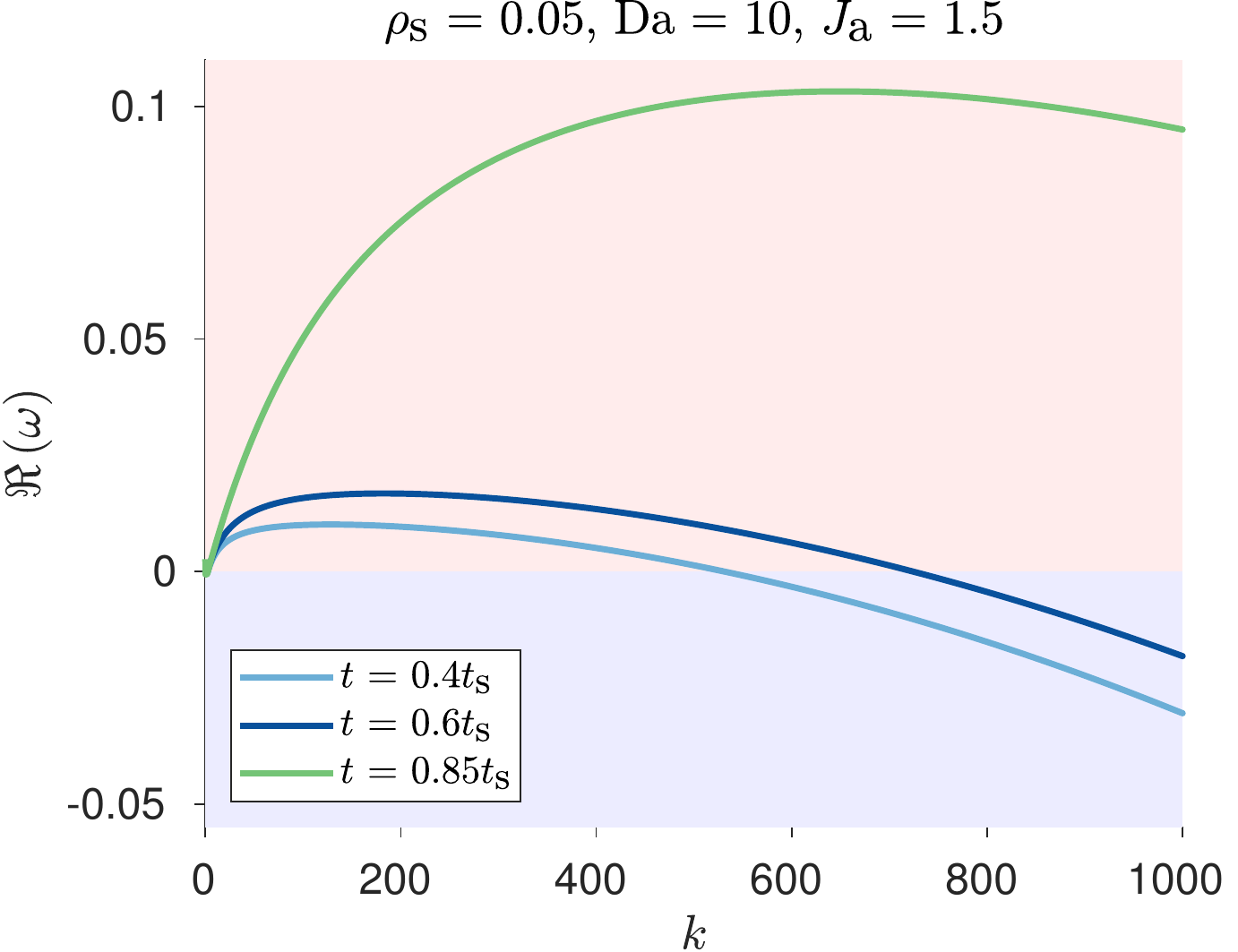}
  \includegraphics[scale=0.38]{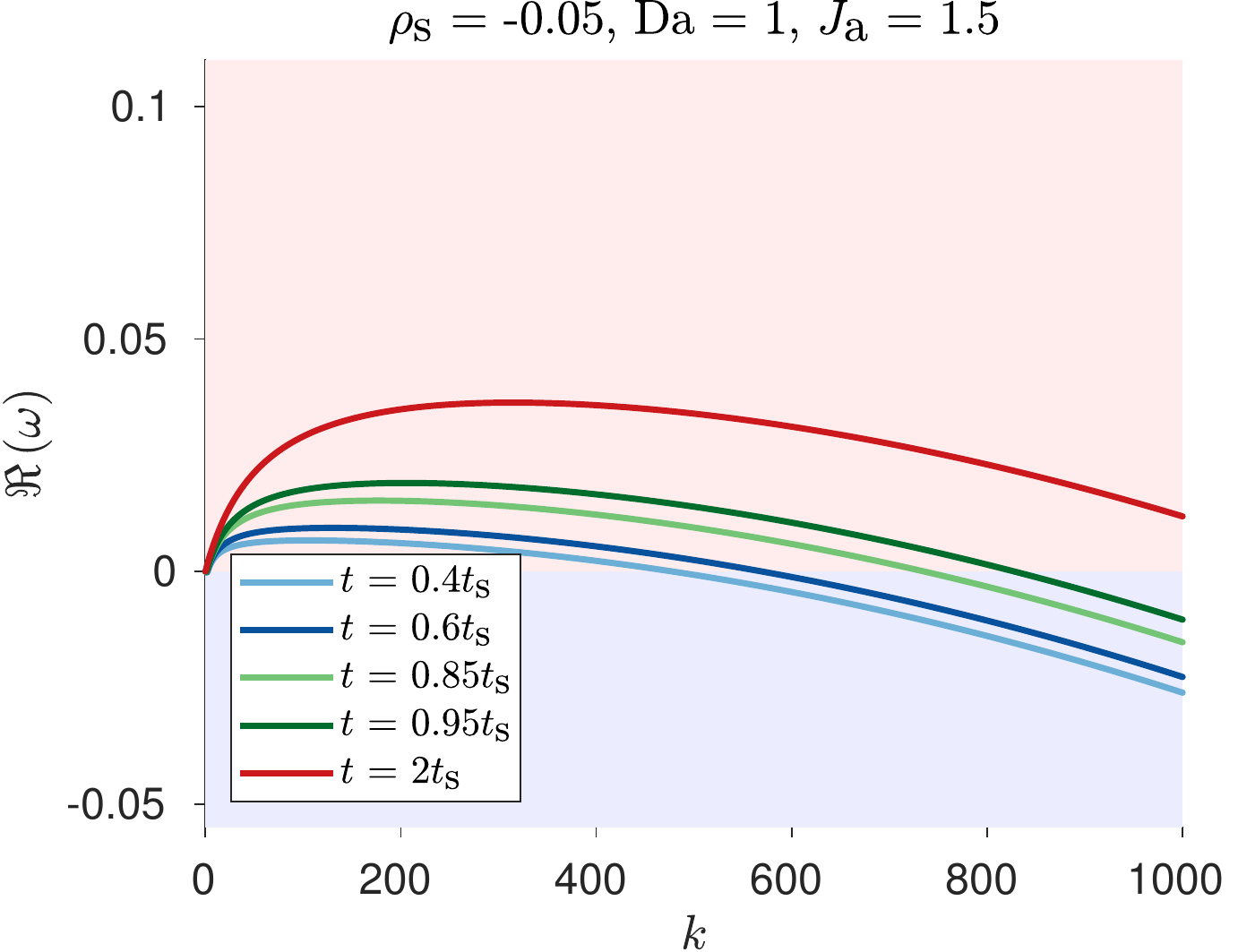}
  \includegraphics[scale=0.38]{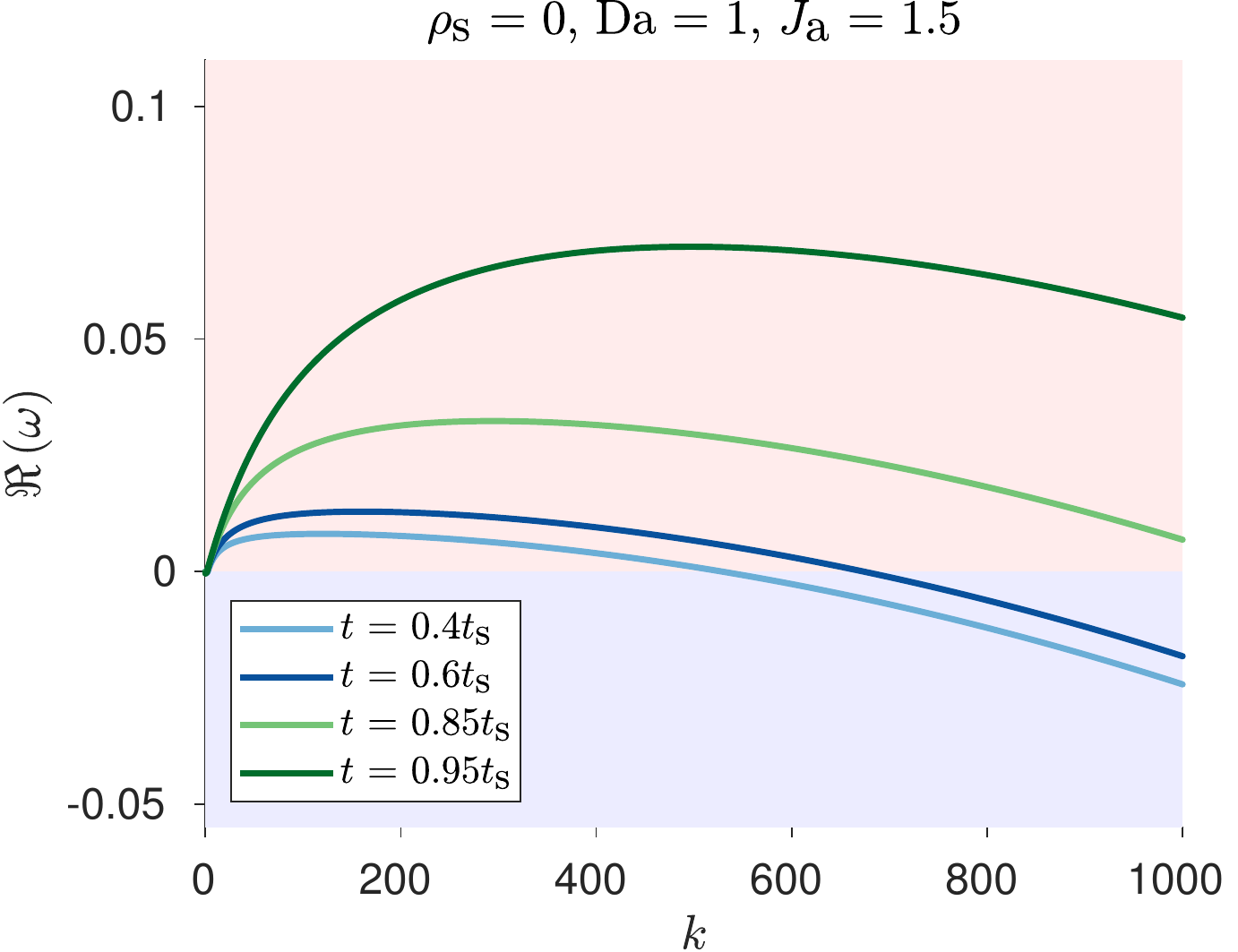}
  \includegraphics[scale=0.38]{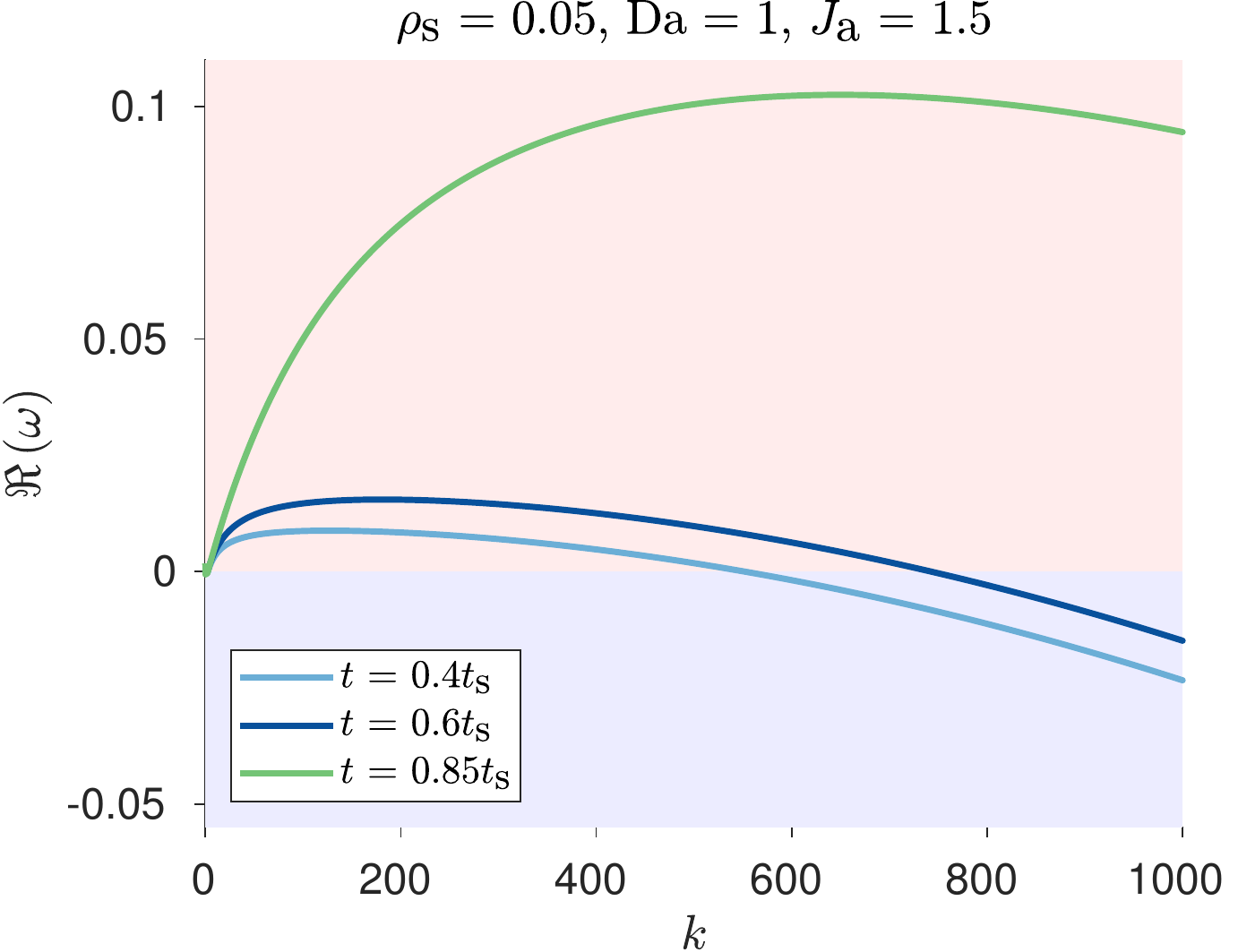}
  \includegraphics[scale=0.38]{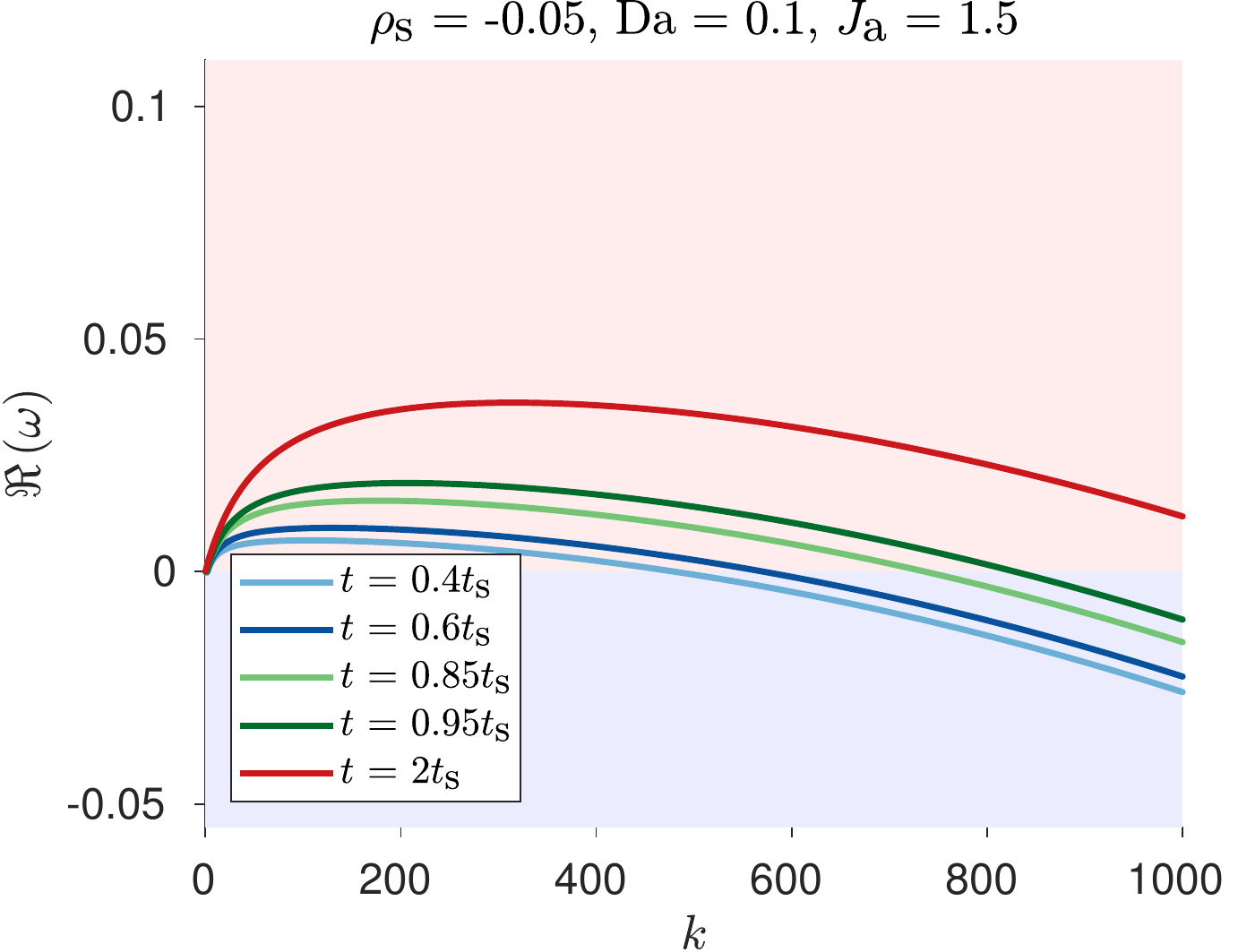}
  \includegraphics[scale=0.38]{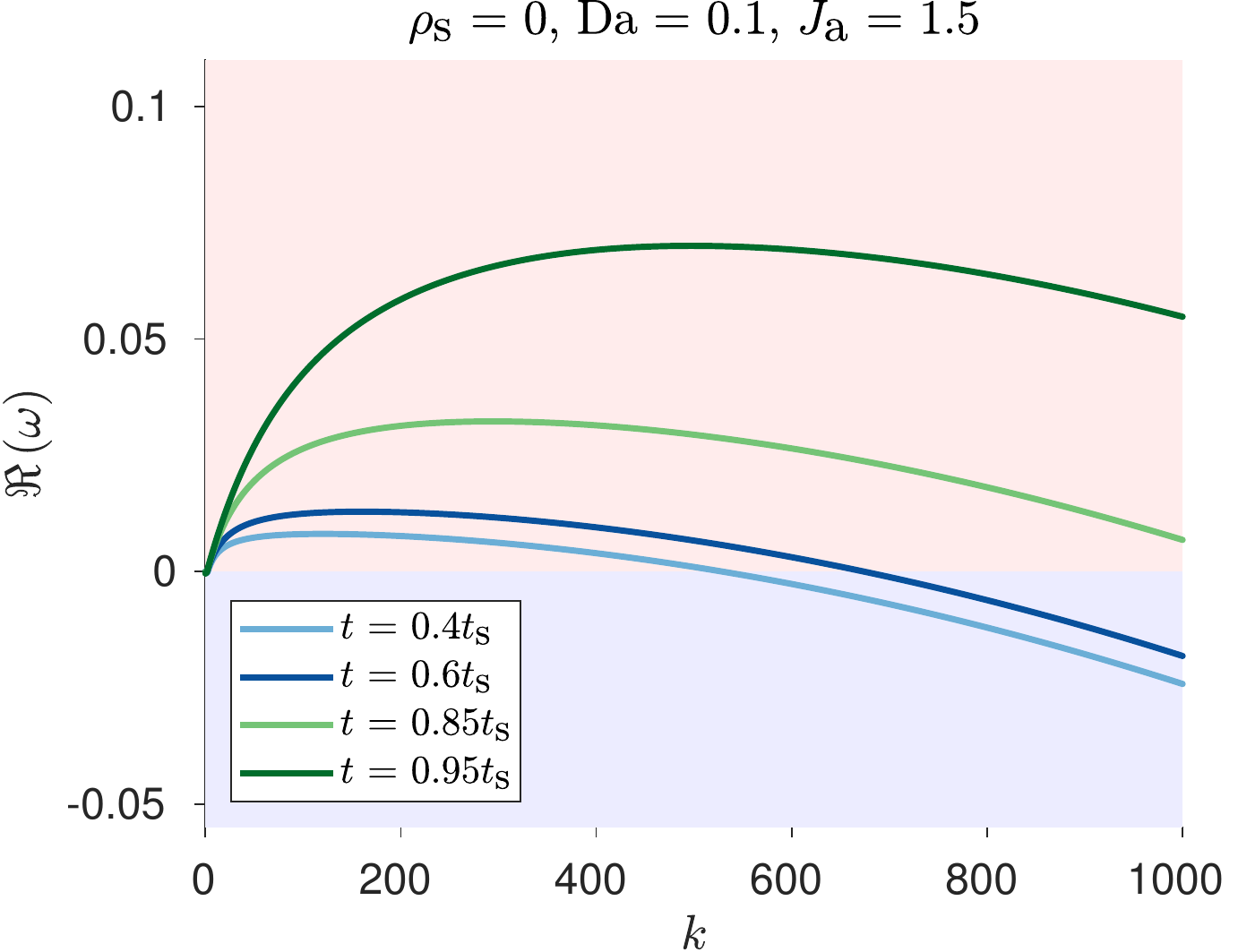}
  \includegraphics[scale=0.38]{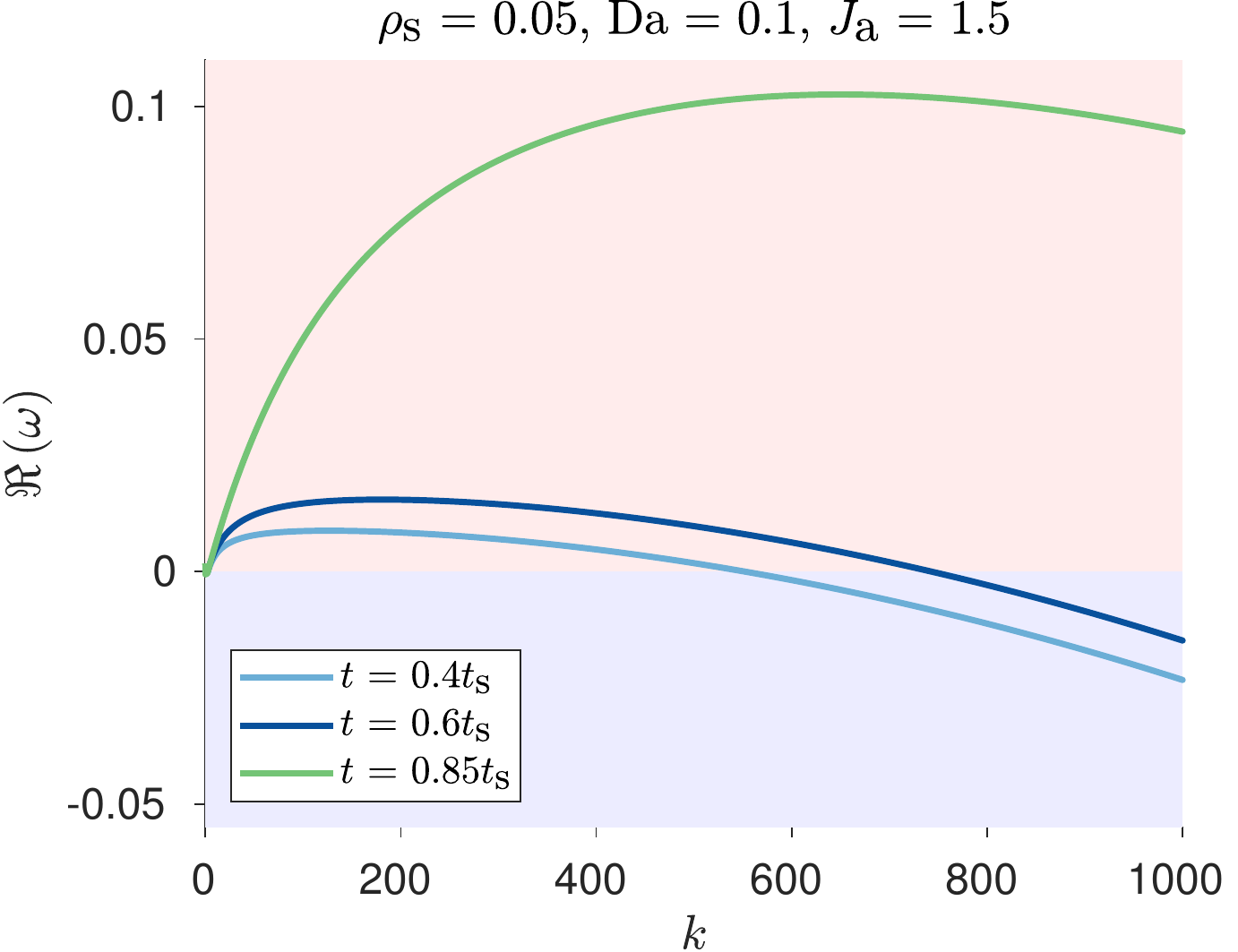}
  \caption{Plots of numerical $\Re\left(\omega\right)$ against $k$ for various $\frac{t}{t_\textn{s}}$ values for $\rho_\textn{s} \in \left \{-0.05, 0, 0.05\right \}$, $\textn{Da} \in \left \{0.1, 1, 10\right \}$ and $J_\textn{a} = 1.5$ (overlimiting current). $\rho_\textn{s}$ increases from left to right and $\textn{Da}$ increases from bottom to top. Blue lines correspond to early times $t = 0.4t_\textn{s}$ and $t = 0.6t_\textn{s}$, green lines correspond to times near Sand's time $t = 0.85t_\textn{s}$ and $t = 0.95t_\textn{s}$, and red line corresponds to time beyond Sand's time $t = 2t_\textn{s}$. For each color, intensity increases in the direction of increasing $t$.}\label{fig:Re(omega) against k for various t/t_s values and J_a = 1.5}
\end{figure}

We now examine the dispersion relation $\omega\left(k\right)$ by plotting numerically computed $\Re\left(\omega\right)$ against $k$ for various $\frac{t}{t_\textn{s}}$ values for $\rho_\textn{s} \in \left \{-0.05, 0, 0.05\right \}$, $\textn{Da} \in \left \{0.1, 1, 10\right \}$ and $J_\textn{a} = 1.5$ (overlimiting current) in Figure~\ref{fig:Re(omega) against k for various t/t_s values and J_a = 1.5}. In Figure~\ref{fig:Re(omega) against k for various t/t_s values and J_a = 1.5}, $\rho_\textn{s}$ increases from left to right and $\textn{Da}$ increases from bottom to top. Generally for all the parameters considered, the $\omega$ curve, in particular the $k_\maxn$, $\omega_\maxn$ and $k_\textn{c}$ points, increases and ``moves in the northeast direction'' as $t$ increases; qualitatively, the ``total amount of instability'' increases with $t$. For $\rho_\textn{s} = -0.05 < 0$, when compared to $\rho_\textn{s} = 0$ and $\rho_\textn{s} = 0.05$, the $\omega$ curve is the smallest at a given $t$ because of a smaller base state electric field $E_0$. The $\omega$ curve also remains bounded at all $t$ and eventually reaches a steady state that is almost attained near $t = 2t_\textn{s}$ because $E_0$ at the cathode behaves in the same fashion. In sharp contrast, for the classical case of $\rho_\textn{s} = 0$ near $t_\textn{s}$, the $\omega$ curve grows dramatically because of the rapidly increasing $E_0$ at the cathode, which eventually diverges at $t_\textn{s}$ and in turn causes the $\omega$ curve to diverge at $t_\textn{s}$ too. Compared to this classical case, for $\rho_\textn{s} = 0.05 > 0$, because $E_0$ at the cathode is larger at a given $t$ and diverges earlier than $t_\textn{s}$, the $\omega$ curve accordingly grows even more rapidly at earlier times and diverges earlier than $t_\textn{s}$. Therefore, by bounding the electric field at the cathode, the presence of a negative background charge confers additional stabilization to the system beyond what is provided by surface energy effects, although it does not completely stabilize the system as there are still regions of positive growth rate in the dispersion relation. On the other hand, for the classical case of zero background charge, the system rapidly destabilizes near Sand's time and ultimately diverges at Sand's time because of the diverging electric field at the cathode, which is also demonstrated in~\cite{elezgaray_linear_1998}. Relative to this classical case, the presence of a positive background charge destabilizes the system even further by generating an electric field at the cathode that is larger at a given time and diverges earlier than Sand's time, resulting in higher growth rates at earlier times and in finite time divergence earlier than Sand's time.

We observe that increasing $\textn{Da}$ generally increases $\omega$ but this effect is very insignificant because the application of an overlimiting current implies that the system is always diffusion-limited regardless of what $\textn{Da}$ is. Hence, in this regime of diffusion-limited electrodeposition under an overlimiting current, specific details of the electrochemical reaction kinetics model are not important in influencing the dispersion relation as long as the model includes the surface energy stabilizing effect, which typically occurs in the functional form of $\gamma k^2$.

In the interest of space, plots of numerically computed $\Re\left(\omega\right)$ against $k$ for $J_\textn{a} = 1$ (limiting current) and $J_\textn{a} = 0.5$ (underlimiting current) are not shown here but are given in Figures 1 and 2 in Section V of Supplementary Material respectively. Since the system is still always diffusion-limited for $J_\textn{a} = 1$, the trends observed for $J_\textn{a} = 1$ are qualitatively similar to our previous discussion for $J_\textn{a} = 1.5$, except that the $\omega$ values are smaller because a smaller applied current density results in a smaller electric field at the cathode. For $J_\textn{a} = 0.5$, because the applied current density is underlimiting, Sand's time is not defined and at the cathode, the bulk electrolyte concentration does not vanish and the electric field does not diverge at any $t$. Therefore, the $\omega$ curve remains bounded at all $t$ and reaches a steady state eventually. Moreover, $\omega$ generally increases with $\textn{Da}$, and this increase is especially pronounced when $\textn{Da}$ increases from $1$ to $10$; this effect was also observed by Sundstr\"om and Bark~\cite{sundstrom_morphological_1995} who focused their analysis on underlimiting currents. This increase in $\omega$ is not directly caused by $E_0$ because $E_0$ does not change appreciably despite the increase in $\textn{Da}$ (refer to Figures 3 to 5 in Section VI of Supplementary Material). Rather, as discussed in Section~\ref{sec:Electrochemical reaction kinetics}, the system becomes diffusion-limited when $\textn{Da} \gg 1$, causing the surface perturbations to destabilize faster.

\begin{figure}
  \centering
  \includegraphics[scale=0.37]{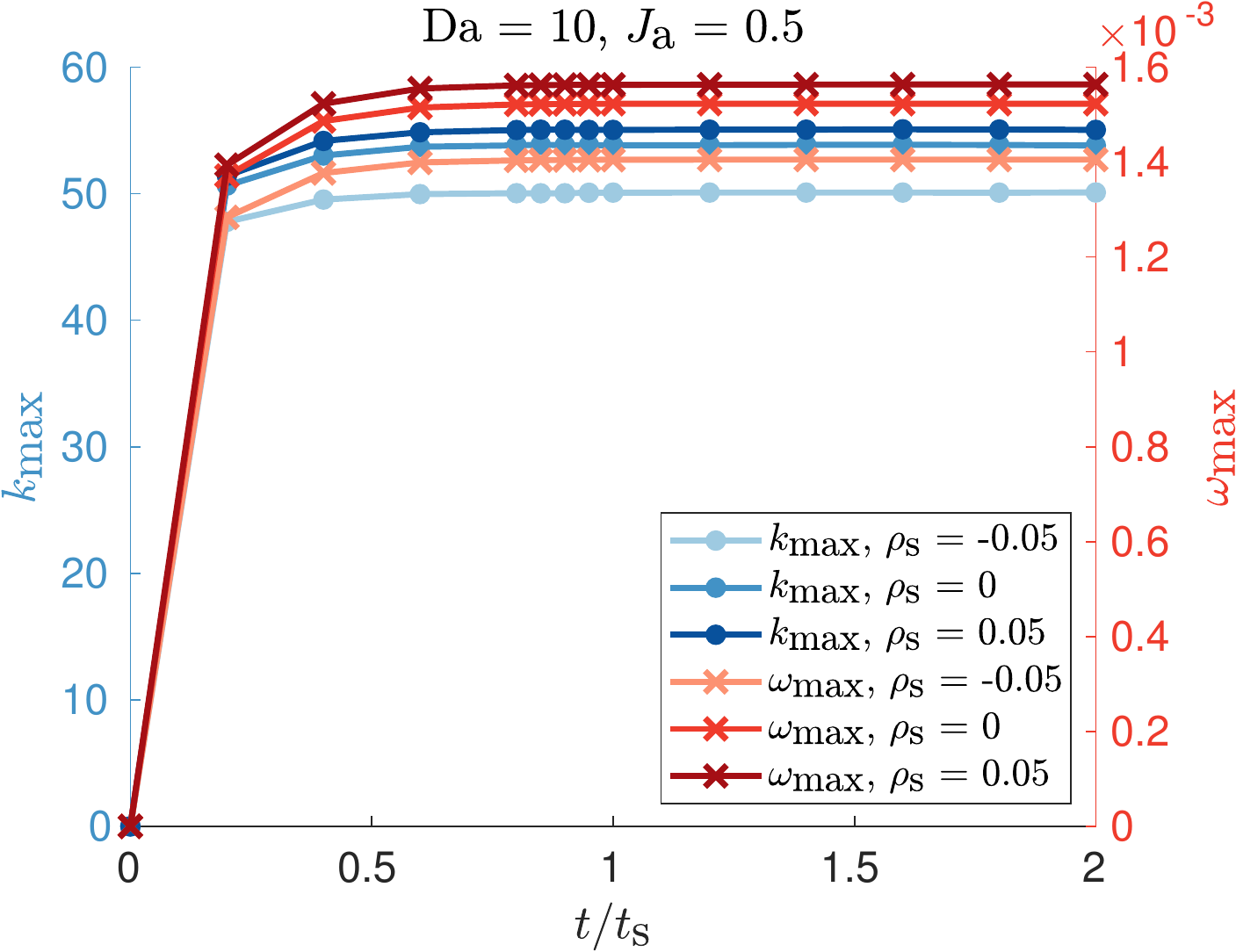}
  \includegraphics[scale=0.37]{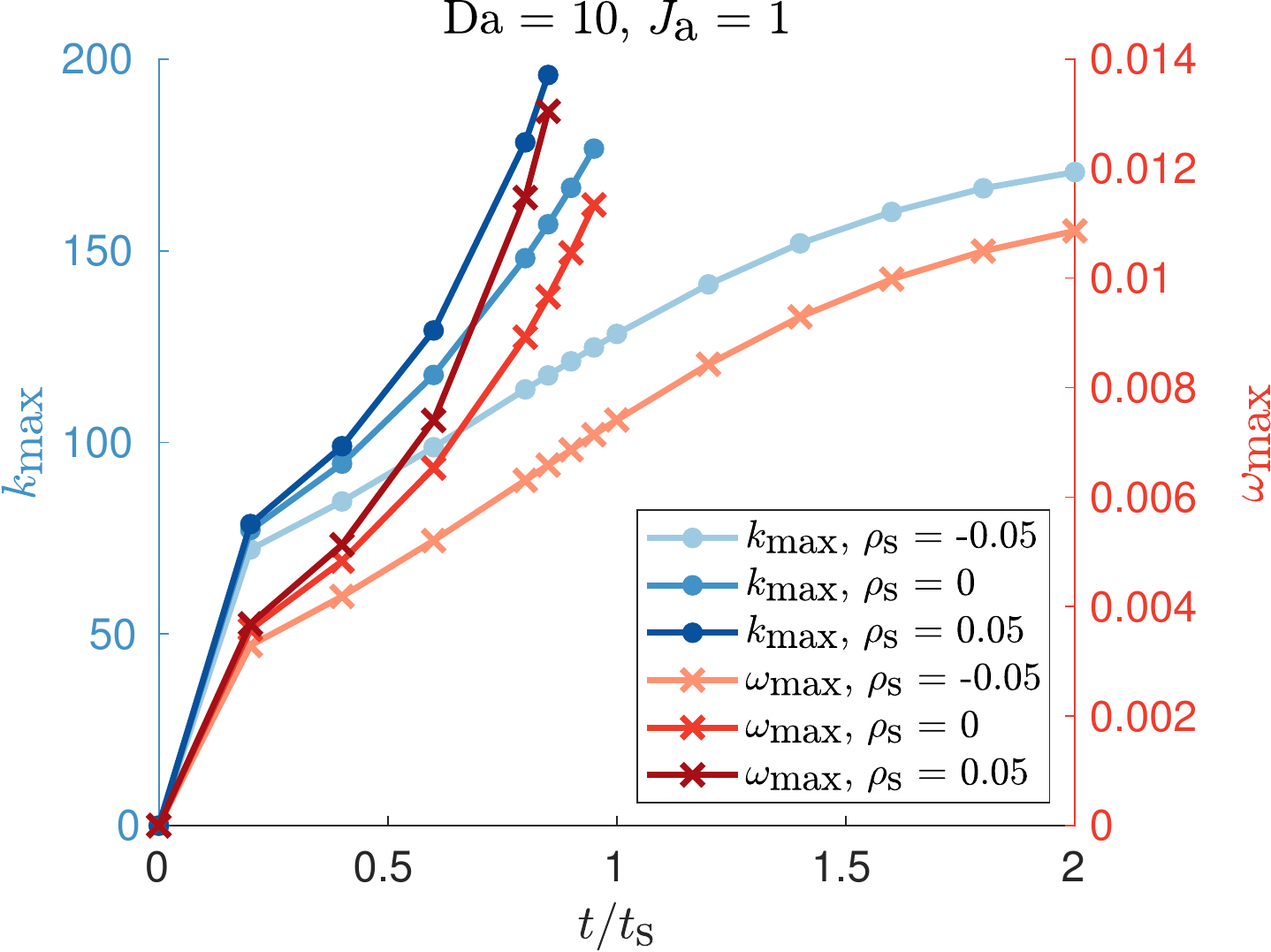}
  \includegraphics[scale=0.37]{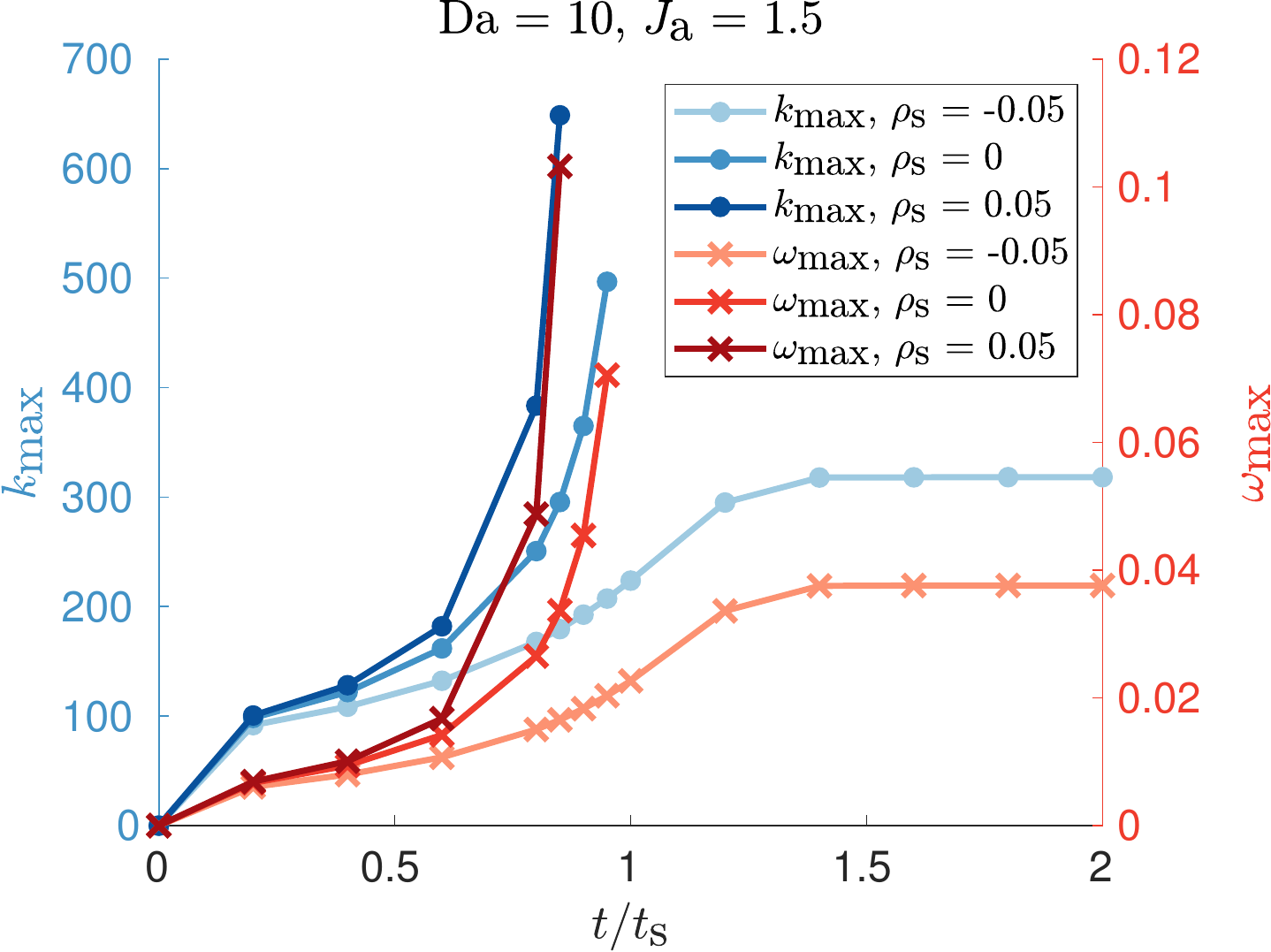}
  \includegraphics[scale=0.37]{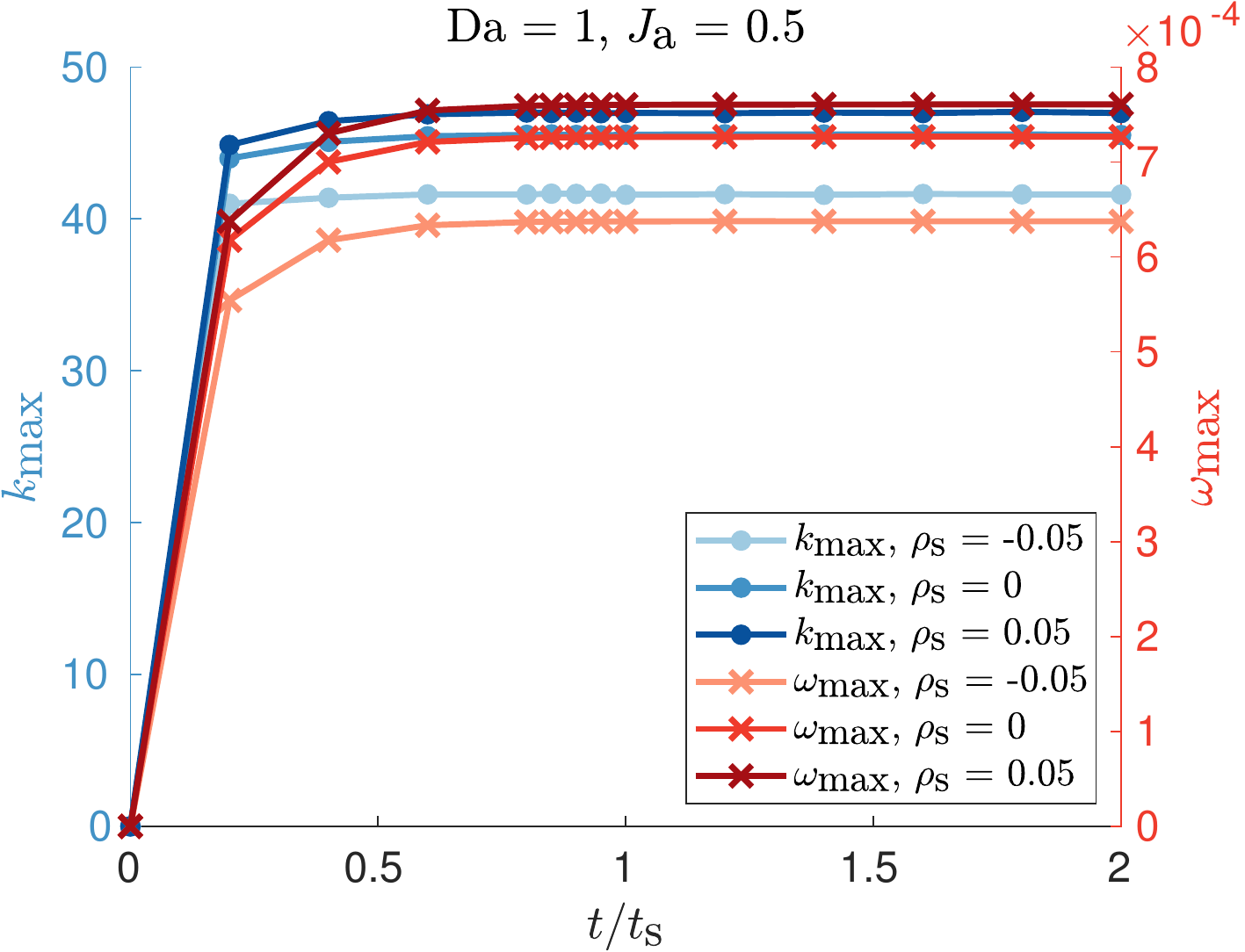}
  \includegraphics[scale=0.37]{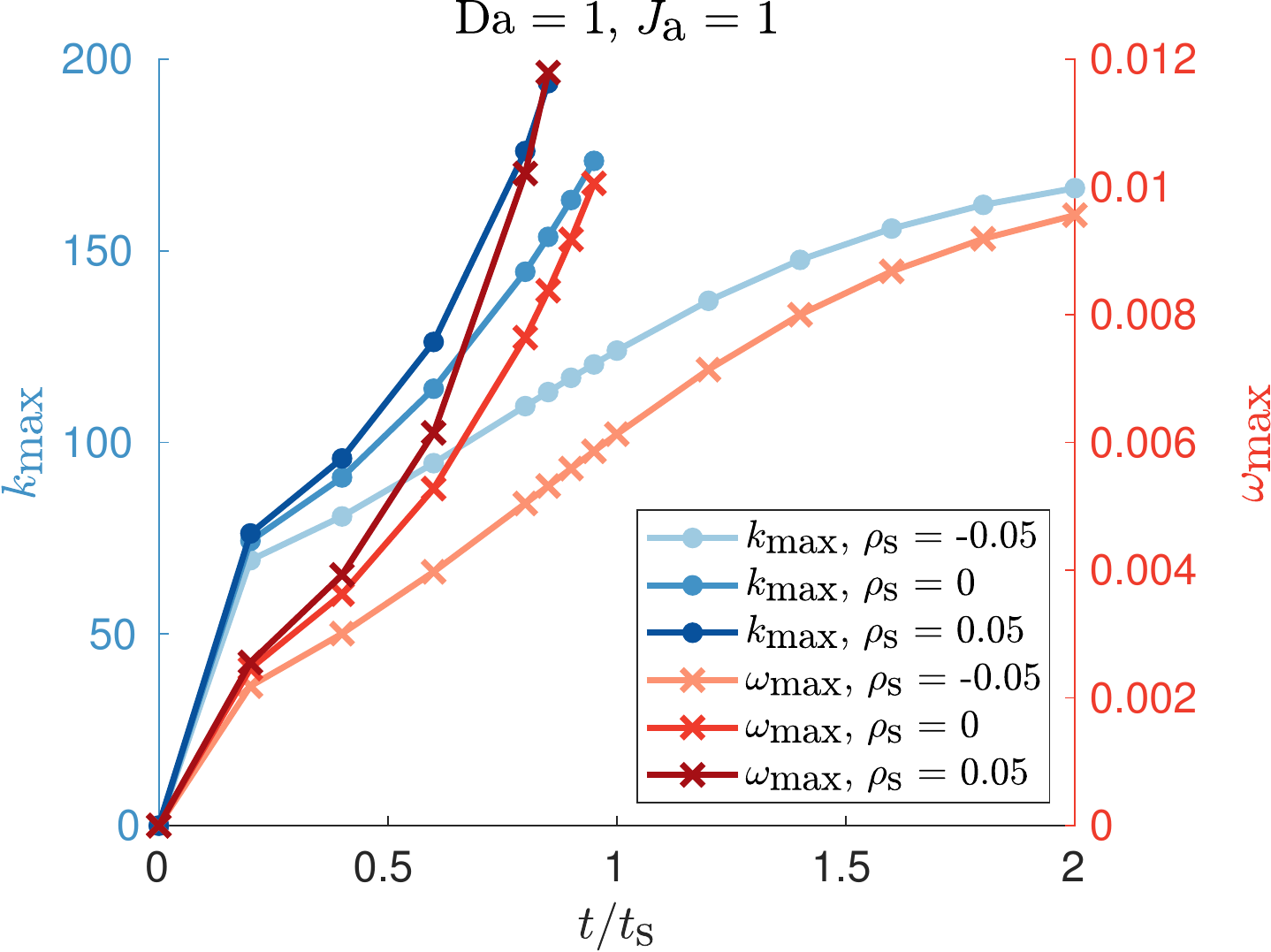}
  \includegraphics[scale=0.37]{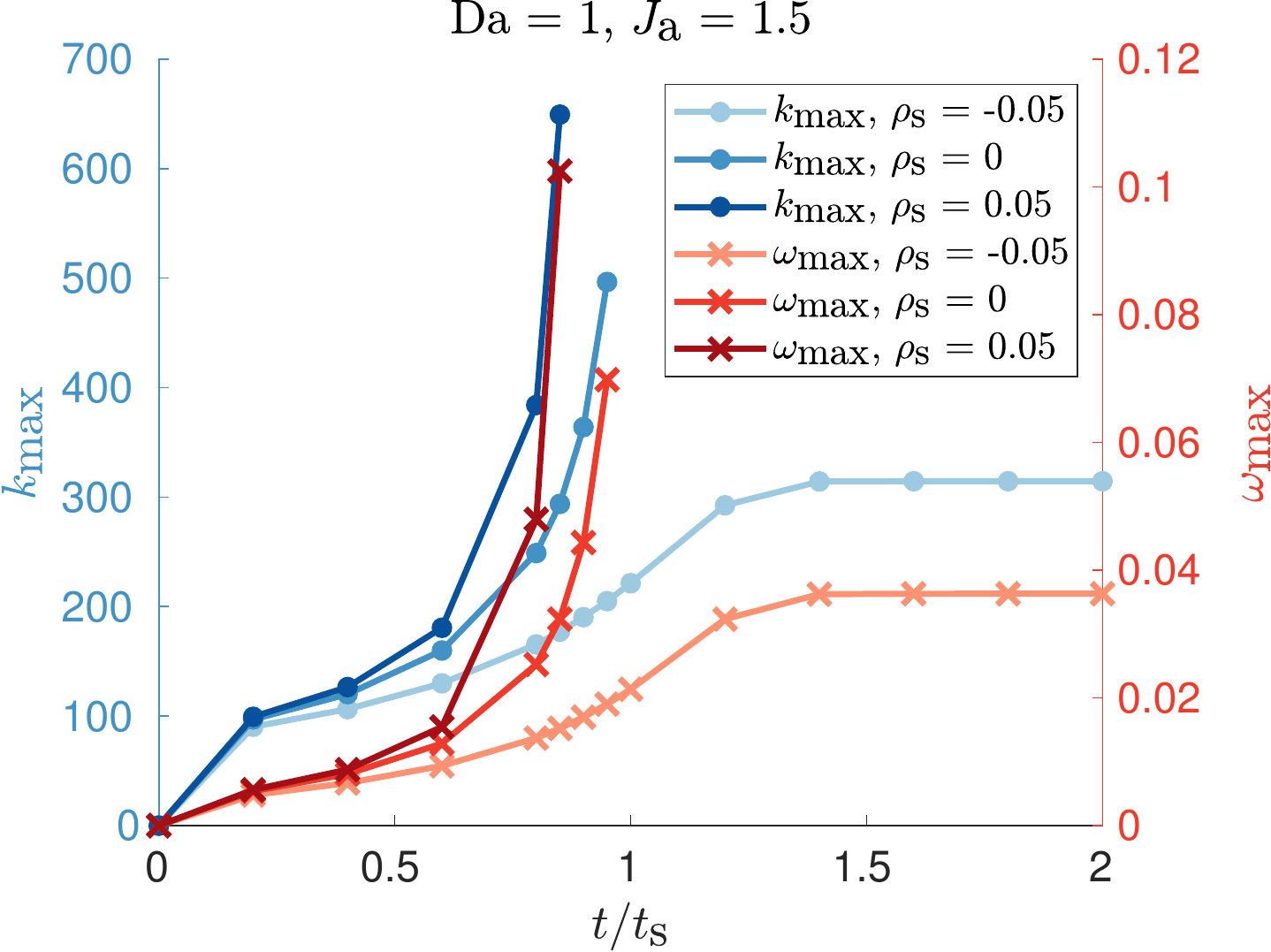}
  \includegraphics[scale=0.37]{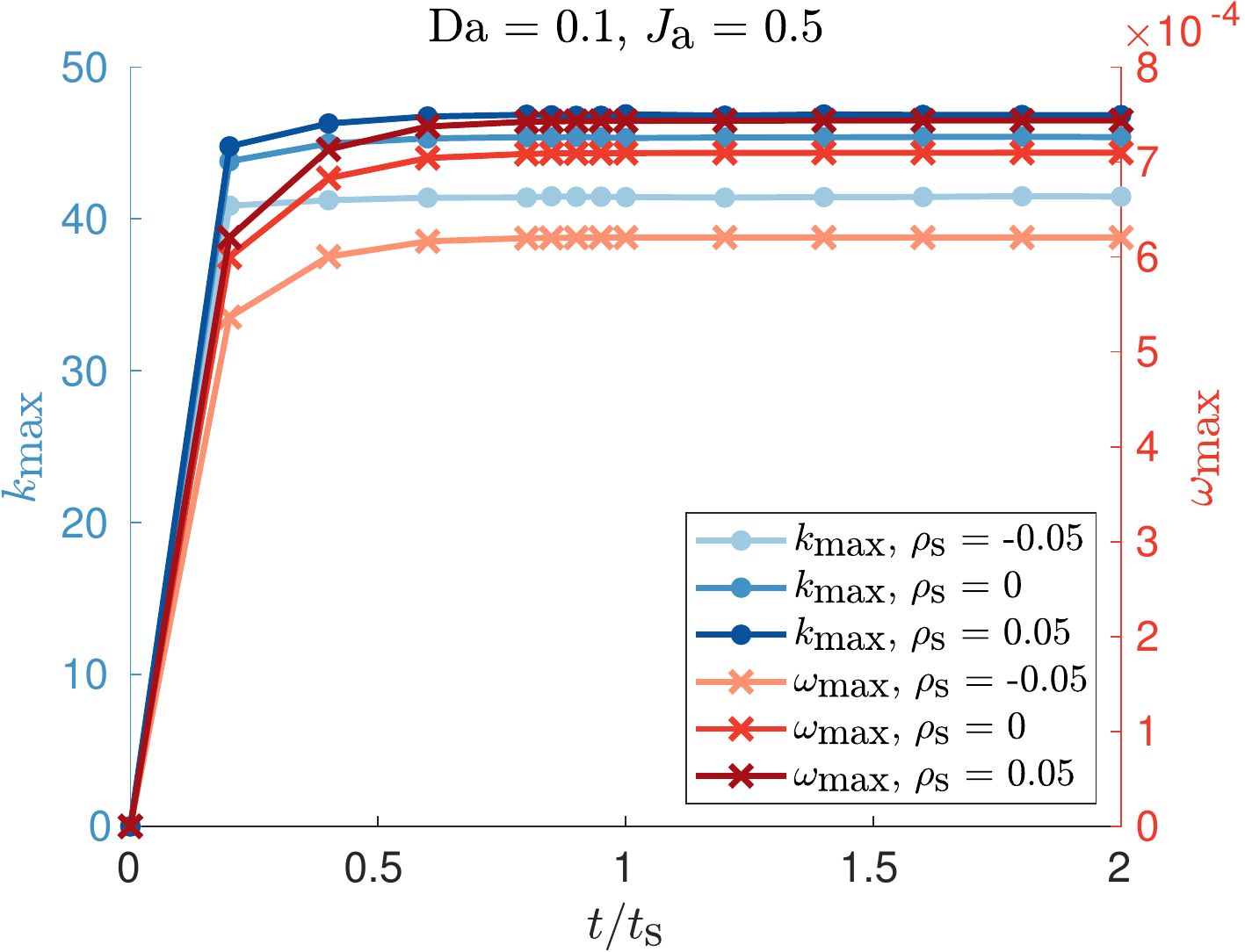}
  \includegraphics[scale=0.37]{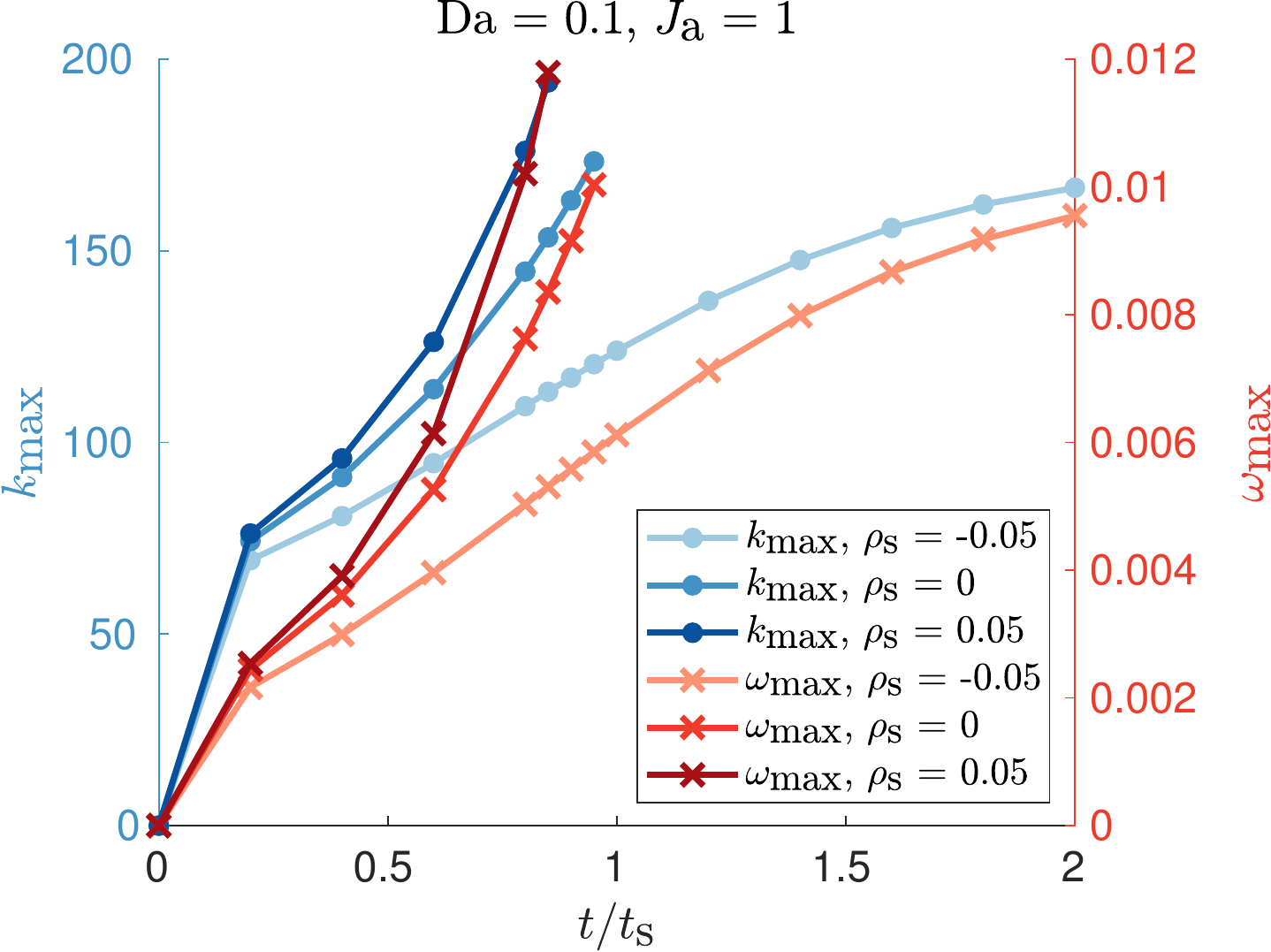}
  \includegraphics[scale=0.37]{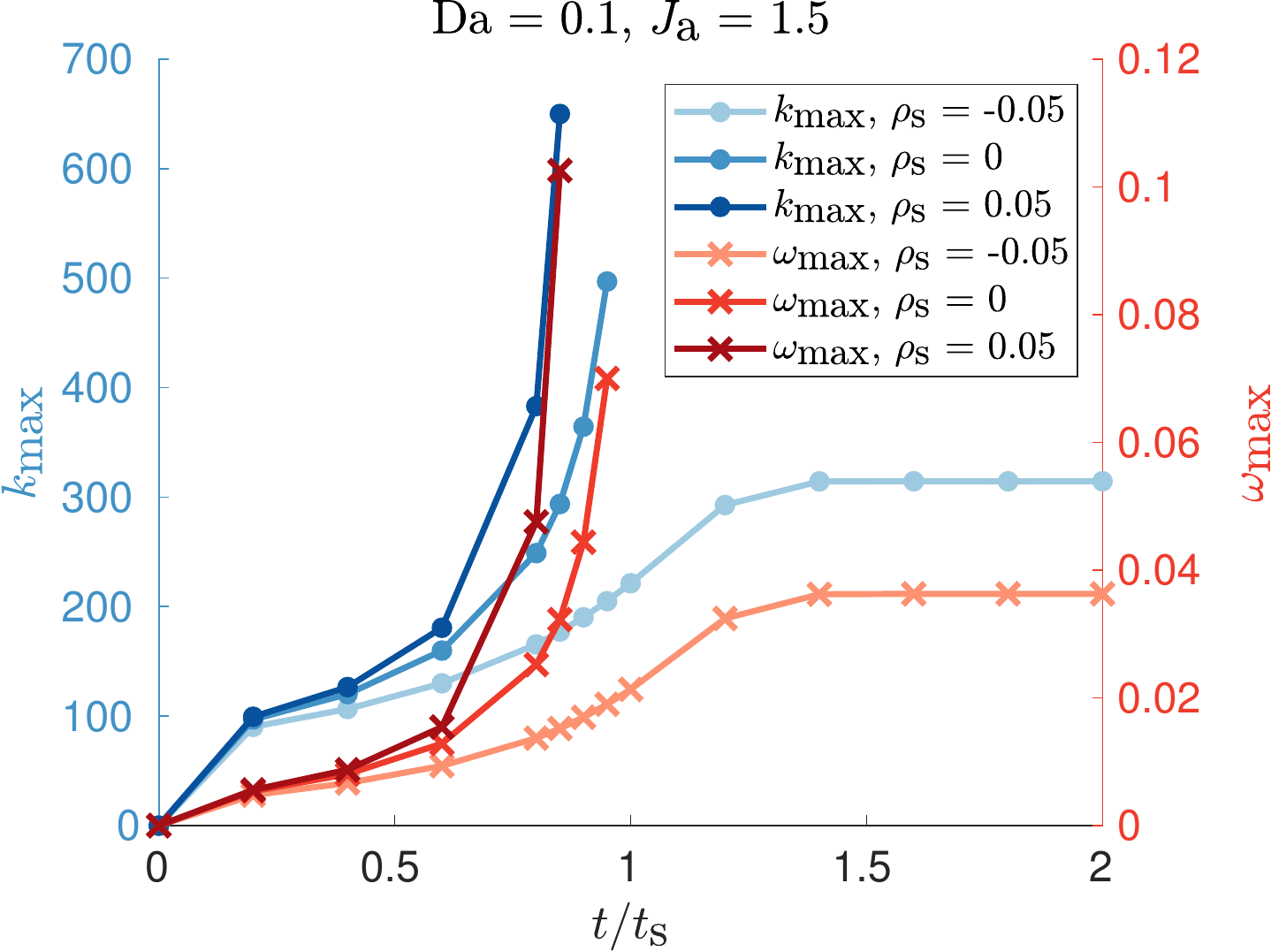}
  \caption{Plots of numerical $k_\maxn$ and $\omega_\maxn$ against $\frac{t}{t_\textn{s}}$ for $\rho_\textn{s} \in \left \{-0.05, 0, 0.05\right \}$, $\textn{Da} \in \left \{0.1, 1, 10\right \}$ and $J_\textn{a} \in \left \{0.5, 1, 1.5\right \}$. $\rho_\textn{s}$ increases from left to right and $\textn{Da}$ increases from bottom to top.}\label{fig:k_max and omega_max against t/t_s}
\end{figure}

\begin{figure}
  \centering
  \includegraphics[scale=0.4]{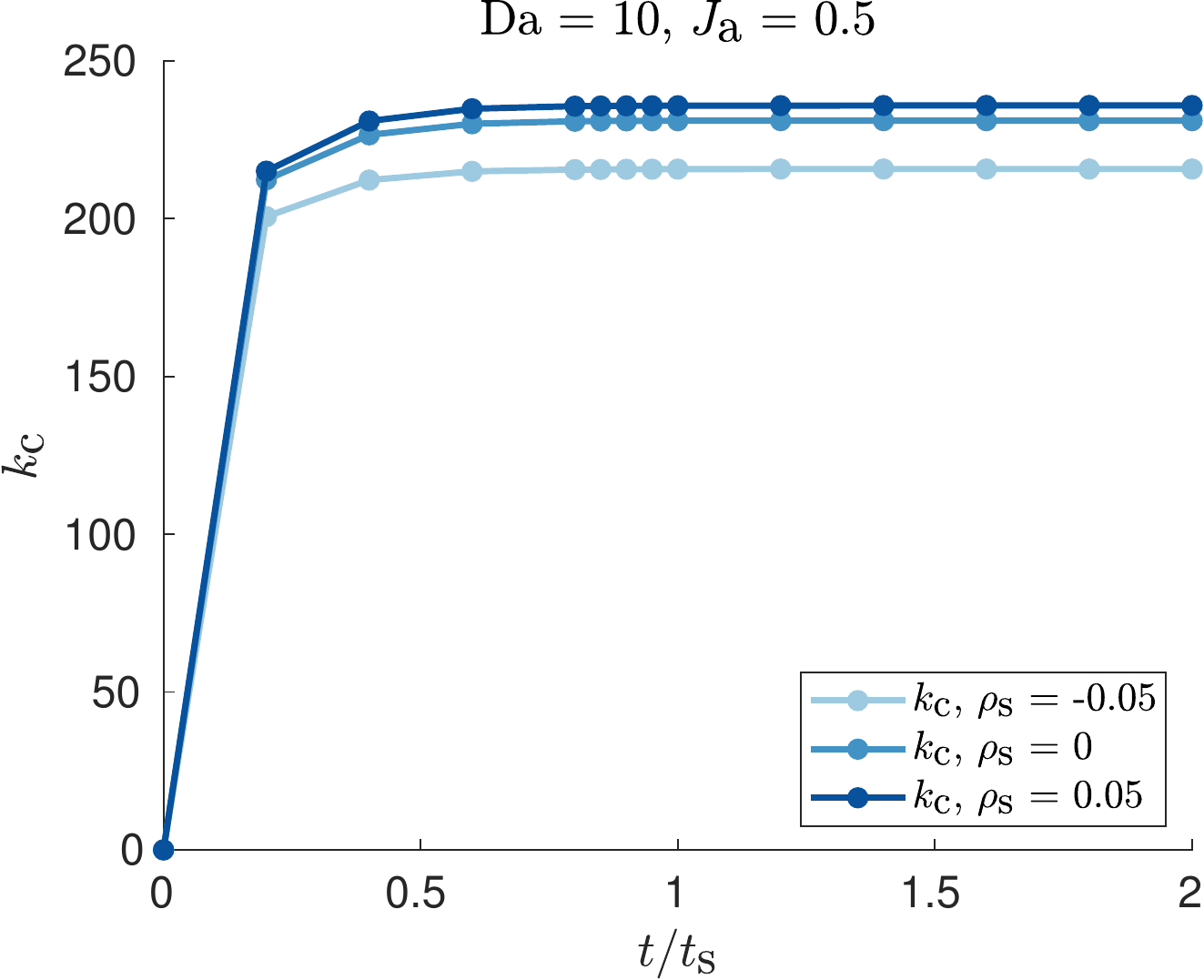}
  \includegraphics[scale=0.4]{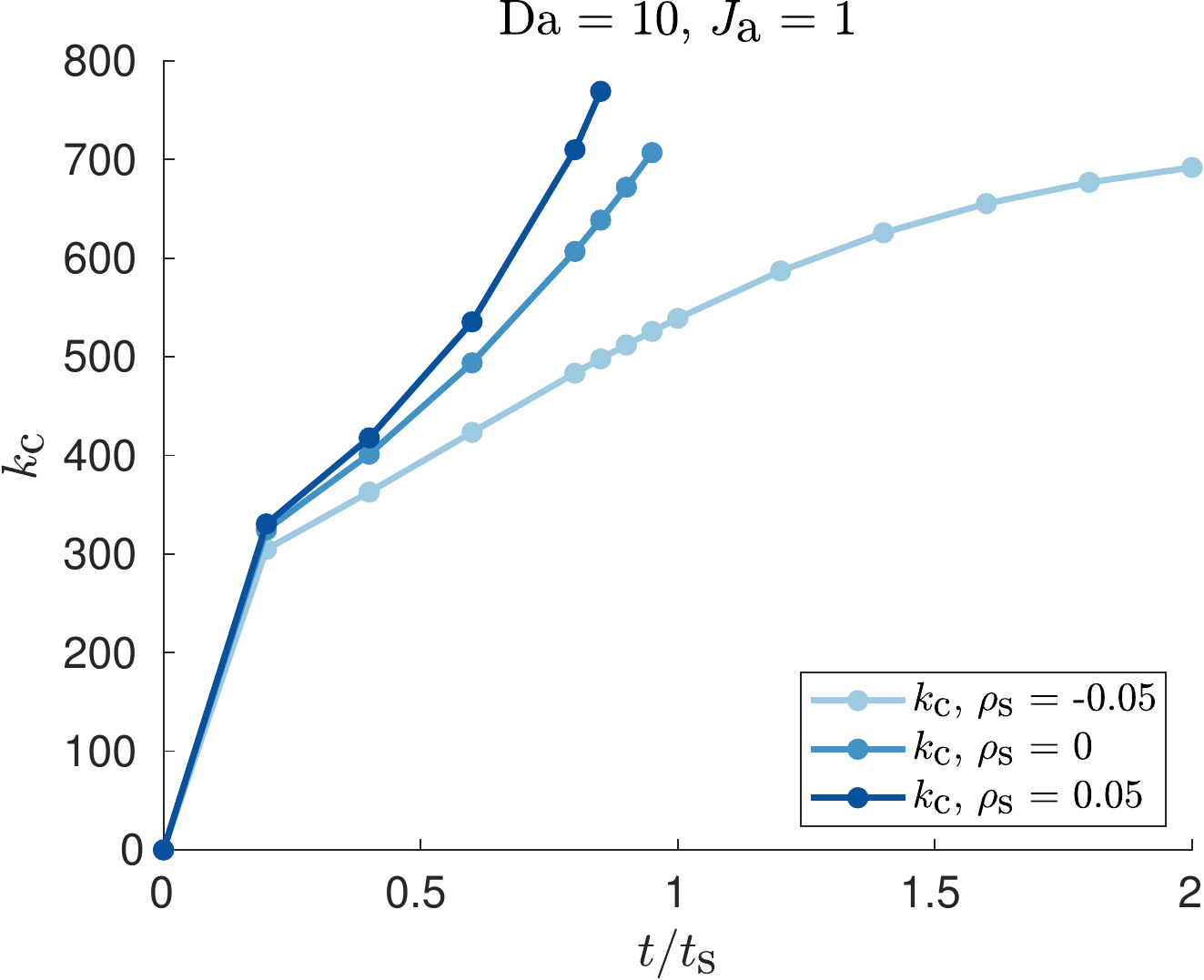}
  \includegraphics[scale=0.4]{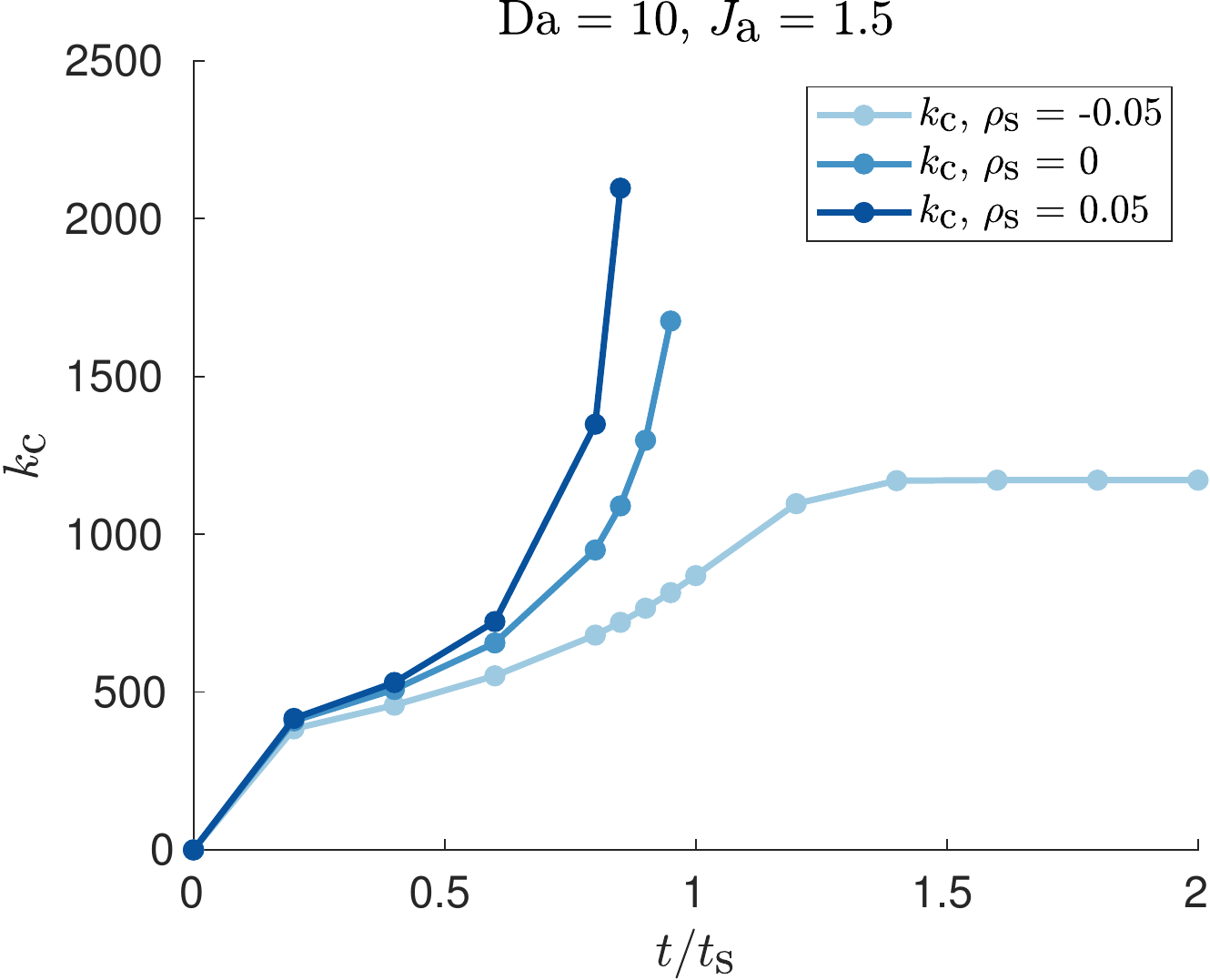}
  \includegraphics[scale=0.4]{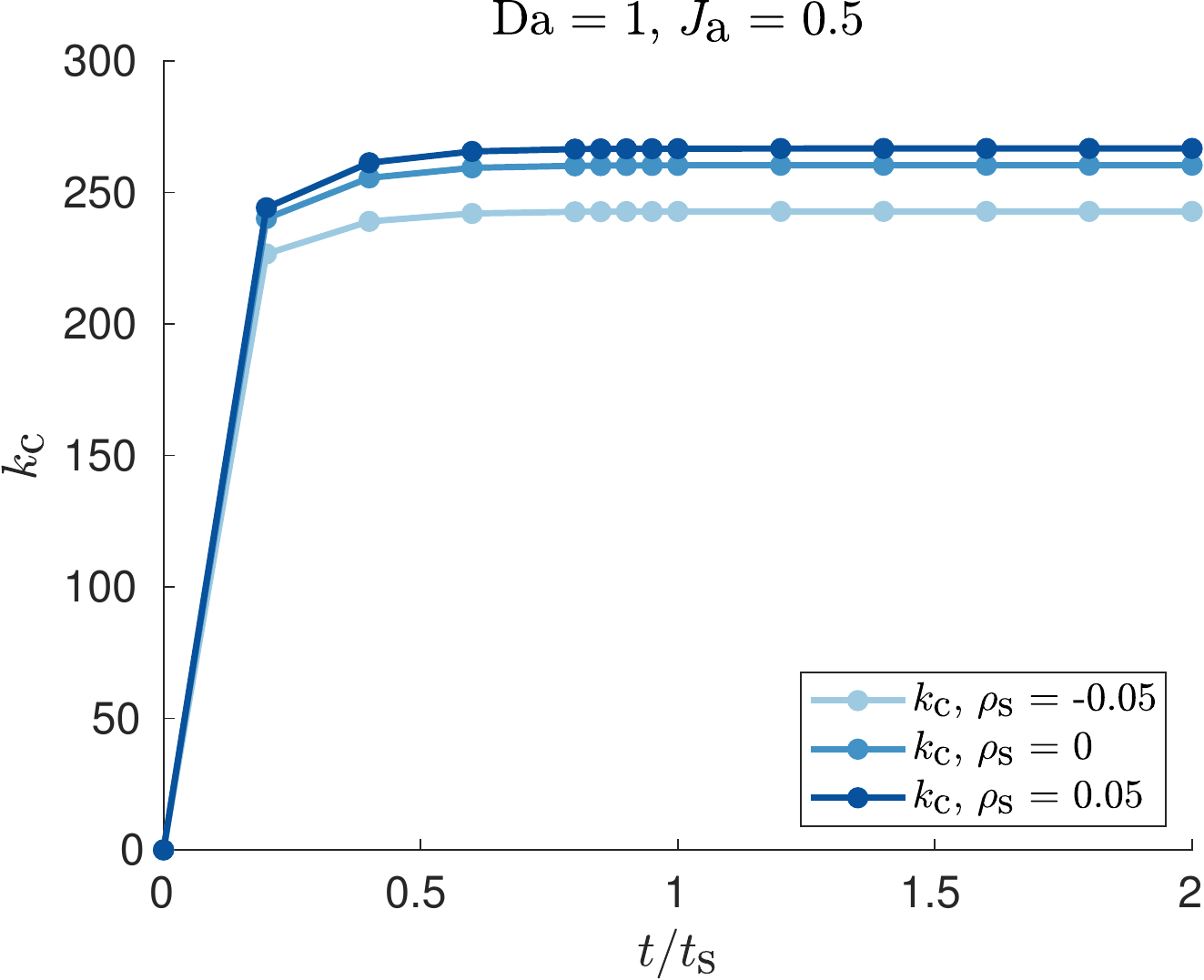}
  \includegraphics[scale=0.4]{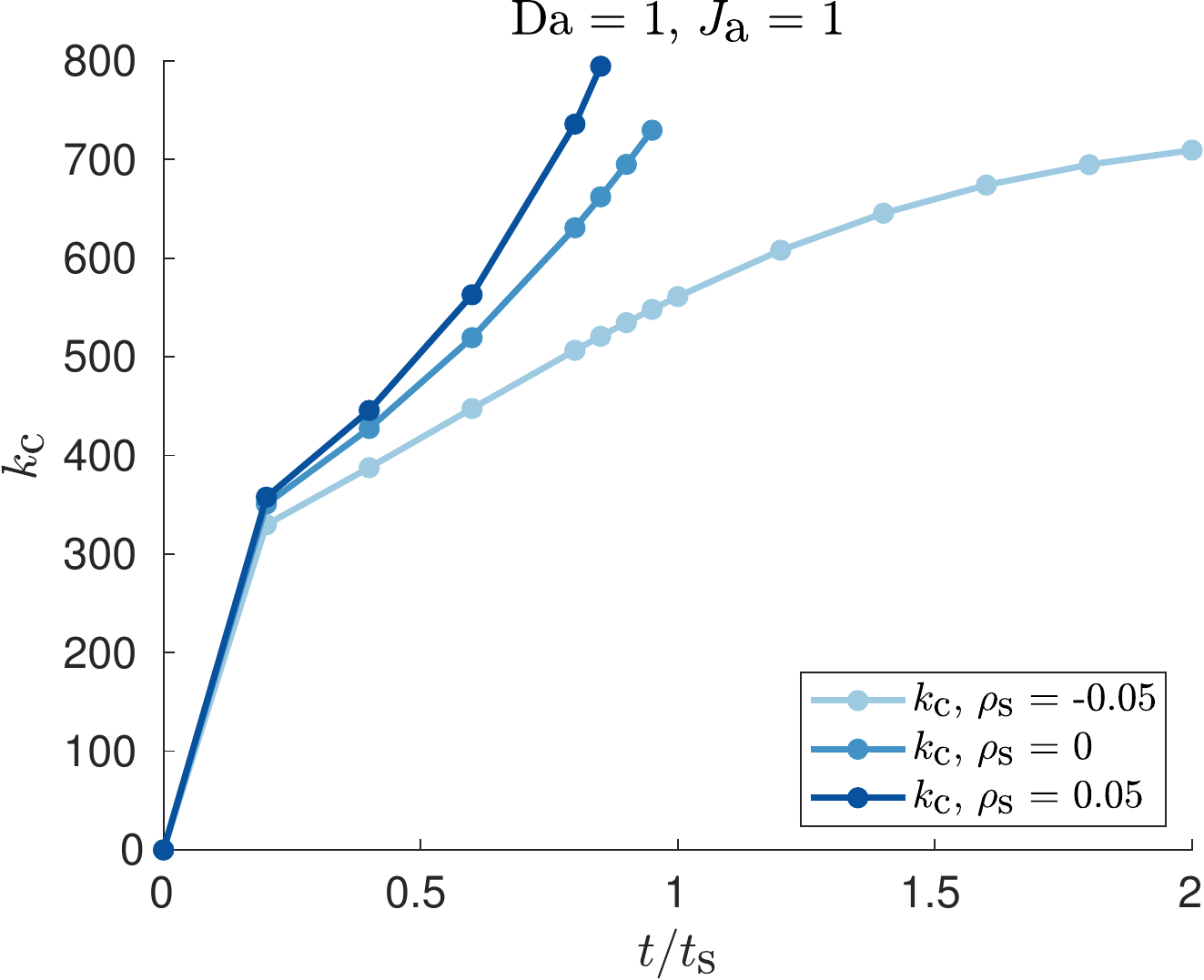}
  \includegraphics[scale=0.4]{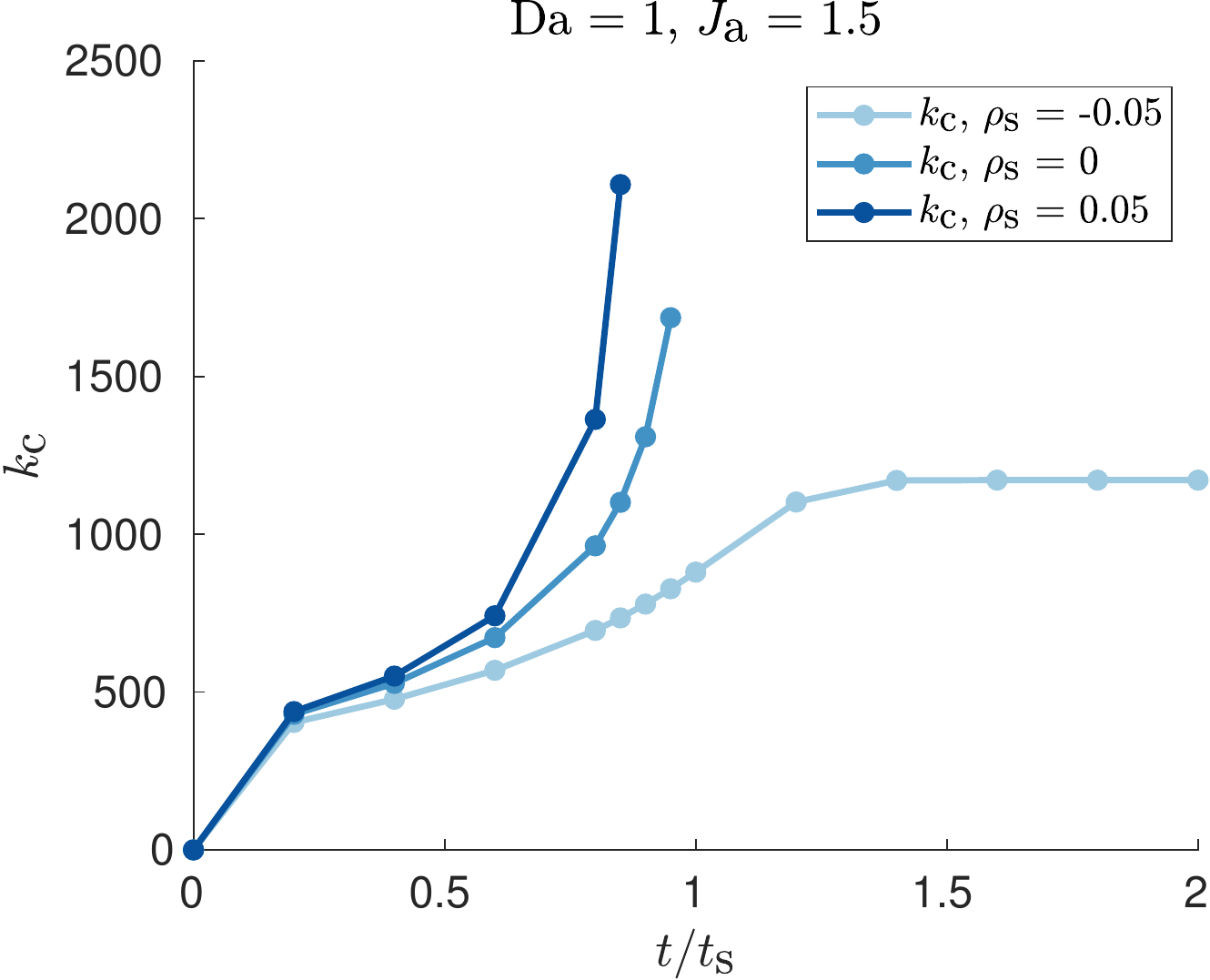}
  \includegraphics[scale=0.4]{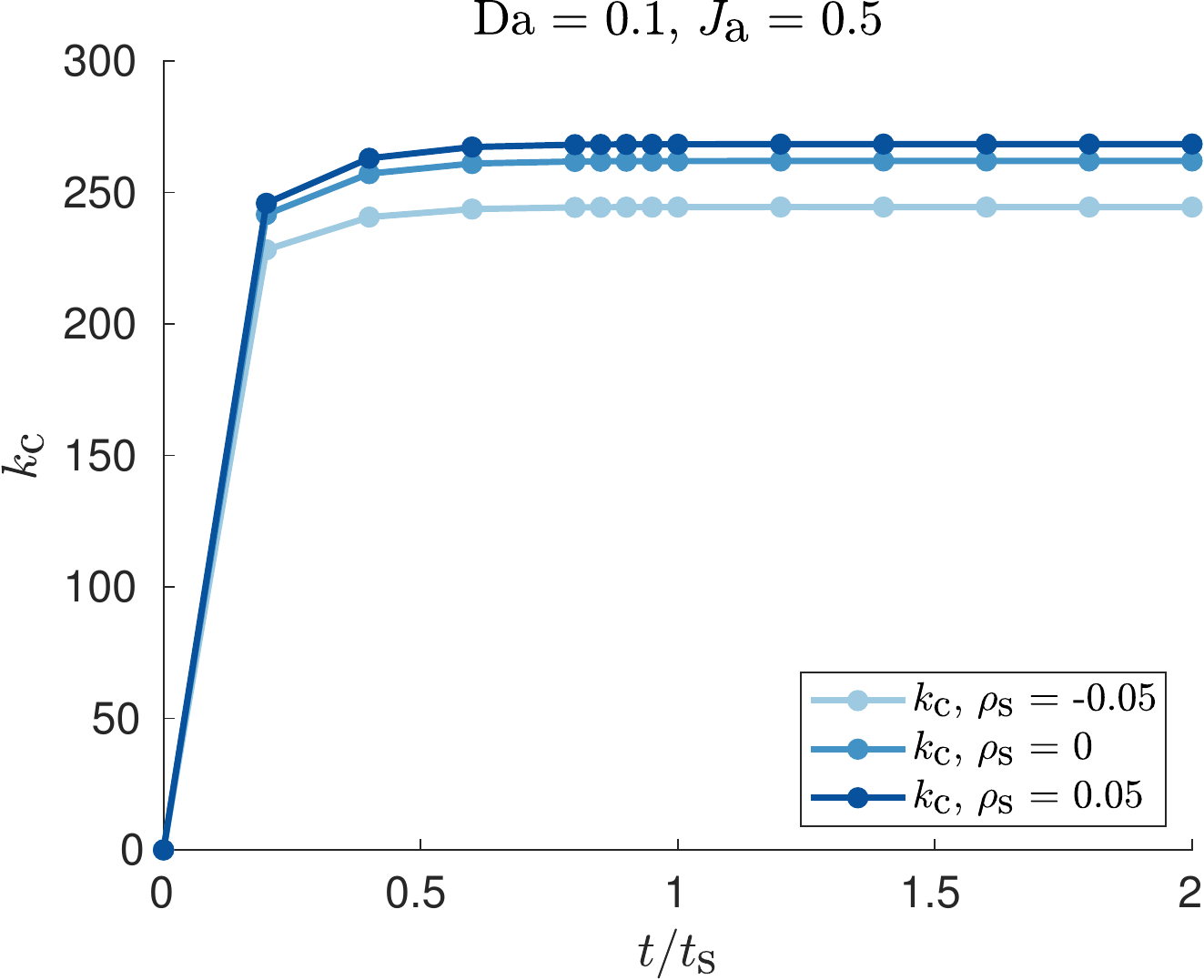}
  \includegraphics[scale=0.4]{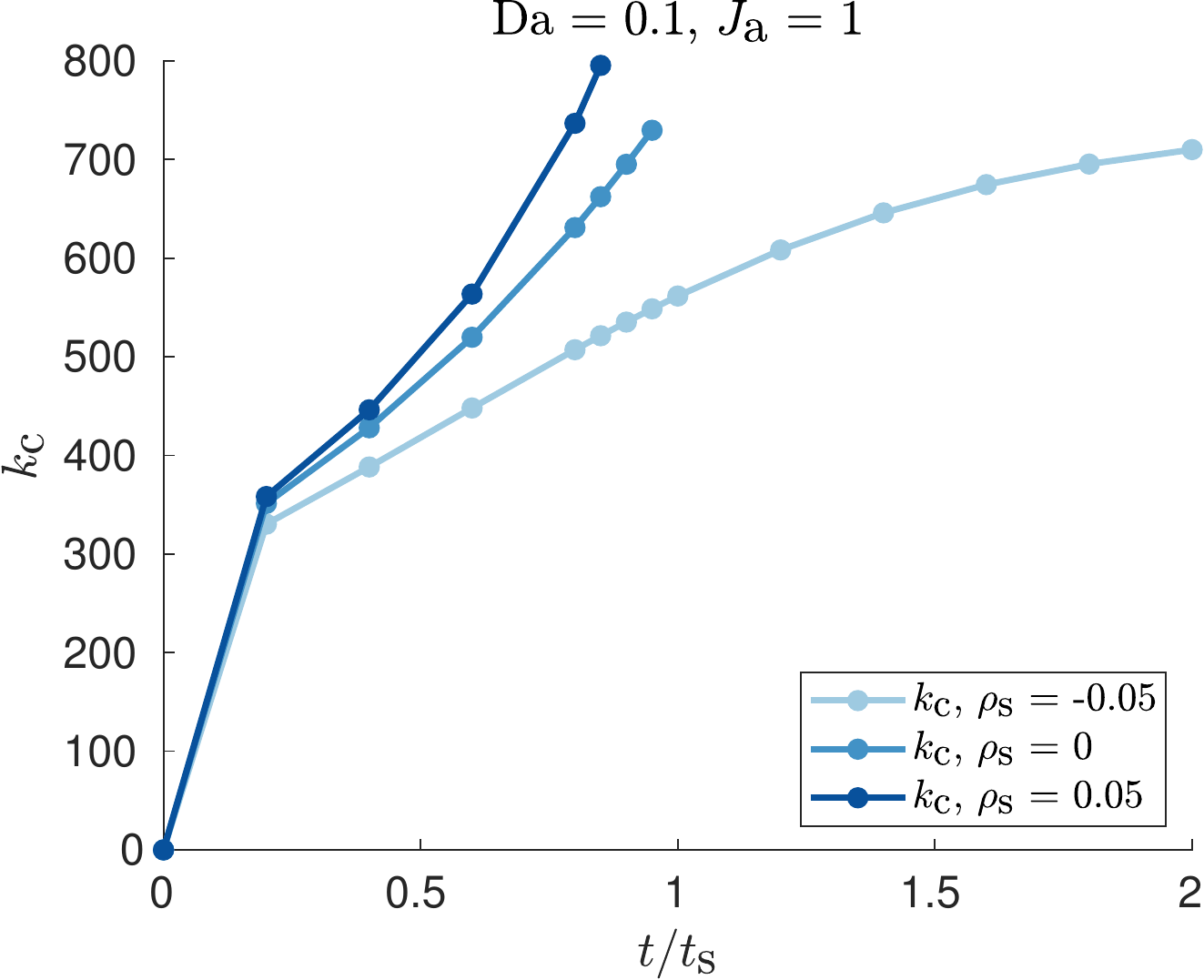}
  \includegraphics[scale=0.4]{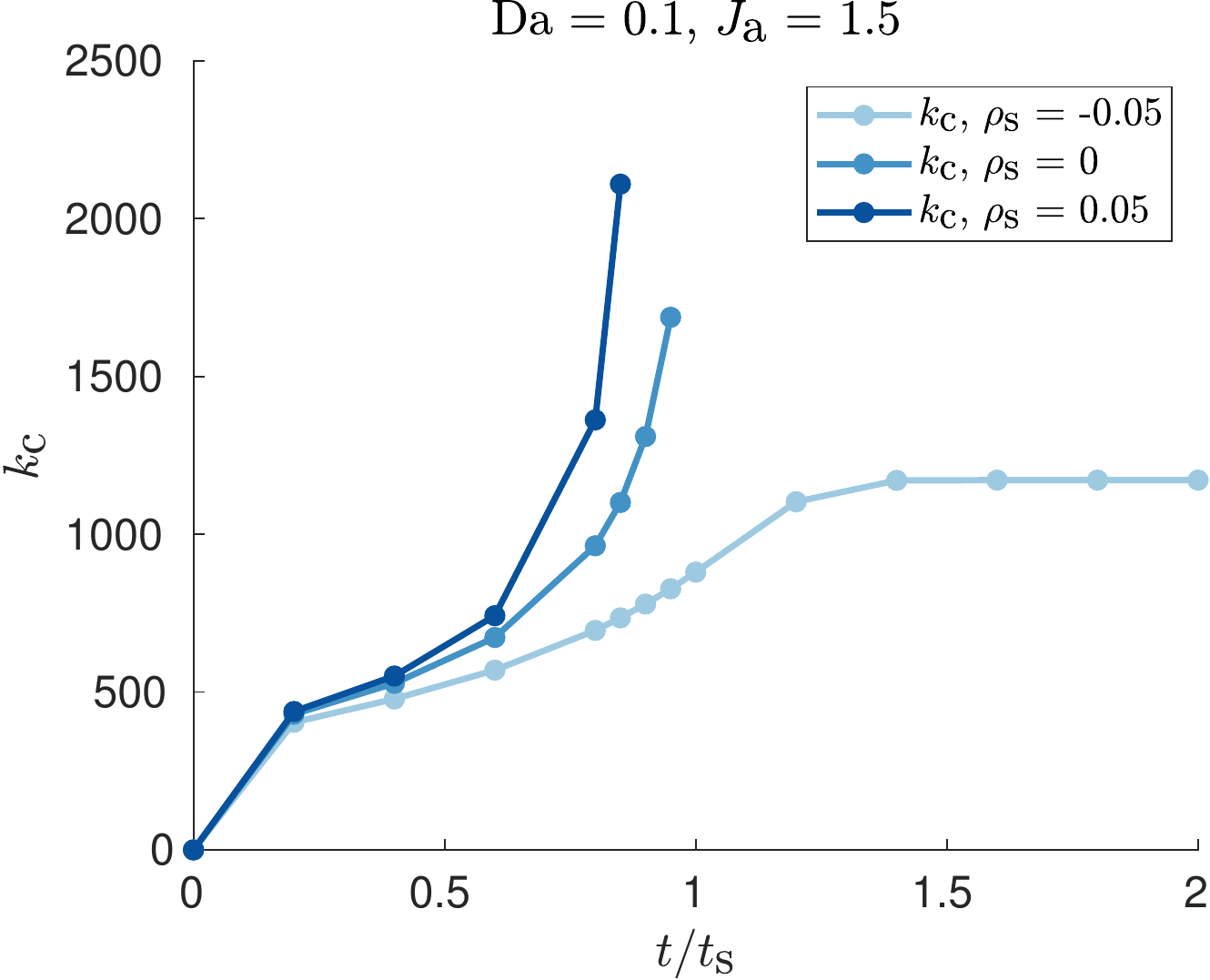}
  \caption{Plots of numerical $k_\textn{c}$ against $\frac{t}{t_\textn{s}}$ for $\rho_\textn{s} \in \left \{-0.05, 0, 0.05\right \}$, $\textn{Da} \in \left \{0.1, 1, 10\right \}$ and $J_\textn{a} \in \left \{0.5, 1, 1.5\right \}$. $\rho_\textn{s}$ increases from left to right and $\textn{Da}$ increases from bottom to top.}\label{fig:k_c against t/t_s}
\end{figure}

As discussed in Section~\ref{sec:Approximations}, at each $t$ point, each $\omega$ curve exhibits a global maximum $\left \{k_\maxn, \omega_\maxn\right \}$ and a critical wavenumber $k_\textn{c}$, which is where the curve crosses the horizontal axis $\omega = 0$. The $\left \{k_\maxn, \omega_\maxn\right \}$ and $k_\textn{c}$ points provide a succinct way to summarize the most physically significant features of the $\omega\left(k\right)$ curve for all the parameter ranges we have explored thus far. Therefore, for $\rho_\textn{s} \in \left \{-0.05, 0, 0.05\right \}$, $\textn{Da} \in \left \{0.1, 1, 10\right \}$ and $J_\textn{a} \in \left \{0.5, 1, 1.5\right \}$, we plot numerically computed $k_\maxn$ and $\omega_\maxn$ against $\frac{t}{t_\textn{s}}$ in Figure~\ref{fig:k_max and omega_max against t/t_s} and numerically computed $k_\textn{c}$ against $\frac{t}{t_\textn{s}}$ in Figure~\ref{fig:k_c against t/t_s}. For $J_\textn{a} \geq 1$, we observe that the $k_\maxn$ and $\omega_\maxn$ curves diverge near $t_\textn{s}$ for $\rho_\textn{s} \geq 0$ but level off to constant values past $t_\textn{s}$ for $\rho_\textn{s} < 0$, therefore these curves appear as if they are ``fanning out''. In contrast, for $J_\textn{a} < 1$, the $k_\maxn$ and $\omega_\maxn$ curves level off past $t_\textn{s}$ for all values of $\rho_\textn{s}$ as the system eventually reaches a steady state when an underlimiting current is applied. The $k_\textn{c}$ curves have the same qualitative shape as the $k_\maxn$ curves except that they are larger, as expected. The effects of $\textn{Da}$ and $J_\textn{a}$ on the $k_\maxn$, $\omega_\maxn$ and $k_\textn{c}$ values, which are previously discussed in the context of the dispersion relation, are also clearly reflected in Figures~\ref{fig:k_max and omega_max against t/t_s} and~\ref{fig:k_c against t/t_s}.

In an effort to make the electrode surface less unstable at overlimiting current, we focus on $\rho_\textn{s} < 0$ to determine how much additional stabilization a negative $\rho_\textn{s}$ confers to the surface as it gets increasingly more negative. Subsequently, we plot numerically computed $k_\maxn$, $\omega_\maxn$ and $k_\textn{c}$ against $\frac{t}{t_\textn{s}}$ for $\rho_\textn{s} \in \left \{-1, -0.75, -0.5, -0.25, -0.05\right \}$, $\textn{Da} = 1$ and $J_\textn{a} = 1.5$ in Figure~\ref{fig:k_max, omega_max and k_c against t/t_s for negative rho_s}. While a more negative $\rho_\textn{s}$ generally decreases $k_\maxn$, $\omega_\maxn$ and $k_\textn{c}$, it is clear that there are diminishing returns to the amount of additional stabilization achieved. It also appears that complete stabilization is not possible as $\omega_\maxn$ remains positive even for $\rho_\textn{s} = -1$, albeit at a small value. In practice, it is probable that a sufficiently small and positive $\omega_\maxn$ value can be deemed to be small enough for considering an electrode surface ``practically stable'', but experiments that measure and correlate $\omega_\maxn$ with observations of metal growth need to be performed in order to determine this threshold $\omega_\maxn$ value.

\begin{figure}
  \centering
  \includegraphics[scale=0.4]{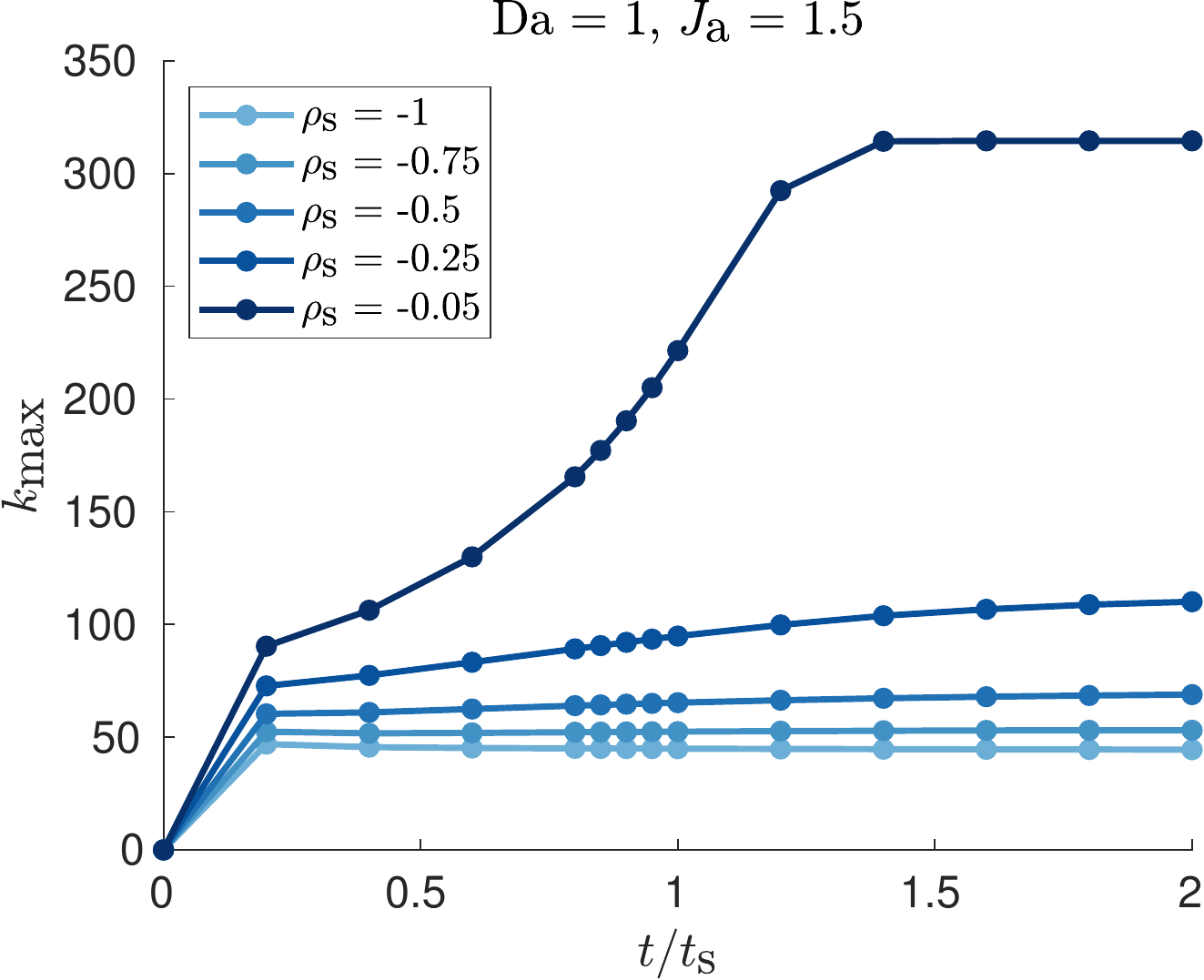}
  \includegraphics[scale=0.4]{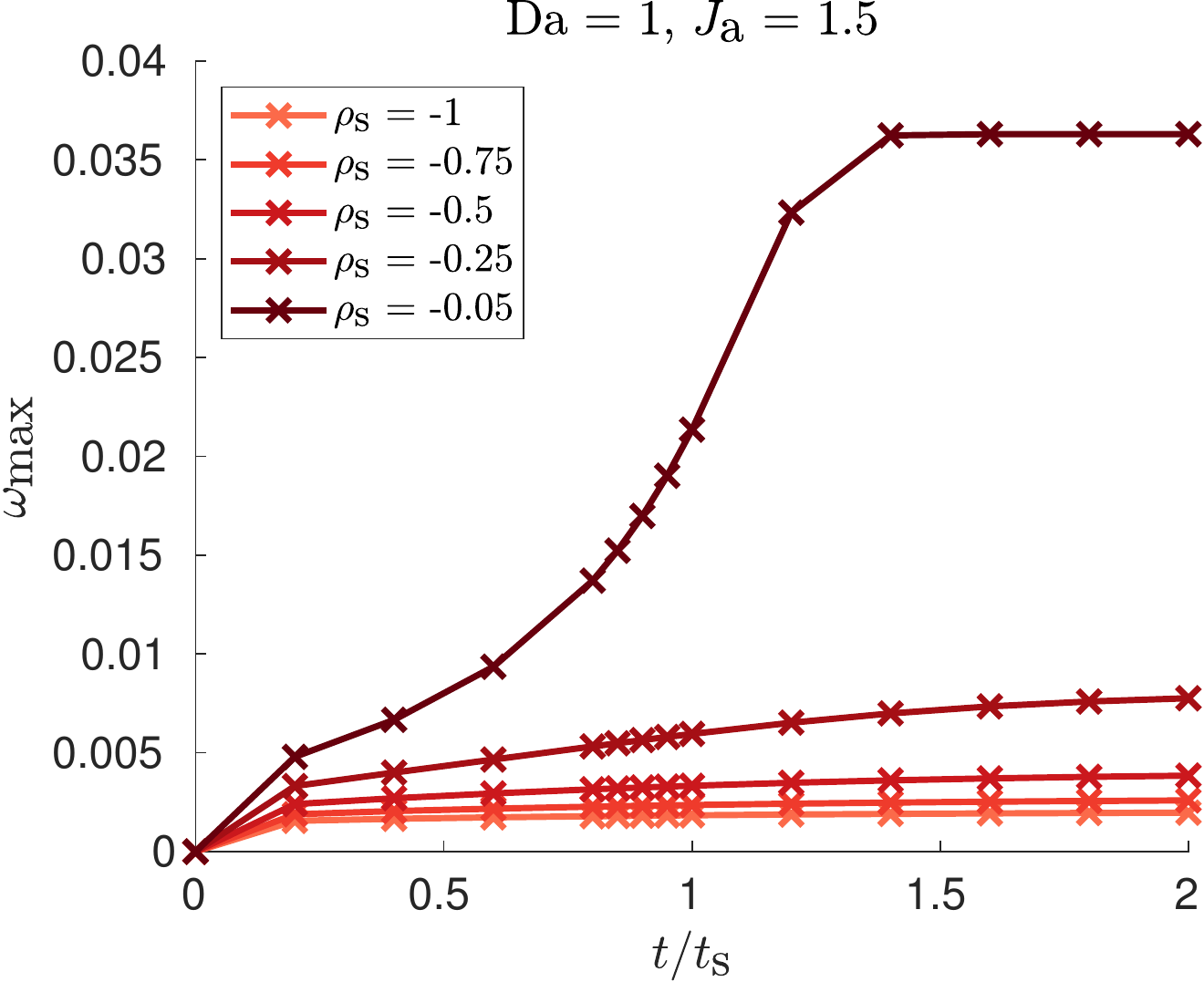}
  \includegraphics[scale=0.4]{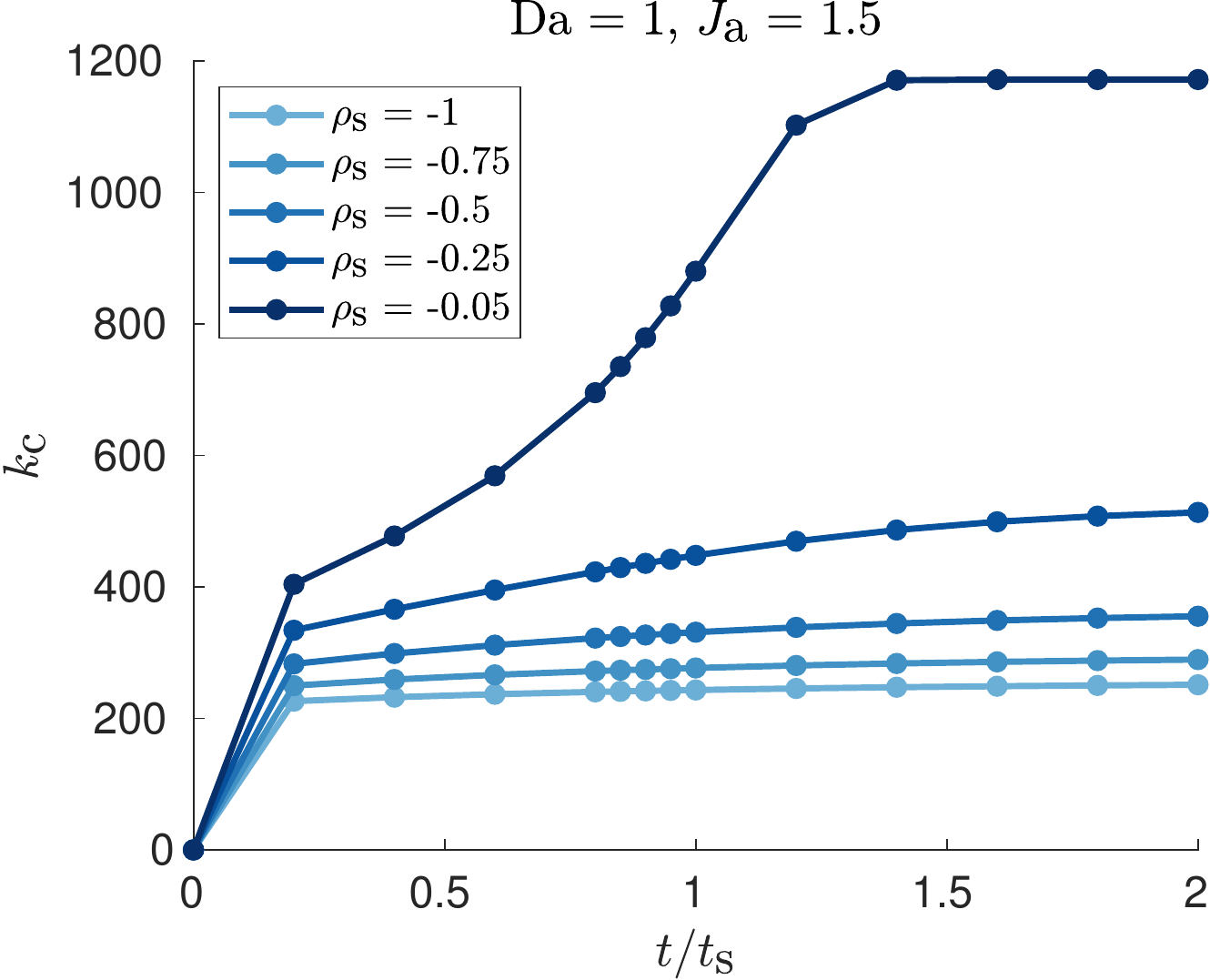}
  \caption{Plots of numerical $k_\maxn$, $\omega_\maxn$ and $k_\textn{c}$ against $\frac{t}{t_\textn{s}}$ for $\rho_\textn{s} \in \left \{-1, -0.75, -0.5, -0.25, -0.05\right \}$, $\textn{Da} = 1$ and $J_\textn{a} = 1.5$ (overlimiting current) for investigating additional stabilization of electrode surface conferred by increasingly negative $\rho_\textn{s}$ values.}\label{fig:k_max, omega_max and k_c against t/t_s for negative rho_s}
\end{figure}

\subsection{Comparison between numerical and approximate solutions}\label{sec:Comparison between numerical and approximate solutions}

To illustrate how well the approximations given by Equations~\ref{eq:omega} and~\ref{eq:k_c} work for the parameter ranges considered, we plot numerical and approximate values of $k_\maxn$, $\omega_\maxn$ and $k_\textn{c}$ against $\frac{t}{t_\textn{s}}$ for $\rho_\textn{s} \in \left \{-0.05, 0, 0.05\right \}$, $\textn{Da} = 1$ and $J_\textn{a} = 1.5$ in Figure~\ref{fig:Numerical and approximate k_max, omega_max and k_c against t/t_s for J_a = 1.5}. In the interest of space, these plots for other values of $\textn{Da}$ and $J_\textn{a}$ are provided in Figures 6 to 11 of Section VII of Supplementary Material. For all parameter ranges considered, the agreement between numerical and approximate values of $k_\maxn$, $\omega_\maxn$ and $k_\textn{c}$ is excellent, giving us confidence that the approximations are useful for rapidly and accurately computing $k_\maxn$, $\omega_\maxn$ and $k_\textn{c}$. This confirms that $k_\maxn$ and $k_\textn{c}$ are large enough that Equations~\ref{eq:omega} and~\ref{eq:k_c}, which have assumed that $k$ is sufficiently large, are accurate for approximating them. We will therefore use Equations~\ref{eq:omega} and~\ref{eq:k_c} extensively in Sections~\ref{sec:Application to copper electrodeposition} and~\ref{sec:Pulse electroplating and pulse charging} that follow.

\begin{figure}
  \centering
  \includegraphics[scale=0.37]{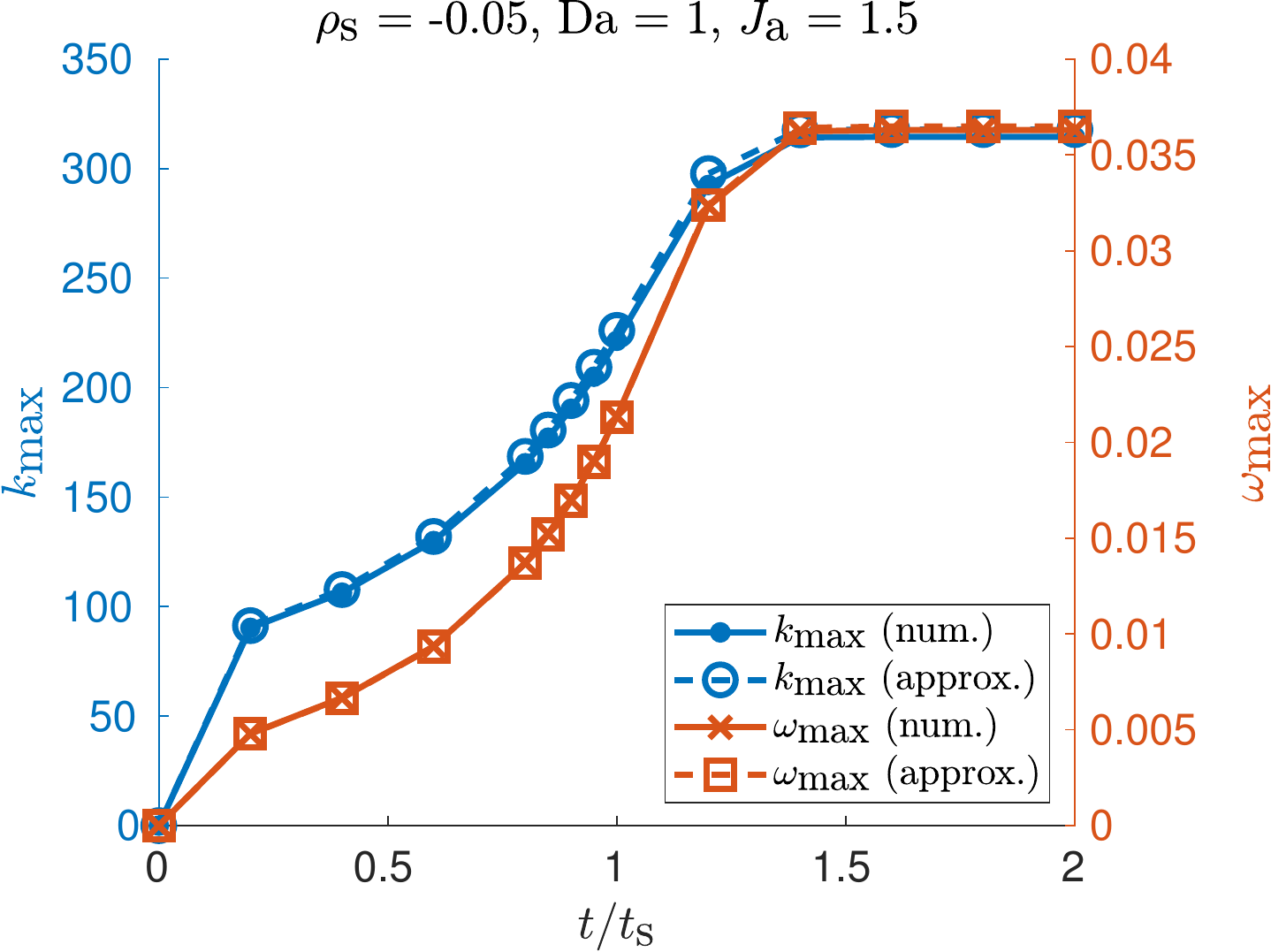}
  \includegraphics[scale=0.37]{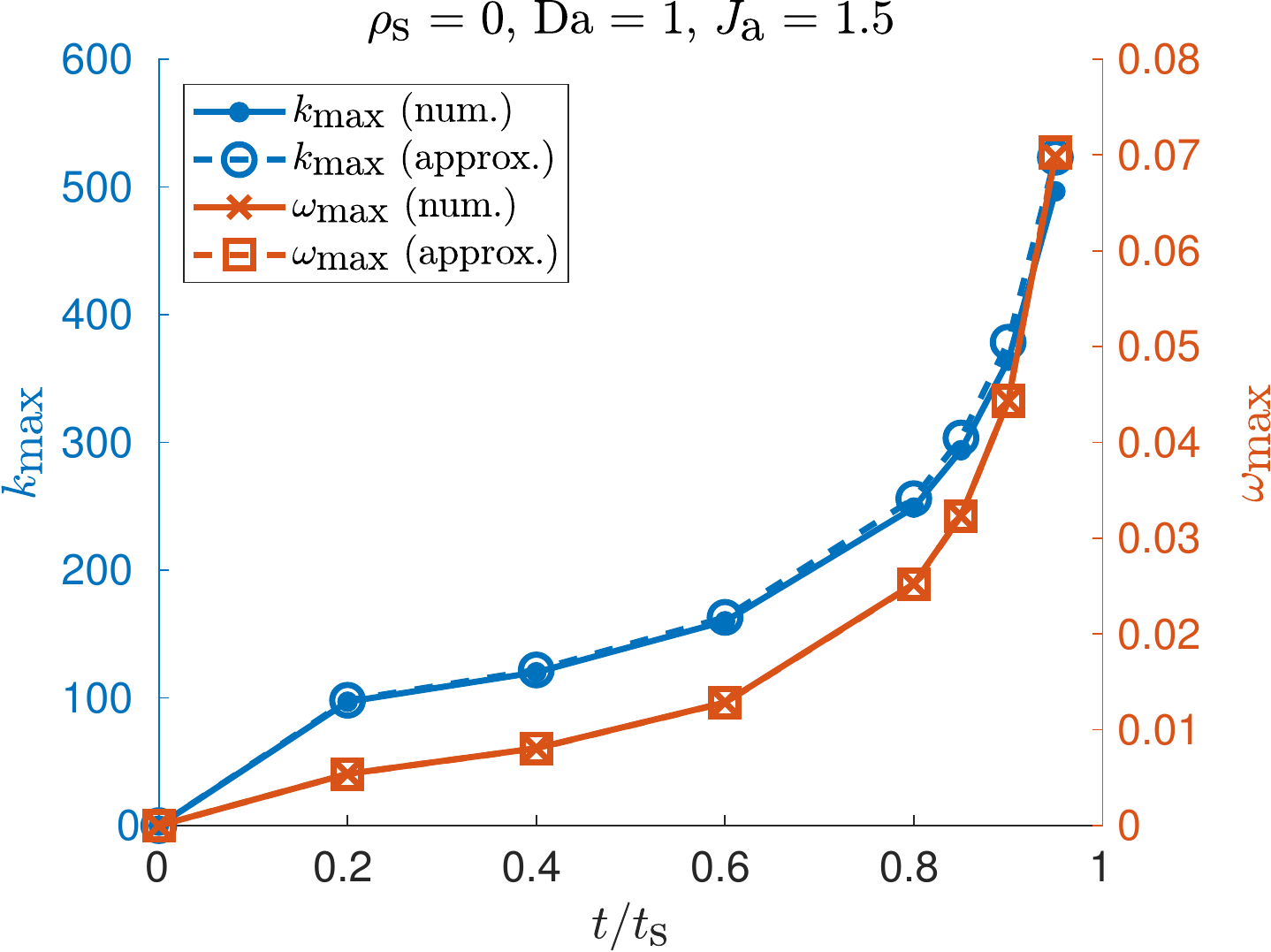}
  \includegraphics[scale=0.37]{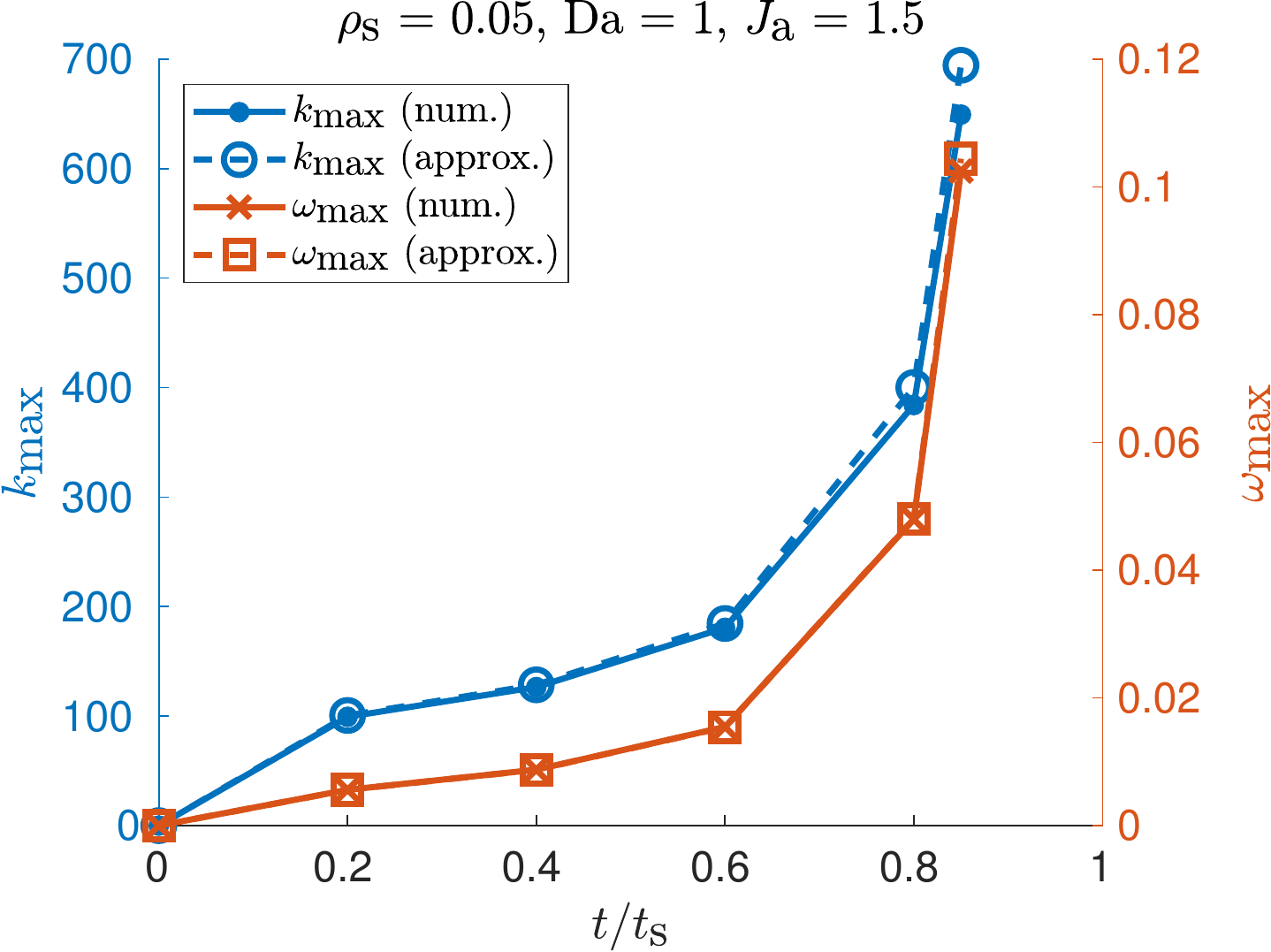}
  \includegraphics[scale=0.37]{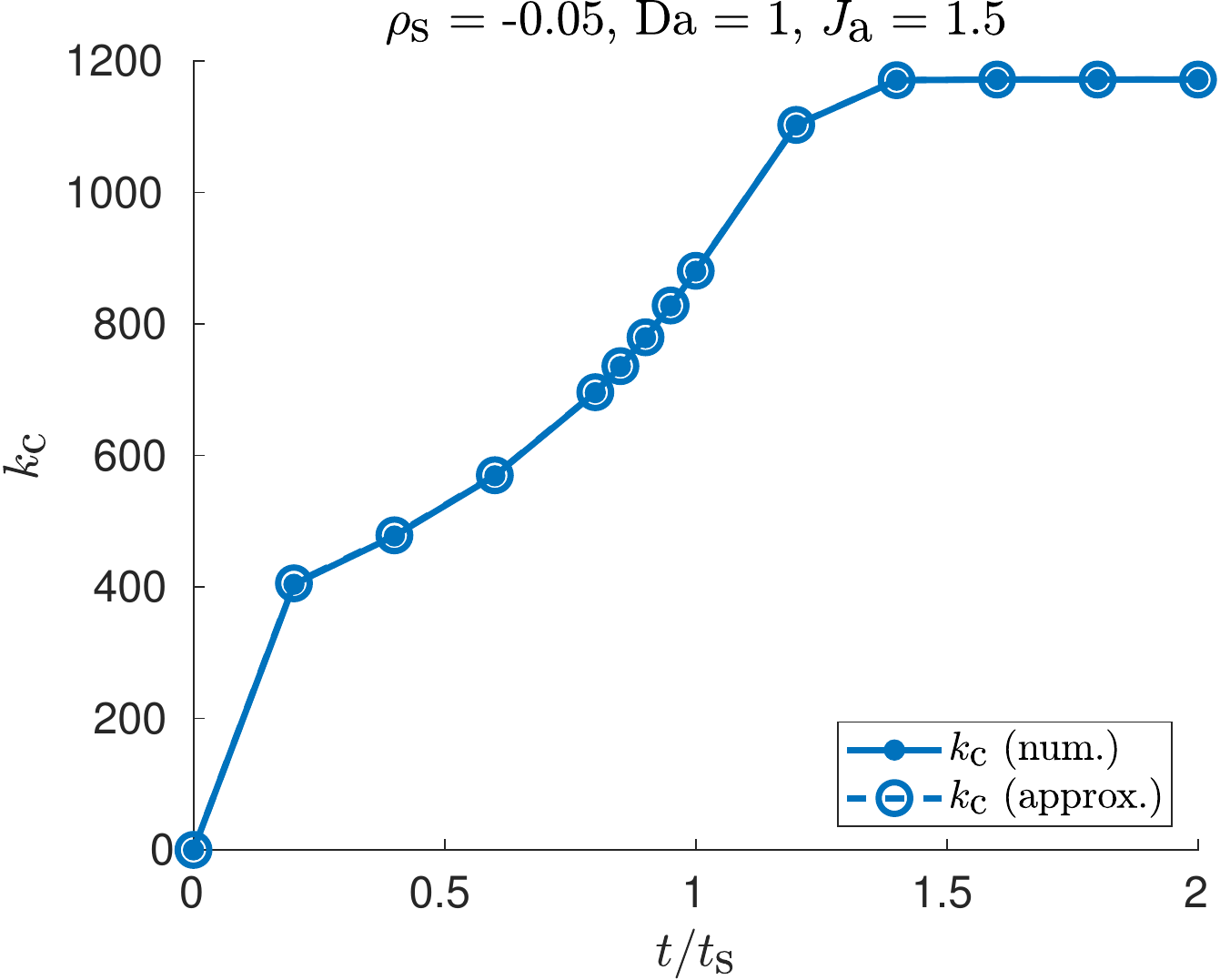}
  \includegraphics[scale=0.37]{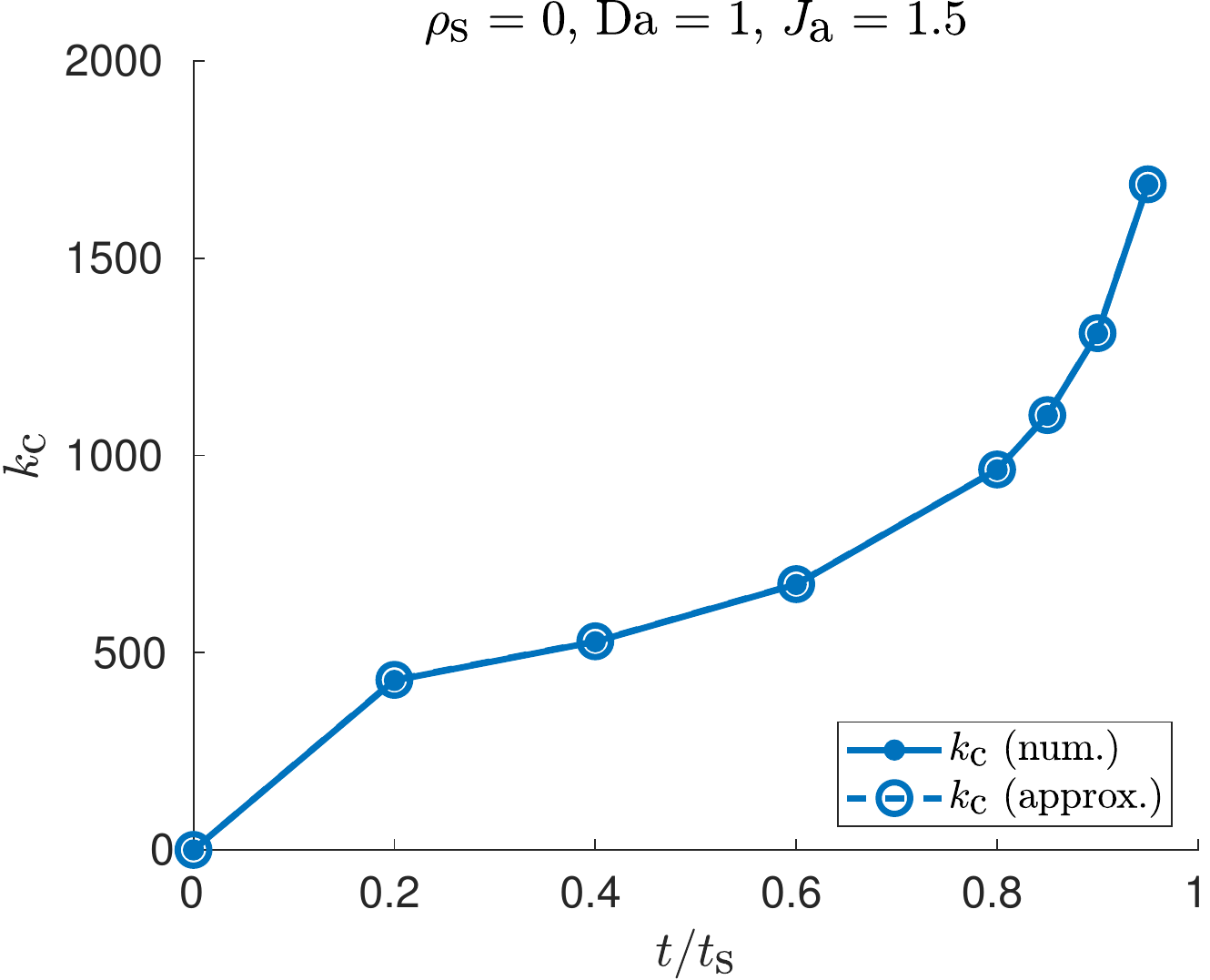}
  \includegraphics[scale=0.37]{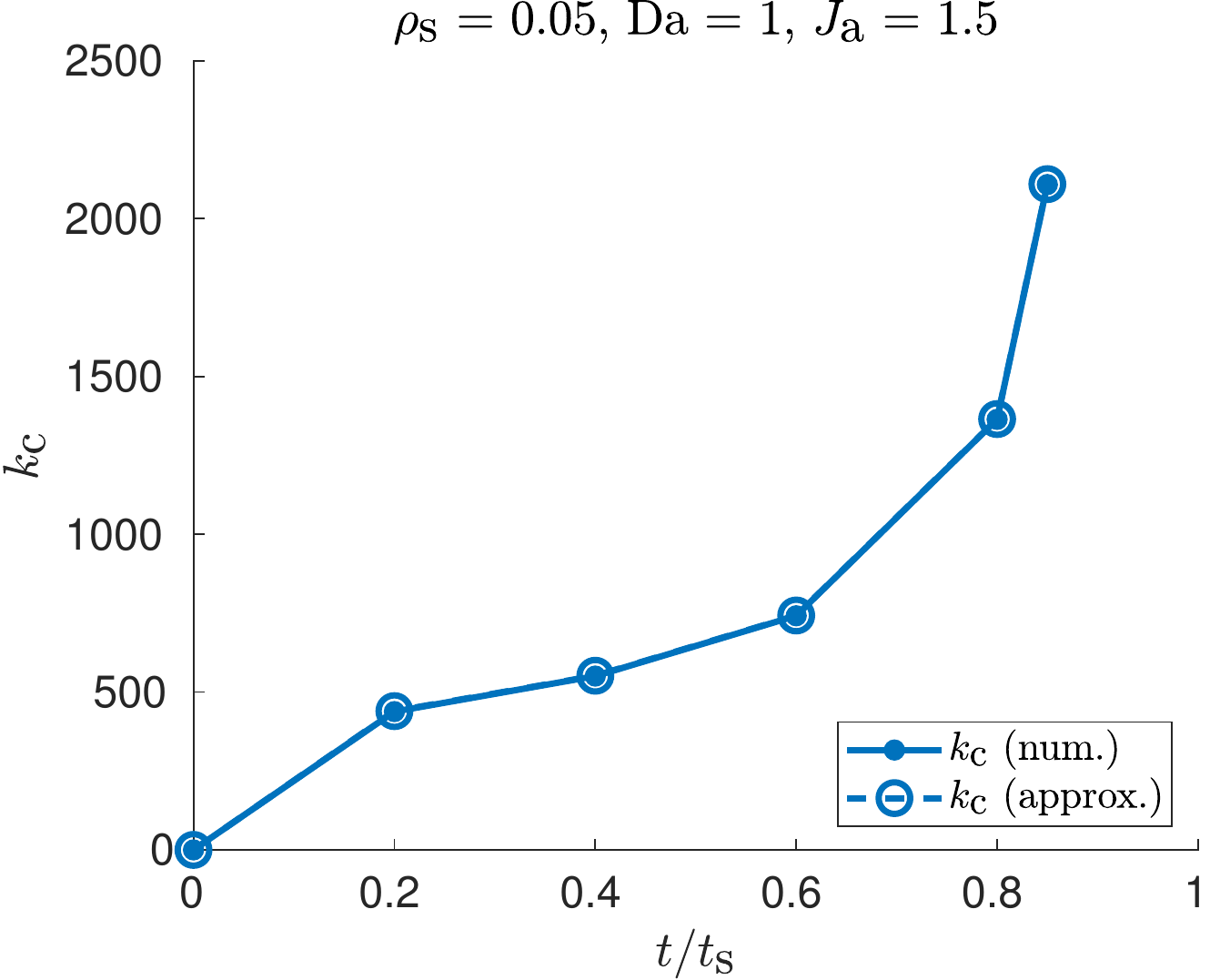}
  \caption{Plots of numerical and approximate values of $k_\maxn$, $\omega_\maxn$ and $k_\textn{c}$ against $\frac{t}{t_\textn{s}}$ for $\rho_\textn{s} \in \left \{-0.05, 0, 0.05\right \}$, $\textn{Da} = 1$ and $J_\textn{a} = 1.5$ (overlimiting current). Top row: Plots of $k_\maxn$ and $\omega_\maxn$. Bottom row: Plots of $k_\textn{c}$. In the legends, ``num.'' refers to numerical solutions while ``approx.'' refers to approximate solutions.}\label{fig:Numerical and approximate k_max, omega_max and k_c against t/t_s for J_a = 1.5}
\end{figure}

\subsection{Application to copper electrodeposition}\label{sec:Application to copper electrodeposition}

We now apply linear stability analysis to the specific case of copper electrodeposition and electrodissolution and compare it with experimental data~\cite{han_dendrite_2016} to determine how well theory agrees with experiment. Because copper electrodeposition involves the overall transfer of two electrons that are transferred one at a time in a serial manner, we need to first derive the overall expression for the Faradaic current density $J_\textn{F}$.

Assuming that the activity of electrons is $1$ and dilute solution theory is applicable, for a $n$-electron transfer reaction, the dimensionless forms of Equations~\ref{eq:J_F} and~\ref{eq:Nernst equation} are given by
\begin{align}
  J_\textn{F} &= j_0\left \{\exp\left(-\alpha n\eta\right) - \exp\left[\left(1-\alpha\right)n\eta\right]\right \}, \quad j_0 = k_0n\hat{c}_\textn{O}^{1-\alpha}\hat{c}_\textn{R}^\alpha = \textn{Da}n\hat{c}_\textn{O}^{1-\alpha}\hat{c}_\textn{R}^\alpha, \\
  \Delta\phi^\textn{eq} &= \frac{1}{n}\ln\frac{\hat{c}_\textn{O}}{\hat{c}_\textn{R}} + E^\Theta - \frac{2\gamma\mathscr{H}}{n}.
\end{align}
For multistep electron transfer reactions, it is more convenient to work with $\Delta\phi$ instead of $\eta$. Therefore, we rewrite $J_\textn{F}$ in terms of $\Delta\phi$ as
\begin{align}
  J_\textn{F} &= n\left \{k_\textn{c}\hat{c}_\textn{O}\exp\left[-\alpha n\left(\Delta\phi+\frac{2\gamma\mathscr{H}}{n}\right)\right] - k_\textn{a}\hat{c}_\textn{R}\exp\left[\left(1-\alpha\right)n\left(\Delta\phi+\frac{2\gamma\mathscr{H}}{n}\right)\right]\right \}, \label{eq:J_F in terms of Delta_phi} \\
  E^\Theta &= \frac{1}{n}\ln\frac{k_\textn{c}}{k_\textn{a}}, \quad  k_0 = k_\textn{a}^\alpha k_\textn{c}^{1-\alpha},
\end{align}
where $k_\textn{c}$ and $k_\textn{a}$ are the cathodic and anodic rate constants respectively.

The reaction mechanism for copper electrodeposition and electrodissolution is given by~\cite{newman_electrochemical_2004,mattsson_galvanostatic_1959,bockris_mechanism_1962,brown_rate-determining_1965}
\begin{align}
  \textn{Cu}^{2+}\textn{(aq)} + \textn{e}^- &\rightleftharpoons \textn{Cu}^+\textn{(ads)}, \\
  \textn{Cu}^+\textn{(ads)} + \textn{e}^- &\rightleftharpoons \textn{Cu}\textn{(s)},
\end{align}
where (aq), (ads) and (s) indicate aqueous, adsorbed and solid respectively. The first step is assumed to be the rate-determining step while the second step is assumed to be at equilibrium. Applying Equation~\ref{eq:J_F in terms of Delta_phi} to each step, noting that the activity of solid metal is $1$ and rewriting $J_\textn{F}$ in terms of $\eta$, we obtain
\begin{align}
  J_\textn{F} &= j_0\left \{\exp\left(-\alpha_1\eta\right) - \exp\left[\left(2-\alpha_1\right)\eta\right]\right \}, \quad j_0 = 2k_0\hat{c}_+^{1-\frac{\alpha_1}{2}} = 2\textn{Da}\hat{c}_+^{1-\frac{\alpha_1}{2}}, \label{eq:J_F for copper} \\
  \Delta\phi^\textn{eq} &= \frac{1}{2}\ln\hat{c}_+ + E^\Theta - 2\gamma\mathscr{H}, \label{eq:Nernst potential for copper}
\end{align}
where $\alpha_1$ is the charge transfer coefficient of the first step.

Previously in Section~\ref{sec:Electrochemical reaction kinetics}, for a $1$-step $n$-electron transfer metal electrodeposition reaction, the dimensionless forms of Equations~\ref{eq:J_F} and~\ref{eq:Simplified Nernst equation and exchange current density} are given by
\begin{align}
  J_\textn{F} &= j_0\left \{\exp\left(-\alpha n\eta\right) - \exp\left[\left(1-\alpha\right)n\eta\right]\right \}, \quad j_0 = k_0n\hat{c}_+^{1-\alpha}= \textn{Da}n\hat{c}_+^{1-\alpha} \label{eq:Dimensionless simplified J_F}, \\
  \Delta\phi^\textn{eq} &= \frac{1}{n}\ln\hat{c}_+ + E^\Theta - \frac{2\gamma\mathscr{H}}{n}. \label{eq:Dimensionless simplified Nernst equation}
\end{align}
By comparing Equations~\ref{eq:J_F for copper} and~\ref{eq:Nernst potential for copper} with Equations~\ref{eq:Dimensionless simplified J_F} and~\ref{eq:Dimensionless simplified Nernst equation}, we set $n = 2$ and $\alpha = \frac{\alpha_1}{2}$ and replace $\gamma$ with $2\gamma$ in the original set of equations in order to adapt the linear stability analysis for copper electrodeposition.

By carrying out nonlinear least squares fitting on experimental steady state current-voltage relations, we have previously performed parameter estimation~\cite{khoo_theory_2018} for copper electrodeposition in a copper(II) sulfate ($\textn{CuSO}_4$) electrolyte in cellulose nitrate (CN) membranes~\cite{han_dendrite_2016}, which are a random nanoporous medium with well connected pores. The parameters that are estimated are $\rho_\textn{s}$, $\tau$, $\textn{Da}$, $\alpha_1$ and $\epsilon_\textn{p}$ and their fitted values are provided in Table III in~\cite{khoo_theory_2018}. Other parameters specific to the copper electrodeposition reaction, $\textn{CuSO}_4$ electrolyte and CN membranes used are also provided in Tables I and II in~\cite{khoo_theory_2018}. For the surface energy of the copper/electrolyte interface, we use dimensional $\gamma = 1.85\,\textn{J}/\textn{m}^2$ given in Table I in~\cite{nielsen_morphological_2015}.  % chktex 36

\begin{figure}
  \centering
  (a)
  \includegraphics[scale=0.62]{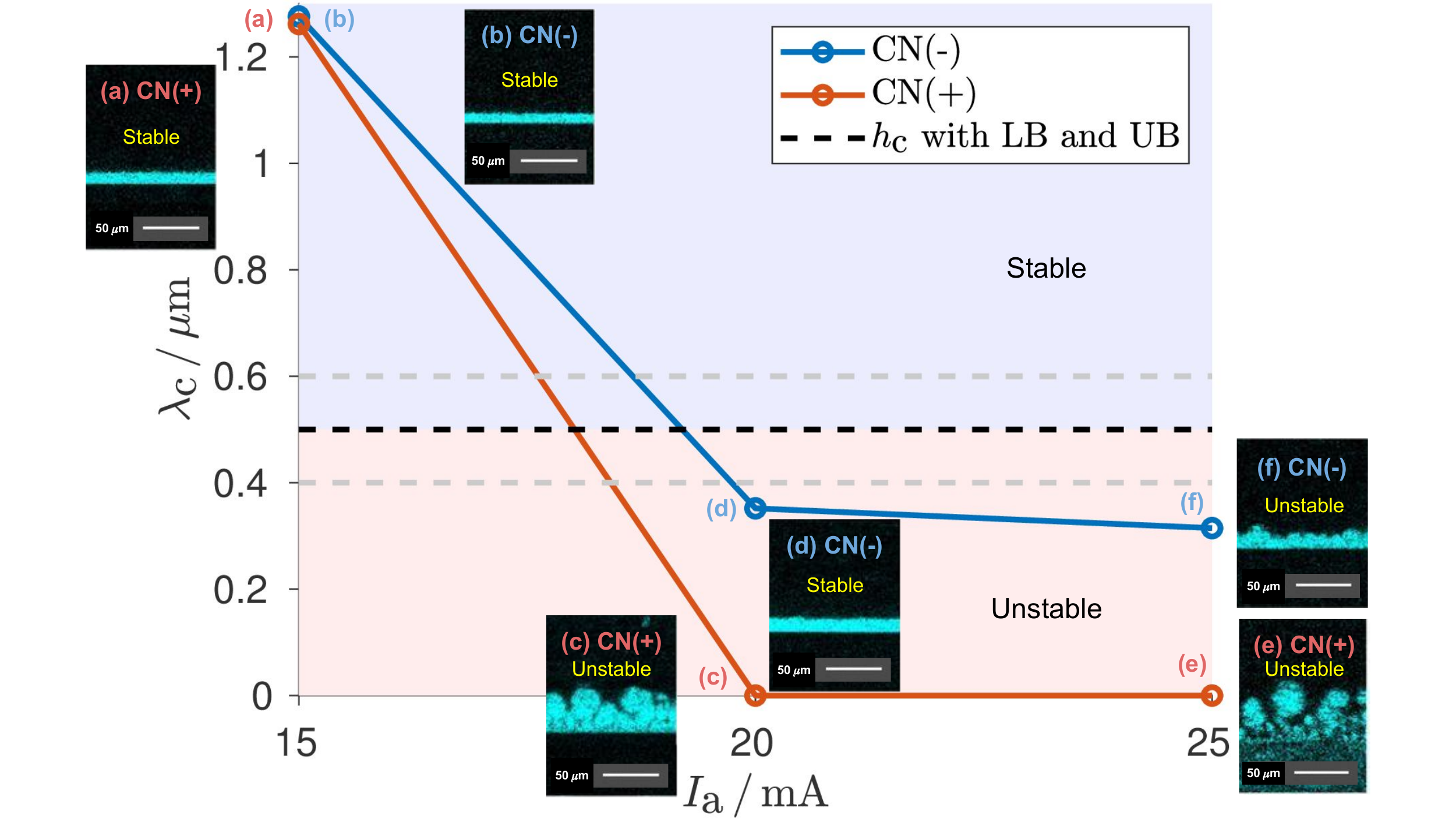} \\
  \vspace{0.25cm}
  (b)
  \includegraphics[scale=0.62]{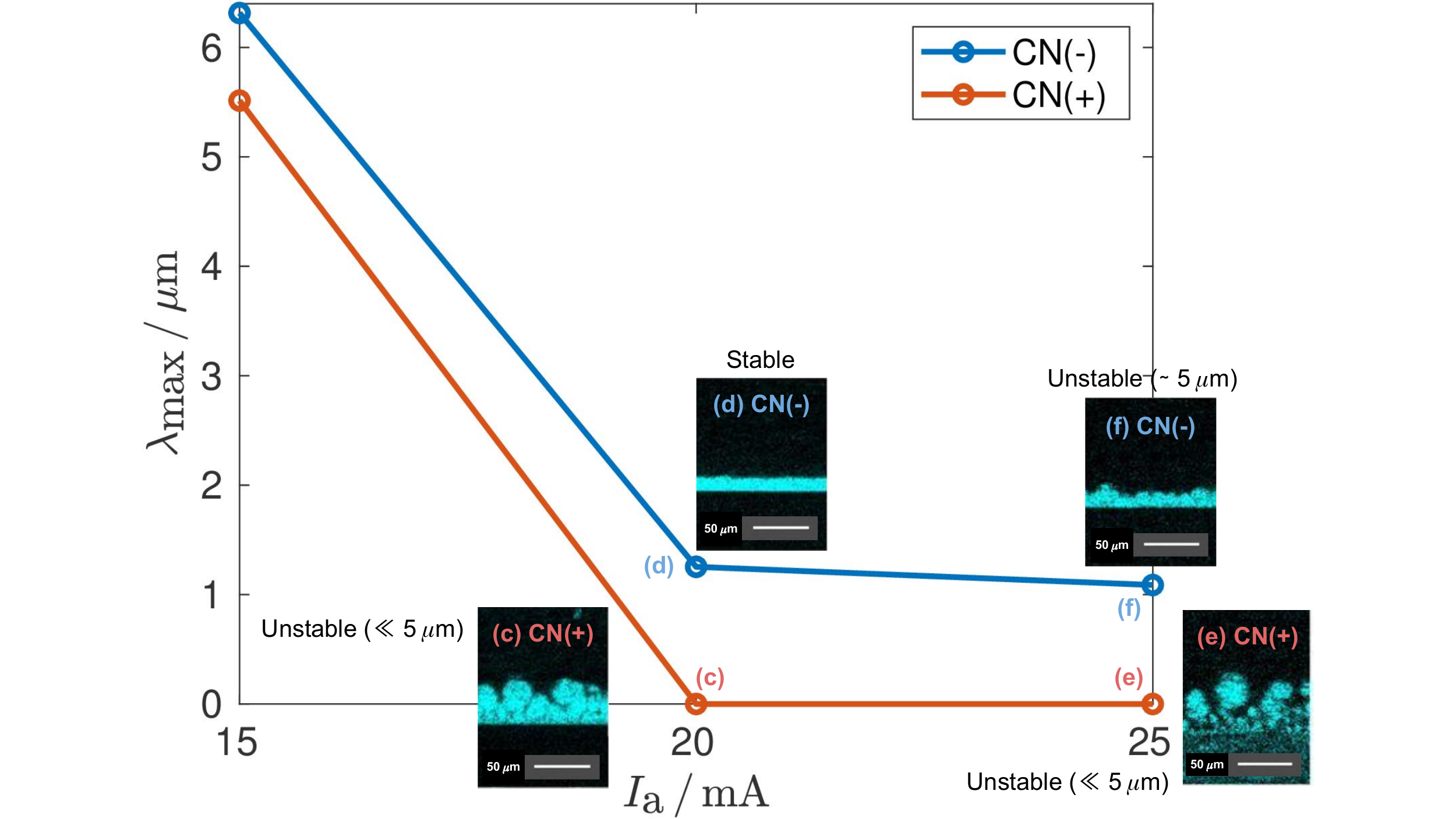}
  \caption{Plots of EDS (energy dispersive X-ray spectroscopy) maps at $2000\,\textn{s}$, and approximate dimensional $\lambda_\textn{c}$ (a) and $\lambda_\maxn$ (b) at steady state against dimensional applied current $I_\textn{a}$ for copper electrodeposition in $100\,\textn{mM}$ copper(II) sulfate ($\textn{CuSO}_4$) electrolyte in cellulose nitrate (CN) membranes. EDS maps in (a) and (b) are magnifications of EDS maps in Figures 6(a) to 6(f) of~\cite{han_dendrite_2016} where the scale bars indicate $50\,\mu\textn{m}$. $h_\textn{c} = 0.5\pm 0.1\,\mu\textn{m}$ in the $\lambda_\textn{c}$ plot in (a) is the characteristic pore size where the mean is indicated by the black dashed line while the lower (LB) and upper (UB) bounds are indicated by gray dashed lines.}\label{fig:Copper}  % chktex 36
\end{figure}

For our analysis here, the specific experimental datasets that we focus on are labeled $\textn{CN}_2(-)$ and $\textn{CN}_2(+)$ in~\cite{khoo_theory_2018}, which correspond to negatively and positively charged CN membranes respectively with a dimensional electrolyte concentration $c_0$ of $100\,\textn{mM}$. We will drop the $2$ subscript for brevity. The morphologies of the electrodeposited copper films, which are visualized by EDS (energy dispersive X-ray spectroscopy) maps, for these $\textn{CN}(-)$ and $\textn{CN}(+)$ membranes at $2000\,\textn{s}$ for dimensional applied currents $I_\textn{a}$ of $15\,\textn{mA}$, $20\,\textn{mA}$ and $25\,\textn{mA}$ are given in Figure~\ref{fig:Copper}(a) that consists of magnifications of EDS maps taken from Figures 6(a) to 6(f) of~\cite{han_dendrite_2016}. At $15\,\textn{mA}$, the copper films for both $\textn{CN}(+)$ and $\textn{CN}(-)$ membranes appear to be uniform and stable. However, at $20\,\textn{mA}$ and $25\,\textn{mA}$, the film for $\textn{CN}(+)$ becomes very unstable and roughens more as the applied current increases. It is difficult to determine quantitatively the instability wavelength using the relatively low resolution EDS maps but it is probably much smaller than $5\,\mu\textn{m}$. In contrast, for $\textn{CN}(-)$, the film still remains uniform and stable at $20\,\textn{mA}$ but slightly destabilizes and roughens at $25\,\textn{mA}$ with an instability wavelength probably on the order of $5\,\mu\textn{m}$. In summary, the onset of overall electrode surface destabilization occurs at $20\,\textn{mA}$ for $\textn{CN}(+)$ with an instability wavelength of much smaller than $5\,\mu\textn{m}$ and at $25\,\textn{mA}$ for $\textn{CN}(-)$ with an instability wavelength of about $5\,\mu\textn{m}$.  % chktex 36

Because the EDS maps are taken at $2000\,\textn{s}$, which is much longer than the diffusion times for $\textn{CN}(-)$ and $\textn{CN}(+)$ of $41.8\,\textn{s}$ and $40.9\,\textn{s}$ respectively, we assume that the system is at steady state. This assumption allows us to use the semi-analytical expressions for the base state variables in~\cite{khoo_theory_2018}, which we have previously discussed in Section~\ref{sec:Base state}, to compute approximate values of $\left \{k_\maxn, \omega_\maxn\right \}$ and $k_\textn{c}$ using Equations~\ref{eq:omega} and~\ref{eq:k_c}. The $\textn{CN}(-)$ dataset has a dimensional limiting current of $18.2\,\textn{mA}$ while the $\textn{CN}(+)$ dataset has a dimensional maximum current, which we have discussed in Section~\ref{sec:Transport}, of $16.9\,\textn{mA}$. Therefore, for $\textn{CN}(-)$, the three applied currents of $15\,\textn{mA}$, $20\,\textn{mA}$ and $25\,\textn{mA}$ correspond to underlimiting, slightly overlimiting and overlimiting currents respectively. On the other hand, for $\textn{CN}(+)$, the model eventually diverges and does not admit a steady state when the applied current $I_\textn{a}$ is above the maximum current $I_\maxn$, therefore we can only obtain finite values of $\left \{k_\maxn, \omega_\maxn\right \}$ and $k_\textn{c}$ for the applied current of $15\,\textn{mA}$ while the model predicts infinite values of $\left \{k_\maxn, \omega_\maxn\right \}$ and $k_\textn{c}$ for the applied currents of $20\,\textn{mA}$ and $25\,\textn{mA}$ due to finite time divergence of the system. Other fitted key dimensionless parameters include $\rho_\textn{s} \approx -0.01$ and $\textn{Da} \approx 2.50$ for $\textn{CN}(-)$ and $\rho_\textn{s} \approx 0.236$ and $\textn{Da} \approx 0.473$ for $\textn{CN}(+)$.

To summarize the model predictions, we plot approximate dimensional values of $\lambda_\textn{c}$ and $\lambda_\maxn$ against the dimensional applied current $I_\textn{a}$ in Figure~\ref{fig:Copper}. In the $\lambda_\textn{c}$ plot in Figure~\ref{fig:Copper}(a), we also indicate the characteristic pore size $h_\textn{c}$ of $0.5\pm 0.1\,\mu\textn{m}$, which is given by twice the pore diameter $d_\textn{p}$ of $250\pm 50\,\textn{nm}$~\cite{han_dendrite_2016}, in order to determine if the model predicts overall electrode surface stabilization. As discussed in Section~\ref{sec:Approximations}, we expect overall electrode surface stabilization if $h_\textn{c} < \lambda_\textn{c}$, which corresponds to the blue shaded region in the $\lambda_\textn{c}$ plot. On the contrary, we expect overall electrode surface destabilization if $h_\textn{c} > \lambda_\textn{c}$, which corresponds to the red shaded region in the $\lambda_\textn{c}$ plot, and the characteristic instability wavelength is $\lambda_\maxn$. Comparing the $\lambda_\textn{c}$ plot with our previous discussion of the onset of overall electrode surface destabilization suggested by the copper film morphologies observed experimentally, we see that the model generally agrees well with experiment; the only disagreement is at $I_\textn{a} = 20\,\textn{mA}$ where the model predicts destabilization for $\textn{CN}(-)$, which has $\lambda_\textn{c} = 0.352\,\mu\textn{m}$, while the EDS map of the copper film at this applied current shows that the film appears to be stable. Nonetheless, this disagreement in theory and experiment is relatively minor since $\lambda_\textn{c} = 0.352\,\mu\textn{m}$ for $\textn{CN}(-)$ at $I_\textn{a} = 20\,\textn{mA}$ is only slightly smaller than the mean of $h_\textn{c}$ of $0.5\,\mu\textn{m}$ and is almost equal to the lower bound of $h_\textn{c}$ of $0.4\,\mu\textn{m}$. In addition, the model predicts $\lambda_\maxn = 0$ (because $k_\maxn \rightarrow \infty$ when $I_\textn{a} > I_\maxn$) at $I_\textn{a} = 20\,\textn{mA}$ and $I_\textn{a} = 25\,\textn{mA}$ for $\textn{CN}(+)$ while it predicts $\lambda_\maxn = 1.09\,\mu\textn{m}$ at $I_\textn{a} = 25\,\textn{mA}$ for $\textn{CN}(-)$. These model predictions of $\lambda_\maxn$ qualitatively agree well with the experimentally observed instability wavelengths at these applied currents that we have previously discussed. Therefore, in conclusion, the theory agrees reasonably well with experimental data, especially given that many assumptions and simplifications are made in the model.  % chktex 36

% \begin{figure}
%   \centering
%   \includegraphics[scale=0.5]{srep28054-f5.jpg}
%   \includegraphics[scale=0.5]{srep28054-f6-cropped.jpeg}
%   \caption{Morphologies of copper films electrodeposited in $100\,\textn{mM}$ copper(II) sulfate ($\textn{CuSO}_4$) electrolyte in cellulose nitrate (CN) membranes at $2000\,\textn{s}$ for different pore surface charges and applied currents visualized by SEM (scanning electron microscopy) images and EDS (energy dispersive X-ray spectroscopy) maps taken from Figures 5 and 6 of~\cite{han_dendrite_2016}. First two rows: SEM images (a, b) and EDS (c, d) maps of copper films electrodeposited in $\textn{CN}(+)$ (a, c) and $\textn{CN}(-)$ (b, d) membranes at $20\,\textn{mA}$; note the scale bars of $50\,\mu\textn{m}$ in the SEM images. Last two rows: EDS maps of copper films electrodeposited in $\textn{CN}(+)$ (a, c, e) and $\textn{CN}(-)$ (b, d, f) membranes at $15\,\textn{mA}$ (a, b), $20\,\textn{mA}$ (c, d) and $25\,\textn{mA}$ (e, f).}\label{fig:Morphologies of electrodeposited copper films}  % chktex 36
% \end{figure}

\subsection{Pulse electroplating and pulse charging}\label{sec:Pulse electroplating and pulse charging}

For many electrochemical applications such as electroplating and charging of metal batteries, which is equivalent to electrodeposition at the metal negative electrode, it is desirable to operate them as quickly as possible at a high current without causing the formation of dendrites that short-circuit the system. To delay or prevent the formation of dendrites, it is common to perform pulse electroplating of metals~\cite{devaraj_pulse_1990,chandrasekar_pulse_2008} or pulse charging of lithium metal batteries (LMBs) and lithium-ion batteries (LIBs)~\cite{li_effects_2001,purushothaman_reducing_2005,purushothaman_rapid_2006,zhang_effect_2006,hussein_review_2011,shen_charging_2012,savoye_impact_2012,mayers_suppression_2012,aryanfar_dynamics_2014} so that there is sufficient time between pulses for the concentration gradients and electric field in the system to relax. For pulse electroplating of metals, it has been empirically observed that the crystal grain size generally decreases with applied current~\cite{devaraj_pulse_1990,chandrasekar_pulse_2008}. Using an applied direct current to perform silver electrodeposition under galvanostatic conditions, Aogaki experimentally observed that the crystal grain size decreases with time~\cite{aogaki_image_1982,aogaki_image_1982-1}, which agrees well with theoretical predictions from linear stability analysis previously done by Aogaki and Makino~\cite{aogaki_theory_1981}. With all these considerations in mind, we apply our linear stability analysis with a time-dependent base state as a tool to investigate how pulse electroplating protocols with high average applied currents, which are inherently time-dependent, affect the linear stability of the electrode surface and the crystal grain size for both zero and negative pore surface charges.

Based on the results in Section~\ref{sec:Application to copper electrodeposition}, we generally expect the characteristic pore size $h_\textn{c}$ to be larger than the critical wavelength $\lambda_\textn{c}$ at high applied currents, therefore the electrode surface is unstable with a characteristic instability wavelength $\lambda_\maxn$. Because a pulse current is applied, $\lambda_\maxn$ varies in time and hence, it would be useful to define an average $\lambda_\maxn$ that averages out the effect of time. In this spirit, we define the average maximum wavenumber $\bar{k}_\maxn$ and the corresponding average maximum wavelength $\bar{\lambda}_\maxn$ as
\begin{align}
  \bar{k}_\maxn &= \frac{\int_0^{t_\textn{f}}k_\maxn\omega_\maxn\ud t}{\int_0^{t_\textn{f}}\omega_\maxn\ud t}, \quad \bar{\lambda}_\maxn = \frac{2\pi}{\bar{k}_\maxn},
\end{align}
where $t_\textn{f}$ is the final time of the pulse and each maximum wavenumber $k_\maxn$ is weighted by its corresponding maximum growth rate $\omega_\maxn$. We expect $\bar{\lambda}_\maxn$ to be on the same order of magnitude as the the crystal grain size that is observed experimentally. As a simple example, we suppose that the pulse electroplating protocol is a periodic pulse wave $J_\textn{a}$ with an ``on'' (charging) time of $\Delta t_\textn{on}$, a ``off'' (relaxation) time of $\Delta t_\textn{off}$, and a period $T$ given by $T = \Delta t_\textn{on} + \Delta t_\textn{off}$. The duty cycle $\gamma_\textn{dc}$ is given by $\gamma_\textn{dc} = \frac{\Delta t_\textn{on}}{T}$ and the average applied current density $\bar{J}_\textn{a}$ over one period is given by $\bar{J}_\textn{a} = J_\textn{a,p}\gamma_\textn{dc}$ where $J_\textn{a,p}$ is the peak applied current density. Hence, for a particular $\bar{J}_\textn{a}$, a smaller $\gamma_\textn{dc}$ implies a larger $J_\textn{a,p}$.

\begin{figure}
  \centering
  \includegraphics[scale=0.41]{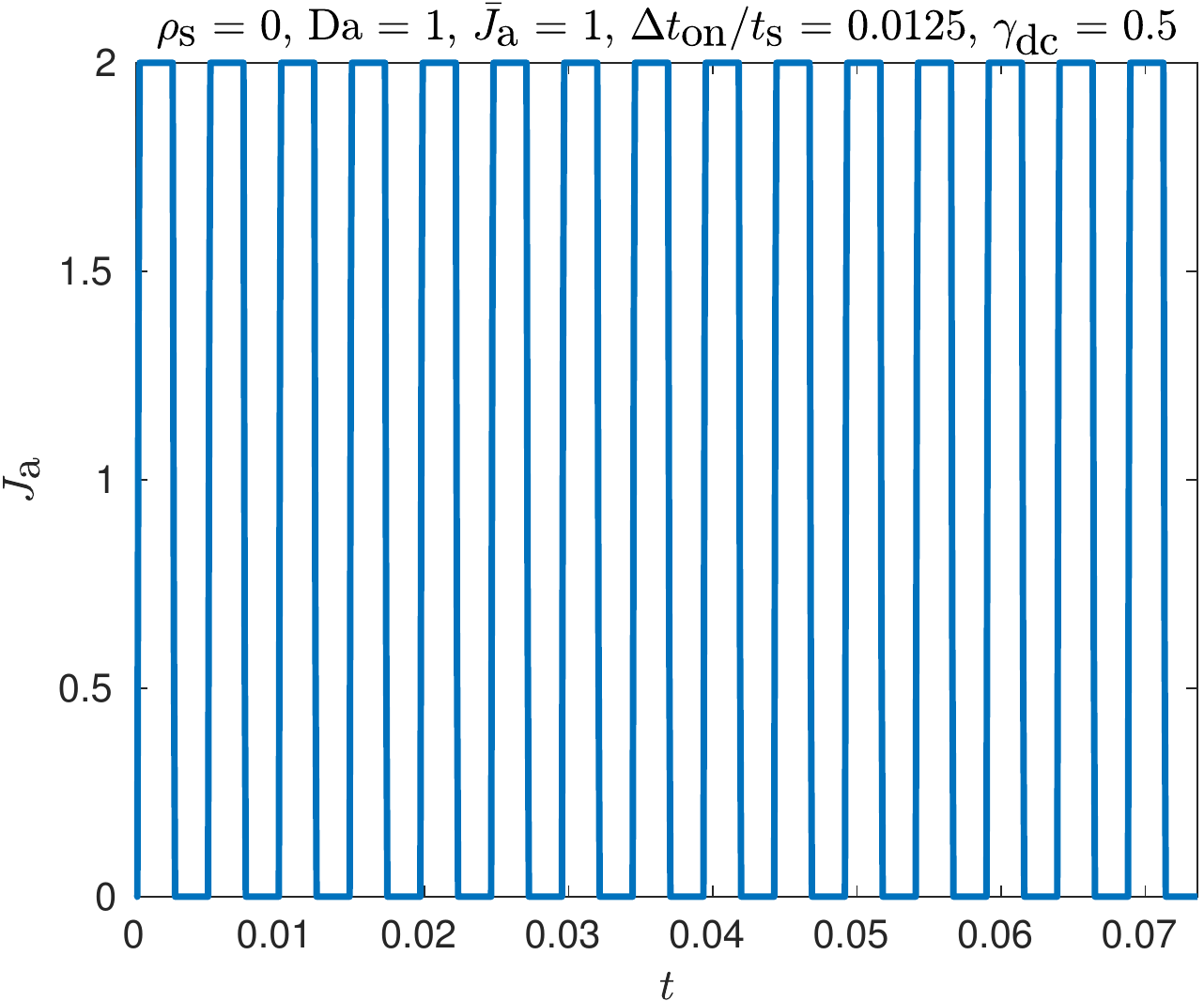}
  \includegraphics[scale=0.41]{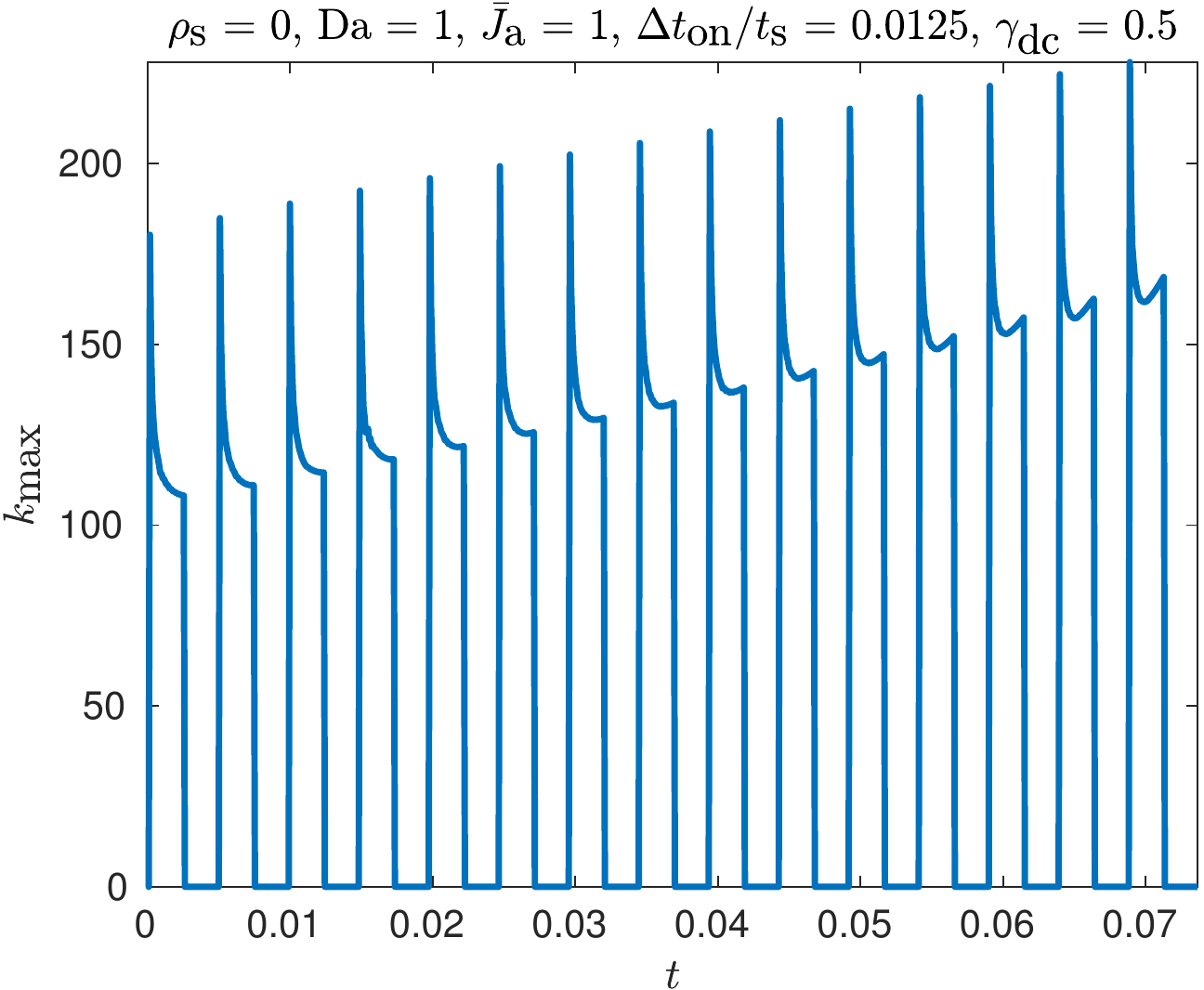}
  \includegraphics[scale=0.41]{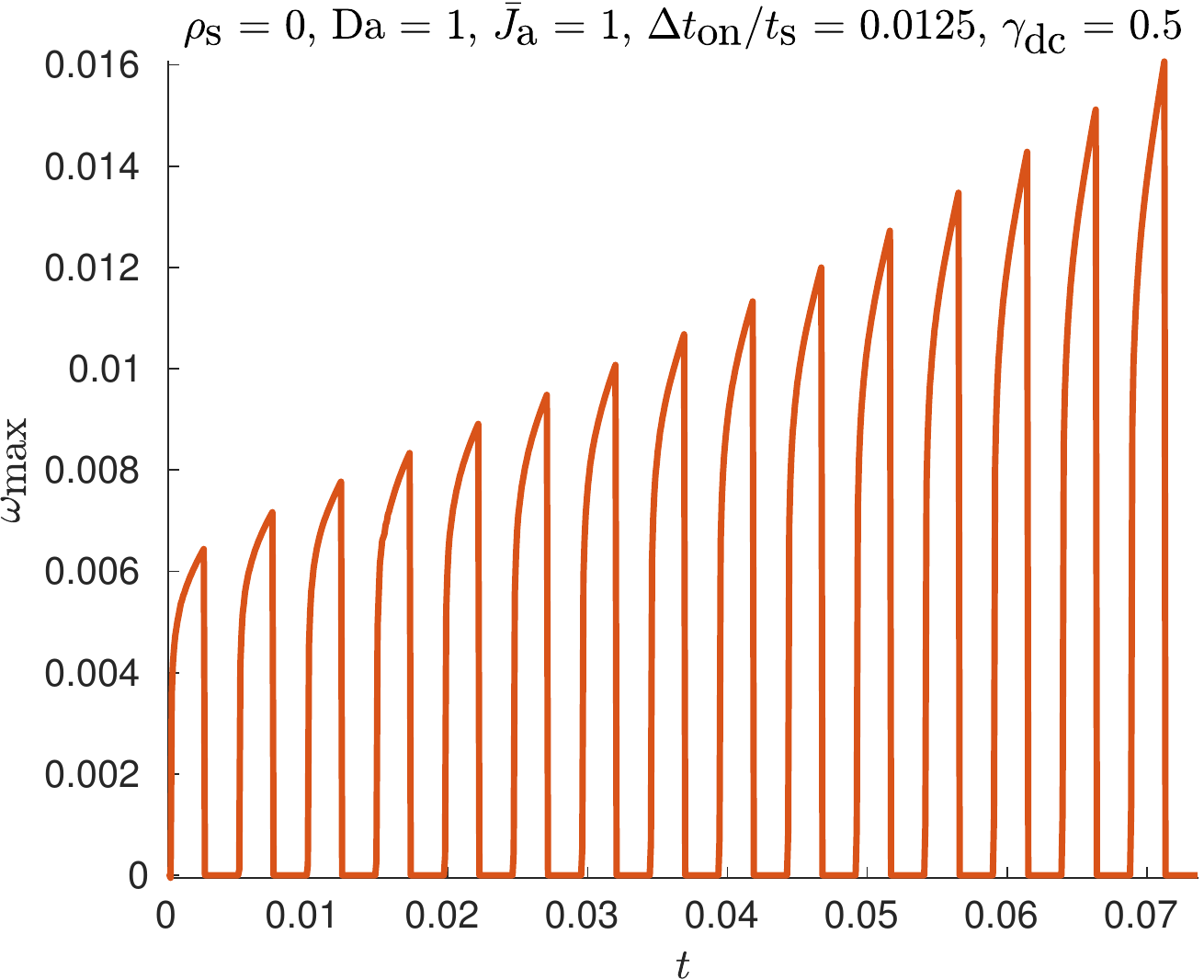}
  \includegraphics[scale=0.41]{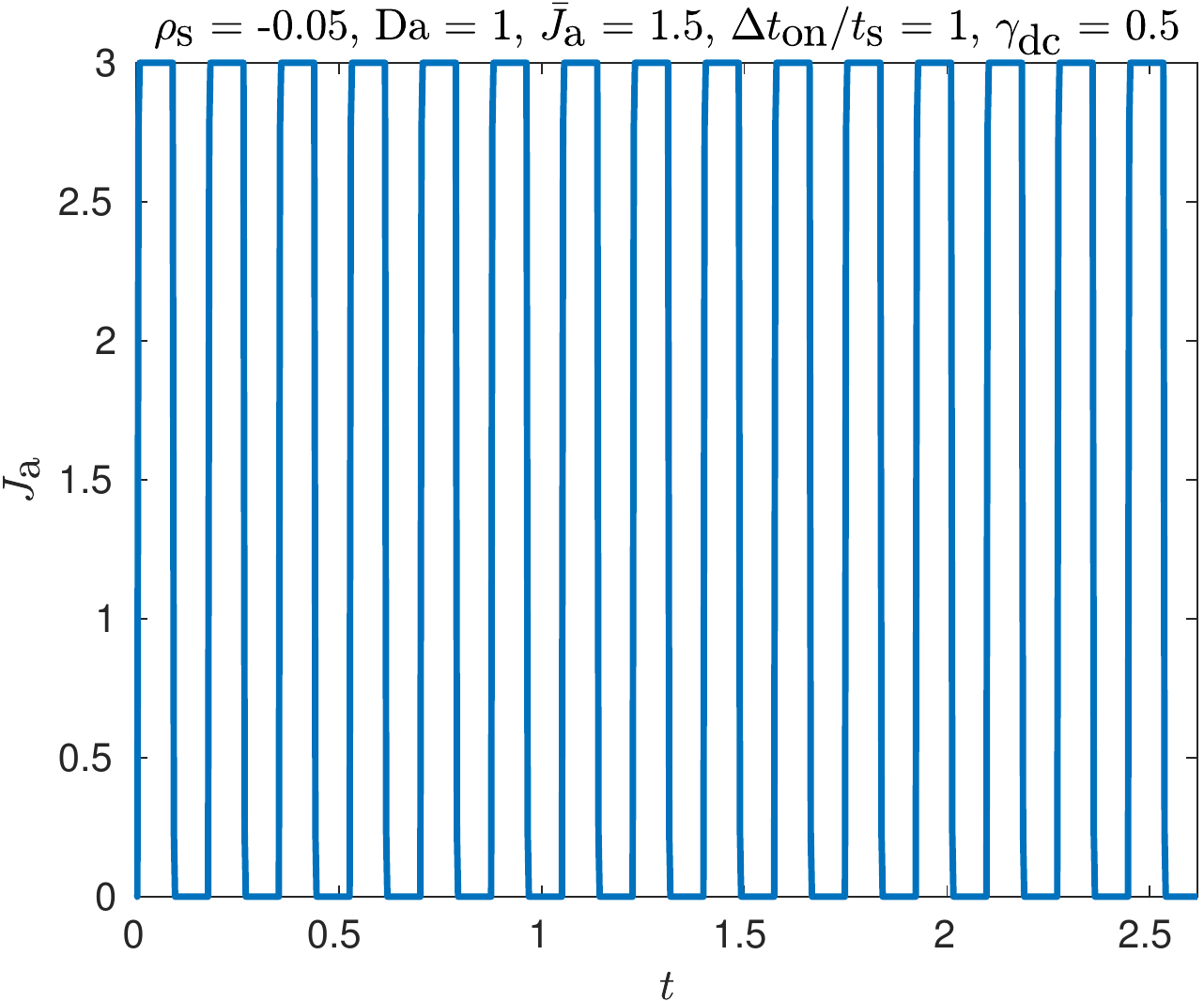}
  \includegraphics[scale=0.41]{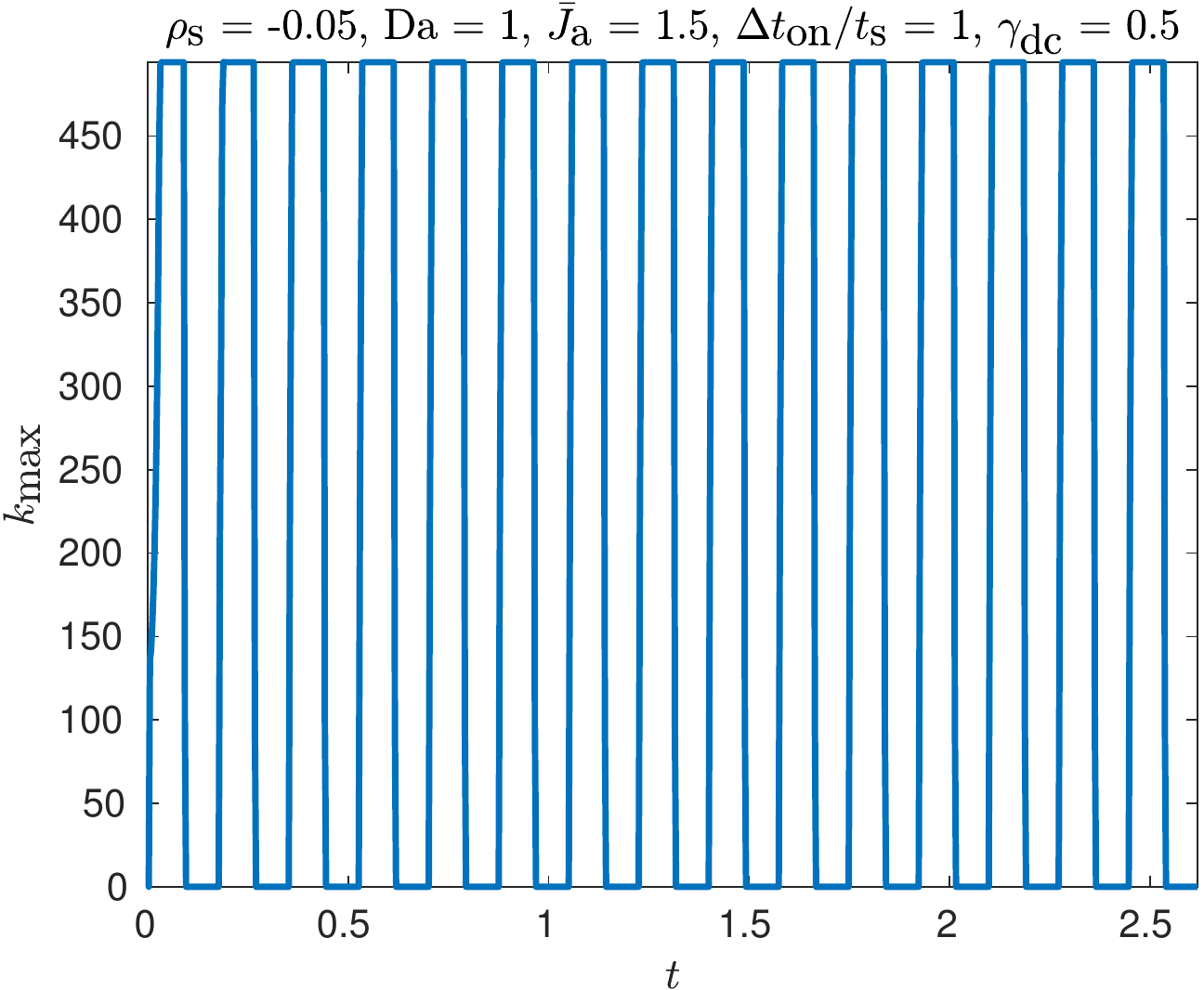}
  \includegraphics[scale=0.41]{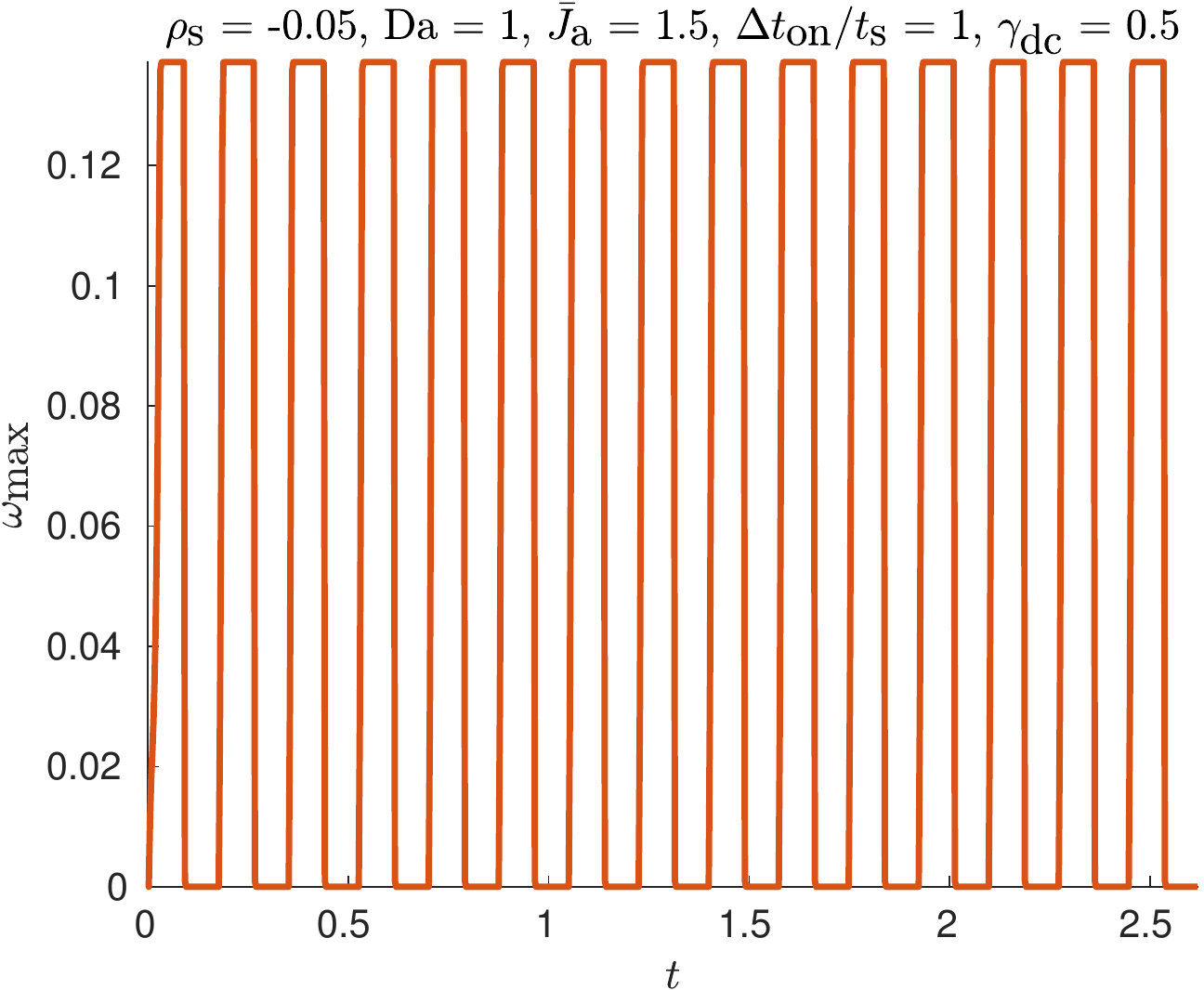}
  \caption{Plots of $J_\textn{a}$, approximate $k_\maxn$ and approximate $\omega_\maxn$ against $t$ with $\textn{Da} = 1$ and $\gamma_\textn{dc} = 0.5$ for pulse electroplating. Top row: $\rho_\textn{s} = 0$, $\bar{J}_\textn{a} = 1$ and $\Delta t_\textn{on} = 0.0125t_\textn{s}$. Bottom row: $\rho_\textn{s} = -0.05$, $\bar{J}_\textn{a} = 1.5$ and $\Delta t_\textn{on} = t_\textn{s}$.}\label{fig:Pulse electroplating}
\end{figure}

\begin{figure}
  \centering
  \includegraphics[scale=0.6]{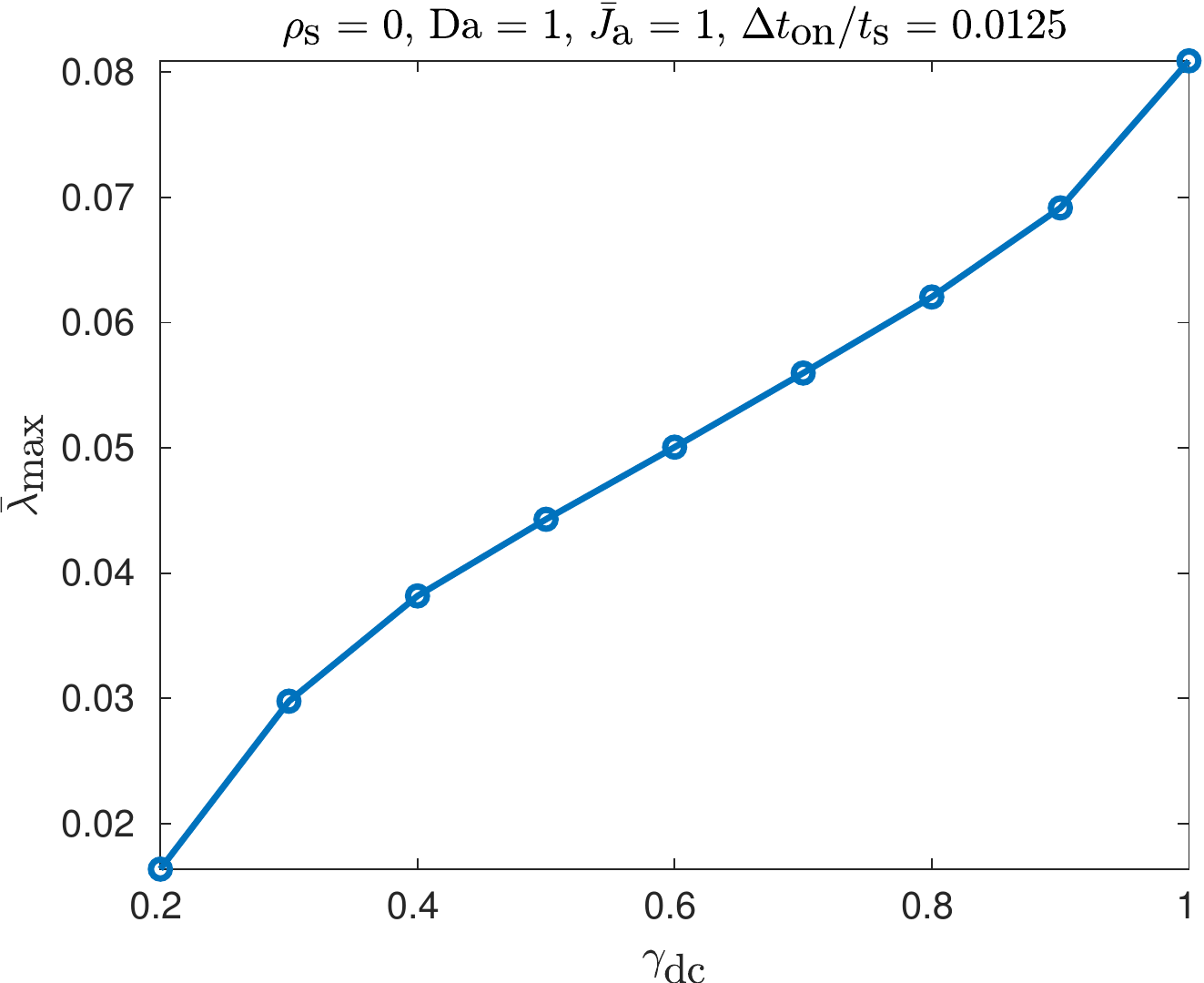}
  \includegraphics[scale=0.6]{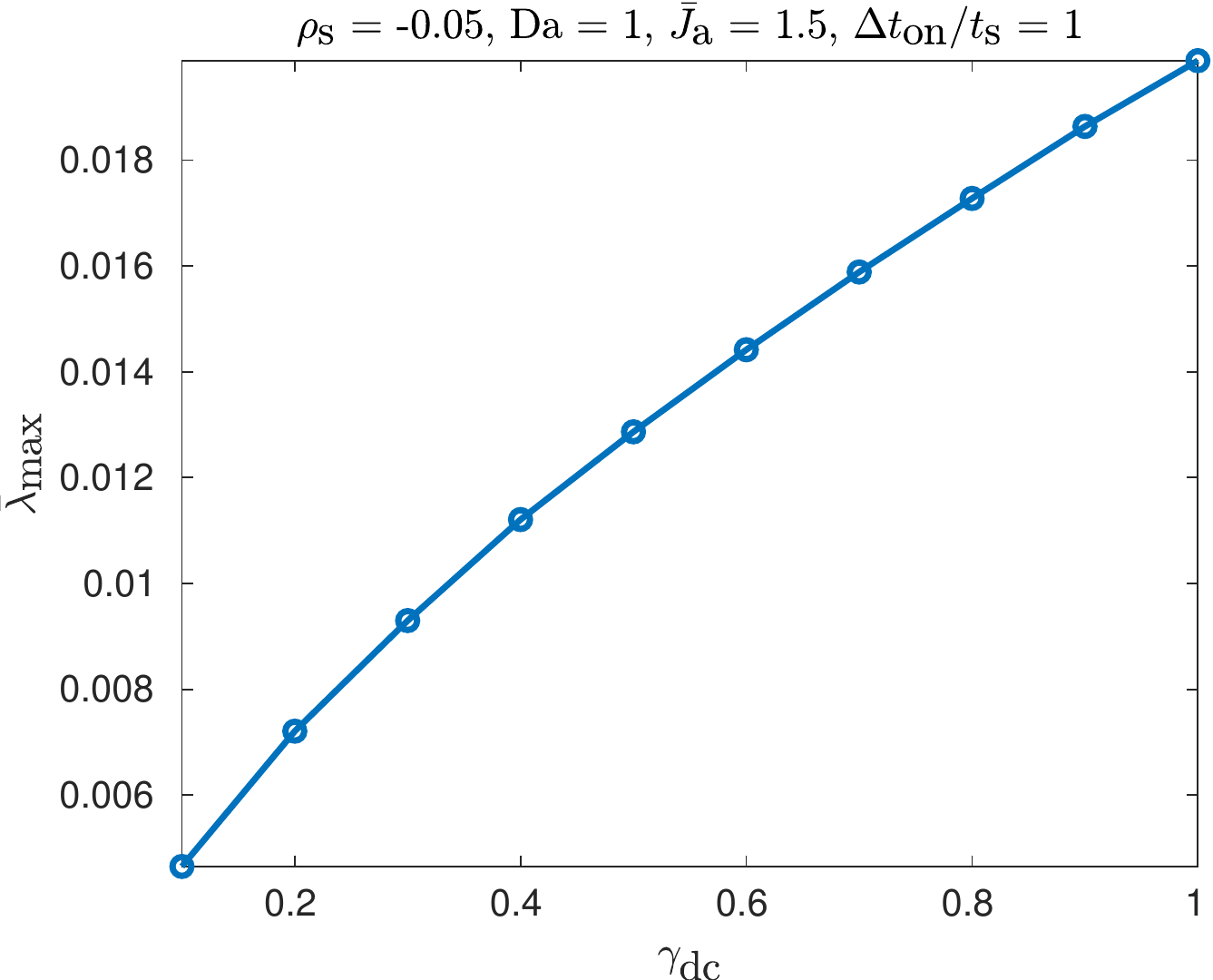}
  \caption{Plots of $\bar{\lambda}_\maxn$ against $\gamma_\textn{dc}$ with $\textn{Da} = 1$. Left: $\rho_\textn{s} = 0$, $\bar{J}_\textn{a} = 1$ and $\Delta t_\textn{on} = 0.0125t_\textn{s}$. Right: $\rho_\textn{s} = -0.05$, $\bar{J}_\textn{a} = 1.5$ and $\Delta t_\textn{on} = t_\textn{s}$.}\label{fig:lambda_max_bar}
\end{figure}

For the classical case of $\rho_\textn{s} = 0$, we pick $\bar{J}_\textn{a} = 1$ and $\Delta t_\textn{on} = 0.0125t_\textn{s}$ and vary $\gamma_\textn{dc}$ from $0.2$ to $1$ (direct current) where the Sand's time $t_\textn{s}$ is calculated based on $\bar{J}_\textn{a}$. $\bar{J}_\textn{a}$, $\Delta t_\textn{on}$ and $\gamma_\textn{dc}$ should be carefully chosen such that $J_\textn{a,p}$ is not too high to deplete the bulk electrolyte at the cathode during the ``on'' cycle so that the system does not diverge at any point in time; this explains why $\gamma_\textn{dc} < 0.2$ for our choice of $\bar{J}_\textn{a} = 1$ and $\Delta t_\textn{on} = 0.0125t_\textn{s}$ cannot be numerically simulated. For $\rho_\textn{s} = -0.05$, we pick $\bar{J}_\textn{a} = 1.5$ and $\Delta t_\textn{on} = t_\textn{s}$ and vary $\gamma_\textn{dc}$ from $0.1$ to $1$ (direct current) to drive the system at an overlimiting average applied current density. We also fix $\textn{Da} = 1$ for both cases and use Equations~\ref{eq:omega} and~\ref{eq:k_c} to compute approximate values of $k_\maxn$ and $\omega_\maxn$. For these choices of parameters, as an illustrative example, we plot $J_\textn{a}$, approximate $k_\maxn$ and approximate $\omega_\maxn$ against $t$ for $\gamma_\textn{dc} = 0.5$ in Figure~\ref{fig:Pulse electroplating}. We note that the large overshoot in $k_\maxn$ at the beginning of each ``on'' cycle for $\rho_\textn{s} = 0$ is caused by the sharp rate of increase of the concentration gradients and electric field as $J_\textn{a}$ rapidly increases from $0$ in the ``off'' cycle to $J_\textn{a,p}$ in the ``on'' cycle. Corresponding to these pulse waves, we plot $\bar{\lambda}_\maxn$ against $\gamma_\textn{dc}$ in Figure~\ref{fig:lambda_max_bar}. For both $\rho_\textn{s} = 0$ and $\rho_\textn{s} = -0.05$, $\bar{\lambda}_\maxn$ increases with $\gamma_\textn{dc}$, which agrees with the empirical observation that the crystal grain size generally decreases with applied current~\cite{devaraj_pulse_1990,chandrasekar_pulse_2008}. The ability to experimentally impart a negative pore surface charge to the nanoporous medium therefore enables pulse electroplating at overlimiting currents for electrodepositing a large amount of charge at a high rate and tuning the desired crystal grain size.

\section{Conclusion}

We have derived the full model that couples the leaky membrane model for ion transport, which is capable of predicting overlimiting current due to surface conduction, with Butler-Volmer reaction kinetics, which describes the metal electrodeposition reaction, and performed linear stability analysis on it with respect to a time-dependent base state. The volume-averaged background charge density can generally be of any sign. As a result, we have generalized previous work on linear stability analysis of electrodeposition carried out in~\cite{sundstrom_morphological_1995,elezgaray_linear_1998,tikekar_stability_2014}. We then performed a boundary layer analysis on the perturbed state in order to derive an accurate approximation for the dispersion relation and a convergence analysis to verify the accuracy and convergence of the full numerical solution of the dispersion relation. By performing parameter sweeps over the volume-averaged background charge density, Damk\"ohler number and applied current density under galvanostatic conditions, we have concluded that a negative background charge significantly stabilizes the electrode surface instability, although it does not completely stabilize it, while a positive background charge further destabilizes this instability. We have also verified that the approximations for the maximum wavenumber, maximum growth rate and and critical wavenumber are very accurate, and applied them to demonstrate good agreement between theory and experimental data for copper electrodeposition in cellulose nitrate membranes~\cite{han_dendrite_2016}. Lastly, we have employed the linear stability analysis as a tool to analyze the dependence of the crystal grain size on duty cycle in pulse electroplating. These results demonstrate the predictive power and robustness of the theory despite its simplicity. Although detailed analysis of the Poisson-Nernst-Planck-Stokes equations for transport in a microchannel by Nielsen and Bruus~\cite{nielsen_concentration_2014} reveals that the leaky membrane model for surface conduction is at best a rough approximation of the real system, the good agreement between theory and experiment that we have demonstrated suggests that the model is applicable in similar electrochemical systems using charged membranes such as shock electrodeposition for information storage applications~\cite{han_resistive_2016} and shock electrodialysis for water treatment~\cite{deng_overlimiting_2013,schlumpberger_scalable_2015,schlumpberger_shock_2016}.

We have made many assumptions and simplifications in the model presented, and relaxing some of them offers opportunities for extending it in useful ways. First, we have ignored surface adsorption, surface diffusion of adsorbed species~\cite{aogaki_morphological_1984,aogaki_morphological_1984-1,aogaki_morphological_1984-2,aogaki_morphological_1984-3} and additional mechanical effects such as pressure, viscous stress and deformational stress~\cite{monroe_effect_2004,monroe_impact_2005,tikekar_stabilizing_2016,ahmad_stability_2017,ahmad_role_2017,natsiavas_effect_2016}, which confer additional stabilization to the electrode surface. Adding these physics and chemistry to the model are likely to result in finite values of the maximum wavenumber, maximum growth rate and critical wavenumber near and at Sand's time under an overlimiting current for zero and positive background charges respectively, as opposed to diverging in our current model. The inclusion of these additional mechanical effects will also extend the applicability of the model to solid electrolytes~\cite{liu_soft_2017} that are used in solid state batteries. Deformation of the porous medium caused by metal growth inside the pores also results in a porosity that varies in both time and space whose effects would be interesting to study. Second, in order to apply the linear stability analysis to lithium metal batteries (LMBs), we would also need to model the solid electrolyte interphase (SEI) layer~\cite{verma_review_2010,cheng_review_2016,peled_reviewsei:_2017,wang_review_2018}, which will certainly increase the complexity of the model but also make it more predictive. Incorporating these two aforementioned extensions into the model may help explain recent experimental studies of lithium growth that have demonstrated that competing SEI reactions and stress effects lead to root growth before Sand's time or below limiting current~\cite{bai_transition_2016,kushima_liquid_2017,bai_interactions_2018}, which is different from tip growth of dendrites under transport limitation that we have focused on in this paper. Third, other chemical mechanisms for overlimiting current such as water splitting~\cite{nikonenko_intensive_2010,nikonenko_desalination_2014} and current-induced membrane discharge~\cite{andersen_current-induced_2012} may be present. These effects are typically highly nonlinear and therefore, we expect them to significantly influence the transient base state and linear stability analysis. Fourth, we should consider the effects of coupling nucleation, which is fundamentally a nonlinear instability unlike spinodal decomposition that is a linear instability, to the current model. Specifically, nucleation may affect the transient base state during initial and early reaction-limited surface growth and create surface roughness on the scale of the characteristic nucleus size, which may in turn influence overall electrode surface stabilization or destabilization when the system reaches transport limitation near or at Sand's time. Fifth, an interesting and useful generalization of the reaction model would be to use the symmetric Marcus-Hush-Chidsey kinetics~\cite{bai_charge_2014,zeng_simple_2014} or asymmetric Marcus-Hush kinetics~\cite{zeng_simple_2015} instead of Butler-Volmer kinetics for modeling electron transfer reactions, which would afford us the reorganization energy as a key system parameter whose influence on the linear stability of the electrode surface can be investigated.

\begin{acknowledgments}

  E. Khoo acknowledges support from the National Science Scholarship (PhD) funded by Agency for Science, Technology and Research, Singapore (A*STAR). H. Zhao and M. Z. Bazant acknowledge support from the Toyota Research Institute through the D3BATT Center on Data-Driven-Design of Rechargeable Batteries.

\end{acknowledgments}

\newpage

\bibliography{bibliography_bibtex}

\end{document}